\pdfoutput=1 

\documentclass[11pt,twoside]{report}
\usepackage{Thesis}
\usepackage{ae,amssymb,amsthm,amsfonts,amsmath}
\usepackage[spanish]{babel}
\usepackage{cancel}
\usepackage[Lenny]{fncychap}
\usepackage{drop}
\usepackage{enumerate}
\usepackage[footnotesize,bf]{caption2}

\begin{document}

\dosjurados
\decimalpoint

\renewcommand{\thetable}{\thechapter.\arabic{table}}%
\renewcommand{\contentsname}{Tabla de contenido}%
\renewcommand{\listtablename}{Índice de tablas}%
\renewcommand{\theequation}{\thechapter.\arabic{equation}}%
\renewcommand{\tablename}{Tabla}

\emblema{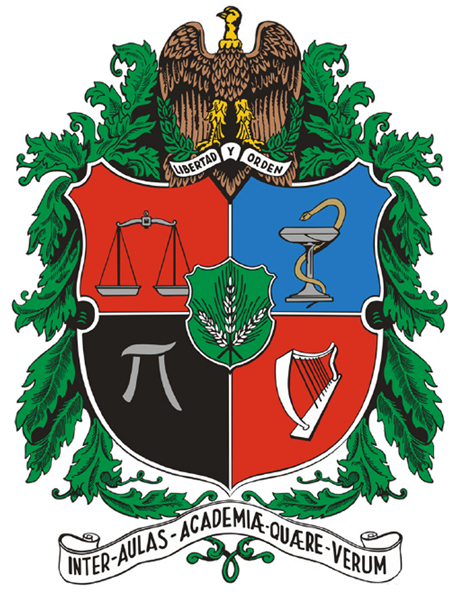}{0.44}

\thispagestyle{empty}%
\vspace*{5cm}%

\hfill
\begin{minipage}{15cm}
\setlinespacing{1}
    \textsl{\hspace{1cm} Nada es imposible, no si puedes imaginarlo, eso es lo que implica ser un científico.}

   \hfill\textsc{Prof. Hubert Farnsworth, Futurama.\\\\\\}

\end{minipage}

\thispagestyle{empty}%
\vspace*{7cm}

{\setlinespacing{2.0}
\begin{center}
    \fontdedicatoria{A la memoria de mi hermano Cristian.}
\end{center}
}

\newpage

{\color{white} . }
\nonumchapter{Resumen}

\noindent La actual expansión acelerada del universo sigue siendo un gran misterio para la física. Varios modelos se han propuesto para su explicación, entre ellos el de la energía oscura, que sin embargo posee problemas tanto de carácter teórico como experimental. Sin embargo, otras aproximaciones, como la de las teorías de gravedad modificada son una interesante alternativa para resolver este problema.
Motivados en este enfoque realizamos un estudio de los modelos cosmológicos en las teorías $f(R)$ que son extensiones naturales de la Relatividad General con funciones arbitrarias del escalar de Ricci. Un capítulo esta dedicado a la obtención de las ecuaciones de campo en el formalismo métrico de las teorías $f(R)$, incluyendo la discusión sobre los términos de frontera. Aplicamos después tales ecuaciones para escribir las ecuaciones dinámicas del universo, aceptando para esto la validez del modelo estándar de la Cosmología, es decir, usando como espacio-tiempo el de Robertson-Walker. La aproximación de sistema dinámico nos permite encontrar criterios de viabilidad para las funciones $f(R)$, exigiéndoles que cumplan con una transición de época de dominio de materia, a una de expansión acelerada. Centramos gran atención en el problema de las distancias cosmológicas en teorías $f(R)$, obteniendo expresiones generales a partir del estudio de la Ecuación de Desvío Geodésico. 
\nonumchapter{Abstract}

\noindent The actual accelerated expansion of the universe continues being a mystery in physics. Some
models had been proposed for this explanations, among them the dark energy, which however
has problems of experimental character as well as theoretical. Other approximations, like modified
gravity theories are an interesting alternative for this problem. Motivated in this approach
we study cosmological models in $f(R)$ theories which are natural extension of General Relativity
with arbitrary functions of the Ricci scalar. One chapter has dedicated to obtain the modified
field equations in the metric formalism of $f(R)$ theories, including the discussion about boundary
terms in the action. Later, we apply these equations in order to describe the dynamics of the
universe, using for this as space-time, the Robertson-Walker metric. The Dynamical system approach
for this equations allow us to find viability criteria for the functions $f(R)$, demanding them
to satisfy a transition between a mater dominated epoch to an accelerated dominating epoch.
We focus our study in the problem of cosmological distances in $f(R)$ theories, obtaining general
expressions from the Geodesic Deviation Equation (GDE).

\nonumchapter{Agradecimientos}%

\noindent En primer lugar quiero agradecer a la Universidad
Nacional de Colombia, sede Bogotá y en especial al Observatorio Astronómico
Nacional, por haberme permitido realizar mis estudios de maestría.
Agradezco al profesor Juan Manuel Tejeiro por darme herramientas necesarias para comenzar mi labor investigativa; también por su apoyo
para la financiación de eventos académicos. También agradezco el apoyo financiero del Programa de Becas Para Estudiantes Sobresalientes de Posgrado de la Universidad Nacional de Colombia.\\

\noindent Agradezco muy especialmente al profesor Leonardo Castañeda Colorado por haberme
aceptado como estudiante y por aceptar dirigir mi tesis. También por sus incontables ayudas, sus consejos y por poner a disposición su gran conocimiento para llevar a cabo la investigación; más que un director se ha convertido en un gran amigo. Quiero agradecer especialmente a todos los miembros del \textit{Grupo de Gravitación y Cosmología}, por sus consejos y por seguir cada semana el proceso de la elaboración de esta tesis; sus ideas, comentarios y bromas hicieron más ameno el desarrollo del trabajo.\\

\noindent Agradezco a mi familia: Mi madre María Eva y a mis hermanos Cristian (Q.E.P.D) y Miguel, quienes me motivaron y brindaron su apoyo incondicional durante toda mi carrera; su compañía y soporte ha sido fundamental para mi, y se convirtieron en el motor que permitió desarrollar esta tesis. A mi mis amigos Pablo, Oswaldo y Wilmar, con quienes he compartido grandes momentos y me han alentado en todos los aspectos.\\

\noindent No alcanzo a nombrar a todas las personas que se cruzaron en este proceso y que de alguna u otra forma aportaron para el desarrollo de este trabajo.
\textit{A todos mil gracias por ser eventos causalmente conectados con mi vida!}.\\

 \vfill\vfill
\noindent Bogot\'a D.C.\hfill \makeatletter \@autor


\nonumchapter{Notación y Convenciones}%

\noindent La convención de signos usada para la métrica es $(-,+,+,+)$, la velocidad de la luz se toma en unidades geometrizadas $c=1$. Los índices griegos $\alpha,\beta, \ldots$  se utilizan para las coordenadas espacio-temporales y van de 0 a 3, los índices latinos $a,b,\ldots$ se utilizan únicamente para las coordenadas espaciales y van de 1 a 3. Letras en negrilla $\mathbf{A}, \mathbf{B}, \ldots$ son utilizados para denotar tensores. \\\\

\noindent
\renewcommand{\arraystretch}{1.1}
\begin{tabular}{ll}
$G$ \hspace{2cm} & Constante de Gravitación Universal ($6.67 \times 10^{-11}$ m$^{2}/$kg s$^{2}$) \\
$\kappa$ \hspace{2cm} & $8\pi G$ \\
$g_{\alpha\beta}$ \hspace{2cm} & Componentes del tensor métrico \\
$g$ \hspace{2cm} & Determinante del tensor métrico \\
$R_{\alpha\beta\gamma\delta}$ \hspace{2cm} & Tensor de Riemann \\
$R_{\alpha\beta}$ \hspace{2cm} & Tensor de Ricci \\
$R$ \hspace{2cm}  & Escalar de Ricci \\
$\Gamma_{\beta\gamma}^{\alpha}$ \hspace{2cm} & Símbolo de Christoffel\\
$G_{\alpha\beta}$ \hspace{2cm} & Tensor de Einstein \\
$\partial_{\alpha}$ \hspace{2cm} & Derivada parcial con respecto a $x^{\alpha}$\\
$\nabla_{\alpha}$ \hspace{2cm} & Derivada covariante\\
$A_{,B}$ \hspace{2cm} & Derivada de $A$ con respecto a $B$\\
$\square$ \hspace{2cm} & d'Alambertiano $\square = \nabla^{\gamma}\nabla_{\gamma}$\\
$\frac{D}{D\nu}$ \hspace{2cm} & Derivada covariante a lo largo de la curva\\
$A_{(\alpha\beta)}$ \hspace{2cm} & Simetrización\\
$A_{[\alpha\beta]}$ \hspace{2cm} & Antisimetrización\\
$\dot{\quad}$ \hspace{2cm} & Derivada temporal\\
$T_{\alpha\beta}$ \hspace{2cm} & Tensor de Energía-Momentum\\
$a(t)$ \hspace{2cm} & Factor de escala \\
$t$ \hspace{2cm} & Tiempo cósmico \\
$z$ \hspace{2cm} & Redshift\\
$H$ \hspace{2cm} & Parámetro de Hubble \\
$k$ \hspace{2cm} & Curvatura espacial del universo\\
\end{tabular}

\newpage
\noindent
\begin{tabular}{ll}
$\rho$ \hspace{2cm} & Densidad de energía\\
$p$ \hspace{2cm} & Presión\\
$w$ \hspace{2cm} & Ecuación de estado $w = p/\rho$\\
$\Lambda$ \hspace{2cm} & Constante cosmológica \\
$d_L$ \hspace{2cm} & Distancia de luminosidad\\
$d_A$ \hspace{2cm} & Distancia diametral angular\\
\end{tabular} 

\nonumchapter{Contribuciones}%

\noindent La investigación presentada en esta tesis fue realizada en el Observatorio Astronómico Nacional de la Universidad Nacional de Colombia, entre Febrero del 2008 a Noviembre del 2010. El contenido de esta tesis ha sido objeto de las dos siguientes publicaciones:
\begin{enumerate}
\item A. Guarnizo, L. Castañeda y J. M Tejeiro. ``Boundary term in metric $f(R)$ gravity:
field equations in the metric formalism''. \textit{Gen. Rel. Grav.} \textbf{42}: 2713-2728 (2010). Preprint en [\href{http://lanl.arxiv.org/abs/1002.0617}{\tt{arXiv:1002.0617}}]. \href{http://www.springerlink.com/content/c756521422207051/}{DOI:10.1007/s10714-010-1012-6}
\item A. Guarnizo, L. Castañeda y J. M Tejeiro. ``Geodesic deviation equation in $f(R)$ gravity''. \textit{Gen. Rel. Grav.} \textbf{43}: 2713-2728 (2011). Preprint en [\href{http://lanl.arxiv.org/abs/1010.5279}{\tt{arXiv:1010.5279}}]. \href{http://www.springerlink.com/content/w41p6g7504x143t3/}{DOI:10.1007/s10714-011-1194-6}
\end{enumerate}
y también presentada en los siguientes eventos:\\\\
2010. \textit{\textbf{Métodos del Análisis Funcional en Relatividad General y Mecánica
Cuántica}}, Universidad de los Andes, Bogotá, Colombia, Mayo 31 a Julio 4 del
2010.\\
Título: \textit{Término de Frontera en Teorías de Gravedad Modificada $f(R)$: Ecuaciones de
campo en el Formalismo Métrico}.\\\\
2010. \textit{\textbf{19th International Conference on General Relativity and Gravitation
(GR19)}}, Universidad Nacional Autónoma de México, México, Ciudad de México,
Julio 5 al 9 del 2010.\\
Titulo: \textit{Non-vacumm Axially Symmetric Solutions in $f(R)$ Gravity}.\\\\
2010. \textit{\textbf{Tercera Reunión Colombo-Venezolana de Relatividad, Campos y Gravitación}}, Universidad Industrial de Santander, Curití, Santander, Colombia, Diciembre 1 al 3 de 2010.\\
Titulo: \textit{Ecuación de Desvío Geodésico en Teorías $f(R)$}.

\logograv{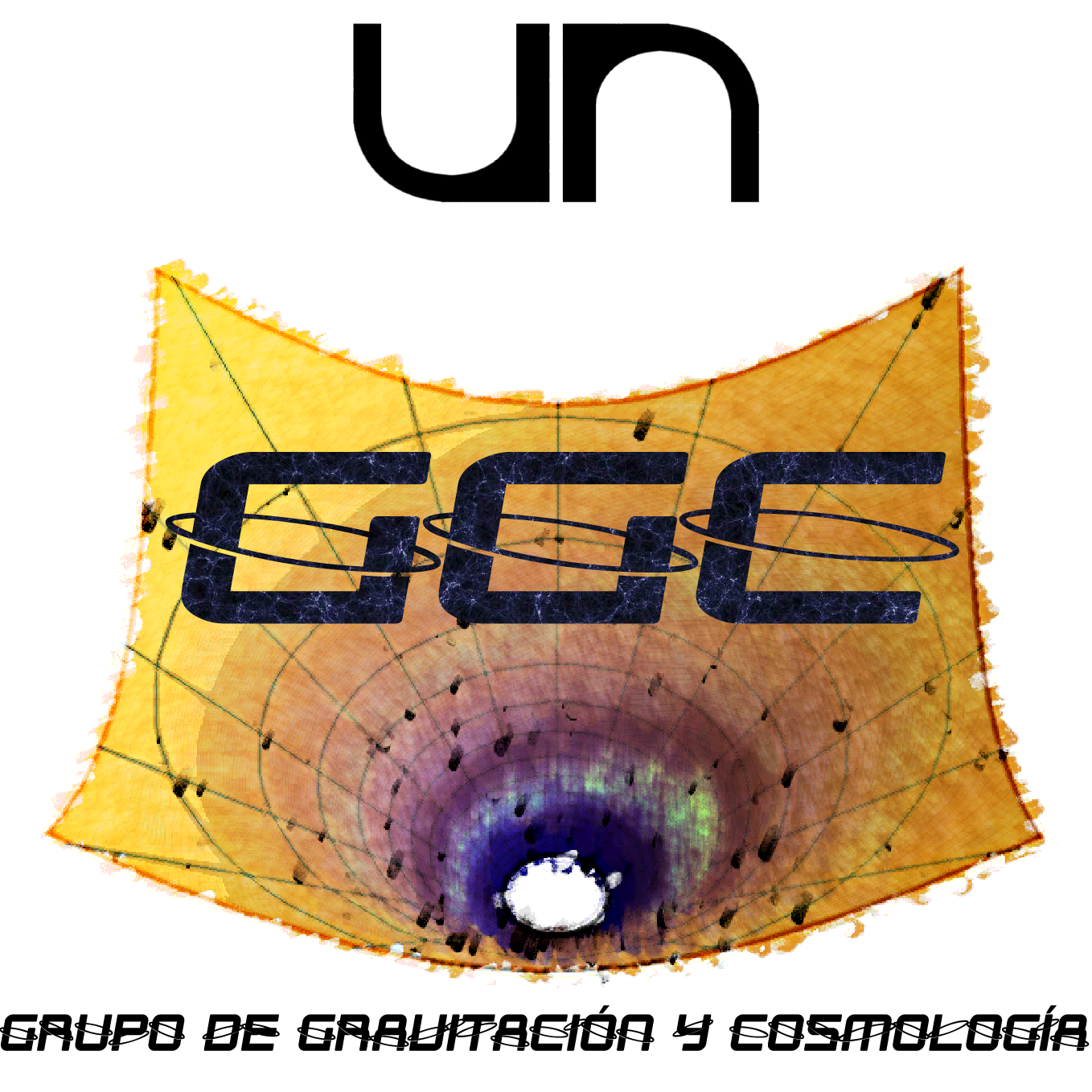}{0.98}

\codigo{189473}
\autor{Alejandro Guarnizo Trilleras}%
\titulo{MODELOS COSMOLÓGICOS EN TEORÍAS DE GRAVEDAD MODIFICADA $\boldsymbol{f(R)}$}%
\universidad{Universidad Nacional de Colombia}%
\direccion{Bogotá, Colombia}%
\departamento{Observatorio Astronómico Nacional}%
\facultad{Facultad de Ciencias}%
\director{Leonardo Castañeda Colorado}%
\fechafinal{11 de Abril de $2011$}%
\theyear{$2011$}%
\gradoautor{Magíster en Ciencias - Astronomía}%
\gradodirector{}%
\gradocodirector{}%
\primerjurado{Juan Manuel Tejeiro Sarmiento}
\segundojurado{Guillermo Alfonso González Villegas}
\cuerpoinicial{1.3}%

\cuerpoprincipal{1.2}%


\chapter{Prefacio}

\drop{D}esde el nacimiento de la Relatividad General, que se ha proclamado como la teoría de gravitación con más aceptación actualmente, se ha buscado una solución que pueda describir la evolución del universo. El modelo estándar de la cosmología esta basado en dos principios fundamentales que nos dicen que a gran escala el universo es estadísticamente homogéneo e isotrópico. La solución de las ecuaciones de campo de Einstein para un universo que satisface estas condiciones nos conducen a la solución de Friedmann-Lama\^{i}tre-Robertson-Walker (FLRW) \cite{Weinberg},\cite{Weinberg2}.\\\\
Einstein intento conciliar el modelo de un universo dinámico introduciendo  \textit{constante cosmológica} $\Lambda$ con el fin de tener un universo estático. Sin embargo, desde las observaciones realizadas por Hubble en 1929 que mostraban que las galaxias se alejaban más rápido en cuanto mas lejos estuvieran, mostrando así un universo en expansión, los cosmólogos y la comunidad científica adoptaban cada vez más la idea de un universo dinámico. Einstein considero entonces la introducción de la constante cosmológica como el error más grande de su vida. \\\\
Hoy en día sabemos que nuestro universo además de expandirse, lo hace de forma acelerada, esto gracias al análisis de los parámetros cosmológicos a partir del estudio de Supernovas de tipo Ia (Sn Ia) \cite{Perlmutter}. Sin embargo no se ha encontrado una explicación satisfactoria para este hecho. Por otro lado las observaciones del proyecto  Wilkinson Microwave Anisotropy Probe (WMAP), en las anisotropías de la Radiación Cósmica de Fondo (CMB), nos predicen que el universo esta compuesto en un 74\% de una misteriosa forma de energía, de la cual no conocemos su naturaleza, y un restante 26\% de materia compuesta en un 22\% por materia oscura y el restante 4\% en materia bariónica. Este tipo de energía lo denominamos \textit{energía oscura}. Se pueden proponer 3 modelos para explicar la expansión acelerada del universo: 1. Retomar la idea de la constante cosmológica $\Lambda$, 2. Energía oscura, 3. Gravedad modificada \cite{Lobo}-\cite{Nojiri2}. \\\\
El modelo de la constante cosmológica provee una explicación para la expansión del universo, aunque posee dos problemas que han llevado a la comunidad científica a buscar nuevas alternativas. A partir de los datos observacionales se puede estimar un valor para la constante cosmológica, el cual difiere con valores obtenidos en teorías cuánticas de campos en 120 órdenes de magnitud; este problema es conocido como \textit{problema de la constante cosmológica}. Otro problema fundamental es el denominado \textit{problema de coincidencia}, y es el por qué actualmente la densidad de energía de la constante cosmológica es comparable a la de la materia \cite{Carroll1}.\\\\
Los modelos de energía oscura vienen de suponer una ecuación de estado de la forma $w \equiv p/\rho$, en donde $p$ es la presión y $\rho$ la densidad de energía. Para un valor de $w < -1/3$ se obtiene una expansión acelerada del universo, como lo muestran las ecuaciones de Friedmann; el caso $w = -1$ corresponde a un fluido de densidad constante que puede asociarse con la constante cosmológica \cite{Sami}. Los modelos en los cuales $-1< w < -1/3$ se denominan de \textit{Quintaesencia} \cite{Capozziello2}. \\\\
La tercera posibilidad consiste en ver los términos adicionales en las ecuaciones de campo de Einstein, que dan la expansión acelerada del universo,  a modificaciones de la gravedad. Una de estas formulaciones son las denominadas \textit{teorías de gravedad modificada $f(R)$} que surgen de una generalización de la acción de Einstein-Hilbert con términos adicionales del escalar de curvatura $R$  \cite{Felice}-\cite{Capozziello}. Existen tres versiones de la gravedad $f(R)$: 1. Formalismo métrico, 2. Formalismo de Palatini y 3. Formalismo métrico-afín; estos tres formalismos se diferencian básicamente en las dependencias de los campos de materia con el tensor métrico y las conexiones, y también en la existencia de una conexión afín en la variedad. \\\\
Puesto que estas teorías se pueden ver como extensiones de la Relatividad General, deben producir los mismos resultados, por ejemplo tener límites Newtonianos correctos a nivel del sistema solar, poseer un problema de Cauchy bien definido y no sufrir de inestabilidades (en el sector de materia por ejemplo existe la llamada inestabilidad de Dolgow-Kawasaki \cite{Faraoni2}, existente en algunos modelos propuestos \cite{Carroll3}; por otro lado algunos modelos muestran que el universo se acelera por siempre o colapsa en un tiempo finito, las cuales se estudian en la denominada inestabilidad de De Sitter \cite{Faraoni3}). Existen varias formas de la gravedad $f(R)$ que predicen de forma acertada la época de inflación y la expansión acelerada del universo \cite{Starobinsky}.\\\\
El objetivo principal del trabajo es obtener modelos cosmológicos en gravedad modificada $f(R)$ que nos den cuenta de la expansión acelerada del universo, y además centrarnos en el problema de las distancias cosmológicas en gravedad $f(R)$, que es el punto central de la tesis. Se estudiará además la teoría de perturbaciones cosmológicas (a primer orden) y el problema de Cauchy en teoriás $f(R)$. Para esto supondremos de nuevo que el universo es homogéneo e isotrópico, y estudiaremos las ecuaciones de Friedmann modificadas en este formalismo.

\newpage

{\color{white} . } 
\chapter{Elementos de Relatividad General}\label{Capitulo2}
\begin{flushright}
\textit{``Most of the fundamental ideas of science are essentially\\
simple, and may, as a rule, be expressed in a\\
language comprehensible to everyone.''\\
Albert Einstein.}
\end{flushright}
\vspace{0.5cm}

\drop{L}a Relatividad General (RG) es la teoría de gravedad más aceptada actualmente. Propuesta por Albert Einstein en 1916, la Relatividad General explica los fenómenos gravitacionales como consecuencias de la curvatura en el espacio-tiempo generada por el contenido de materia presente. De esta manera se pudo realizar una transición de la descripción fenomenológica de la gravedad vista como una fuerza de atracción (Newton) a una propiedad netamente geométrica del espacio-tiempo (Einstein). Aunque la RG cuenta con un fundamento observacional que la hace una de las teorías más sólidas en física, la gravedad sigue siendo aun una de las interacciones más enigmáticas.  Nuestro propósito en este capítulo es introducir muy rápidamente los postulados y ecuaciones de la RG, haciendo sin embargo gran énfasis en los principios variacionales de la RG, las cuales serán la base fundamental de la tesis.

\section{Principios de Relatividad General}

\noindent Como bien es sabido la gravedad fue primeramente descrita por Newton como una fuerza que actúa a distancia y que depende principalmente de las masas de los objetos. Esta descripción permitió realizar grandes avances en la física, especialmente en la astronomía y condujo a valiosos descubrimientos. Aunque la descripción Newtoniana de la gravedad permitía explicar fenómenos que hasta el siglo XVII habían permanecido son resolver (en especial el movimiento de los astros en el sistema solar, sin contar la anomalía en la órbita de Mercurio, y la caída de los cuerpos), presentaba algunas fallas, en especial la imposibilidad de explicar el origen de esa fuerza de atracción  inherente a todo cuerpo con masa.\\\\
La teoría Newtoniana basa sus leyes en la existencia de los sistemas de referencia inerciales (siendo sin embargo la existencia de tales sistemas privilegiados un problema actual en física), tiene un gran problema cuando su fuerza de atracción es considerada, pues presupone una interacción a distancia con una velocidad de información infinita. Esto llevo a Einstein en 1905 a postular la Teoría de la Relatividad Especial (RE) basada en los siguientes postulados \cite{Tejeiro1}:
\textit{
\begin{enumerate}
\item Las leyes de la física son independientes del sistema de referencia inercial.
\item La velocidad de la luz $c$ en el vacío es la misma para todos los observadores inerciales, independiente de la dirección de propagación.
\end{enumerate}}
\noindent El carácter fundamental se basa entonces en la constancia de la velocidad de la luz, razón por la cual la teoría gravitacional de Newton debería ser reformulada. En esto trabajo Einstein entre los años 1905 a 1916 en su Teoría de la Relatividad General (RG).
Una de sus bases fundamentales es el denominado \textit{Principio de Equivalencia de Einstein} que generaliza el Principio de Equivalencia Débil \footnote{La equivalencia entra masa inercial y masa gravitacional de un cuerpo.} y que puede escribirse como:\\\\
\textit{Principio de Equivalencia de Einstein}: No es posible determinar bajo ningún experimento si un objeto se encuentra en caída libre en un campo gravitacional uniforme $g$ o si esta en un movimiento uniformemente acelerado con aceleración $a=g$.\\\\
En donde $g$ es la aceleración del campo gravitacional.  La idea de Einstein fué entonces ver la trayectoria de partículas en caída libre y ver a que efecto se debe su aceleración. El cambio fundamental es el de transformar el concepto de fuerza de atracción al de deformación (o mejor curvatura) del espacio-tiempo, y así una partícula no experimenta una aceleración por una fuerza sino una trayectoria en \textit{geodésicas} en un espacio-tiempo curvo. El ente que realiza esta curvatura en el espacio-tiempo es cualquier forma de materia-energía, de modo que siguiendo \cite{Misner}: \textit{La curvatura le dice a la materia como moverse, pero la materia le dice al espacio como curvarse}.

\section{Postulados y Formulación Matemática de la Relatividad General}
\noindent Daremos entonces un repaso de los elementos básicos de la RG siguiendo \cite{Misner}-\cite{Hawking}. Si bien Einstein logró encontrar una teoría geométrica de la gravitación, su construcción se basa en postulados, que sin embargo fueron desarrollados buscando que las ecuaciones dinámicas den los límites correctos (gravedad Newtoniana y Relatividad Especial). El primer postulado se basa en la descripción del espacio-tiempo:\\\\
\textbf{Postulado 1:} \textit{El espacio-tiempo está descrito por un par $(\mathcal{M},\mathbf{g})$ siendo $\mathcal{M}$ una variedad 4-dimensional y $\mathbf{g}$ una métrica Lorentziana sobre $\mathcal{M}$. }\\\\
La curvatura en la variedad esta descrita por el tensor de curvatura de Riemann $\mathbf{R}$, que en componentes puede escribirse como
\begin{equation}\label{Riemanntens}
R_{\beta\gamma\delta}^{\alpha} = \partial_{\gamma}\Gamma_{\delta\beta}^{\alpha} - \partial_{\delta}\Gamma_{\gamma\beta}^{\alpha} + \Gamma_{\gamma\sigma}^{\alpha} \Gamma_{\delta\beta}^{\sigma} - \Gamma_{\sigma\delta}^{\alpha}\Gamma_{\gamma\beta}^{\sigma},
\end{equation}
con $\Gamma_{\beta\gamma}^{\alpha}$ las conexiones (o símbolos de Christoffel)
\begin{equation}
\Gamma_{\beta\gamma}^{\alpha} = \frac{1}{2}g^{\alpha\sigma}\bigl[\partial_{\gamma} g_{\sigma\beta} + \partial_{\beta} g_{\sigma\gamma}-\partial_{\sigma} g_{\beta\gamma}\bigr],
\end{equation}
en donde $g_{\alpha\beta}$ son las componentes del tensor métrico. Una restricción fundamental que se impone sobre la variedad $\mathcal{M}$ en RG es que sea libre de torsión, lo cual se basa en el siguiente teorema \cite{Nakahara}
\begin{thm}[Teorema Fundamental de la Geometría Riemanniana]
En una variedad Riemanniana $(\mathcal{M},\mathbf{g})$, existe una única conexión simétrica $\Gamma_{\beta\gamma}^{\alpha}$ que es compatible con la métrica $\mathbf{g}$. Esta conexión es llamada la \textbf{conexión de Levi-Civita}.
\end{thm}
\noindent La condición de que la variedad sea libre de torsión se satisface exigiendo que el tensor de torsión de Cartan $S_{\beta\gamma}^{\alpha}$ sea nulo
\begin{equation}\label{cartan}
S_{\beta\gamma}^{\alpha} \equiv \Gamma_{[\beta\gamma]}^{\alpha} =0,
\end{equation}
y la relación entre la conexión $\Gamma_{\beta\gamma}^{\alpha}$ y la métrica $g$ se obtiene imponiendo que la derivada covariante de la métrica sea nula
\begin{equation}
\nabla_{\gamma} g_{\alpha\beta} =0,
\end{equation}
con la definición usual para la derivada covariante
\begin{multline}
\nabla_{\gamma} T_{\beta_1, \cdots \beta_s}^{\alpha_1, \cdots, \alpha_r} = \partial_{\gamma} T_{\beta_1, \cdots \beta_s}^{\alpha_1, \cdots, \alpha_r} + \Gamma_{\gamma\sigma}^{\alpha_1} T_{\beta_1, \cdots \beta_s}^{\sigma, \cdots, \alpha_r} + \cdots + \Gamma_{\gamma\sigma}^{\alpha_r} T_{\beta_1, \cdots \beta_s}^{\alpha_1, \cdots, \sigma} - \Gamma_{\gamma\beta_1}^{\sigma} T_{\sigma, \cdots \beta_s}^{\alpha_1, \cdots, \alpha_r}\\ - \cdots - \Gamma_{\gamma\beta_s}^{\sigma} T_{\beta_1, \cdots \sigma}^{\alpha_1, \cdots, \alpha_r}.
\end{multline}
El postulado de conservación de energía juega un papel fundamental en RG.\\\\
\textbf{Postulado 2:} \textit{Conservación local de la energía: Existe un tensor simétrico $T_{\alpha\beta}=T_{\alpha\beta}(\psi)=T_{\beta\alpha}$ que es función de los campos de materia $\psi$ y sus derivadas tal que} :
\begin{enumerate}[{i.}]
\item \textit{$T_{\alpha\beta}=0$ sobre $\mathcal{U} \subset \mathcal{M}$ si y sólo si $\psi_i=0$ para todo $i$ sobre $\mathcal{U}$.
\item $\nabla_{\beta}T^{\alpha\beta}=0$. }
\end{enumerate}
Finalmente, se encuentra el postulado de las ecuaciones de campo de Einstein, las cuales nos van a dar las relaciones entre la geometría (curvatura) con los campos de materia:\\\\
\textbf{Postulado 3:} \textit{La métrica sobre la variedad espacio-tiempo $(\mathcal{M},\mathbf{g})$ esta determinada por las ecuaciones de campo de Einstein}
\begin{equation}\label{campo}
\boxed{R_{\alpha\beta} - \frac{1}{2}R g_{\alpha\beta} = \kappa T_{\alpha\beta},}
\end{equation}
\textit{siendo $R_{\alpha\beta}$ el tensor de Ricci ($R_{\alpha\beta} = R_{\alpha\eta\beta}^{\eta}$), $R$ el escalar de curvatura ($R = g^{\alpha\beta}R_{\alpha\beta}$), $T_{\alpha\beta}$ el tensor de energía-momentum, y $\kappa = 8\pi G$, $G$ la constante de gravitación universal y usamos unidades de $c=1$}.\\\\
Si definimos el tensor de Einstein $G_{\alpha\beta}$ como
\begin{equation}
G_{\alpha\beta} \equiv R_{\alpha\beta} - \frac{1}{2}R g_{\alpha\beta},
\end{equation}
entonces se tiene una restricción geométrica
\begin{equation}\label{Bianchiden}
\nabla_{\beta}G^{\alpha\beta} =0.
\end{equation}
ecuación conocida como Identidad de Bianchi.\\\\
El movimiento de una partícula es descrito por su trayectoria en el espacio-tiempo, $x^{\alpha}(\lambda)$, donde $\lambda$ es un parámetro. Una partícula libre, i.e., una partícula sobre la que no se ejerce fuerza alguna (otra diferente de la gravedad), satisface la ecuación de la geodésica, la cual
se escribe
\begin{equation}
V^{\mu}\nabla_{\mu}V^{\nu}=0,
\end{equation}
donde $V^{\mu}=\frac{dx^{\mu}}{d\lambda}$ es el vector tangente a la
trayectoria. Esta ecuación puede escribirse en la forma conocida
\begin{equation}\label{eqn-geodesica}
\frac{d^2 x^{\gamma}}{d\lambda^2}+\Gamma_{\alpha\beta}^{\gamma}\frac{d x^{\alpha}}{d \lambda}
\frac{d x^{\beta}}{d \lambda}=0.
\end{equation}
La expresión aplica para los siguientes casos:
\begin{itemize}
\item Partículas con masa, en cuyo caso usualmente se toma como parámetro
$\lambda$ el llamado tiempo propio tal que el correspondiente vector
tangente $V^{\alpha}$ está normalizado: $g_{\alpha\beta}V^{\alpha}V^{\beta}=-1$.
\item Partículas sin masa, en particular el fotón, en cuyo caso el vector
tangente, usualmente denotado por $k^{\alpha}$ es nulo, i.e.,
$g_{\alpha\beta}k^{\alpha}k^{\beta}=0$.
\end{itemize}
Las ecuaciones de campo de Einstein nos determinan entonces la métrica dado un contenido de materia-energía, y las ecuaciones del potencial gravitacional Newtoniano (la ecuación de Poisson) se encuentra como un límite de tales ecuaciones. La relatividad especial es entonces un caso particular de la relatividad general, para la cual $\mathcal{M}$ es una variedad conformalmente plana y $\mathbf{g} = \boldsymbol{\eta}$ con $\boldsymbol{\eta}$ el tensor de Minkowski.
\subsection{Principio Variacional en Relatividad General}
\noindent Un problema que enfrentó Einstein después de escribir su Teoría de la Relatividad General fue el de encontrar sus ecuaciones de campo a partir de un lagrangiano y una principio variacional $\delta S = 0$ con $S$ expresando la acción total. Einstein y Hilbert encontraron, de manera independiente, la acción asociada al campo gravitacional, la cual es denominada como la acción de Einstein-Hilbert. En términos está $S_{EH}$, el término de frontera de Gibbons-York-Hawking  $S_{GYH}$ (el cual fue introducido para relajar las condiciones de frontera como veremos más adelante) \cite{Hawking1},\cite{Hawking2},  y la acción asociada con todos los campos de materia $S_{M}$, la acción total se puede escribir como \cite{Poisson}:
\begin{equation}
S = \frac{1}{2\kappa}\bigl(S_{EH} + S_{GYH}\bigr) + S_{M},
\end{equation}
donde
\begin{equation}\label{eh1}
S_{EH} = \int_{\mathcal{V}} d^4x\, \sqrt{-g}R,
\end{equation}
\begin{equation}
S_{GYH} = 2\oint_{\partial \mathcal{V}} d^3y \, \varepsilon\sqrt{|h|}K,
\end{equation}
aquí $\mathcal{V}$ es un hipervolumen en $\mathcal{M}$, $\partial \mathcal{V}$ su frontera, $h$ el determinante de la métrica inducida, $K$ es la traza de la curvatura extrínseca sobre la frontera $\partial \mathcal{V}$, y $\varepsilon$ es igual a $+1$ si $\partial \mathcal{V}$ is como de tiempo y $-1$ if $\partial \mathcal{V}$ es como es espacio (se asume que $\partial \mathcal{V}$ no es nulo en alguna parte). Las coordenadas $x^{\alpha}$ son usadas para la región finita  $\mathcal{V}$ y $y^{\alpha}$ para la frontera $\partial \mathcal{V}$. Vamos a obtener ahora  las ecuaciones de campo de Einstein variando la acción con respecto a $g^{\alpha\beta}$. Fijamos tal variación a la condición \cite{Wald},\cite{Poisson}
\begin{equation}\label{frontera}
\delta g_{\alpha\beta}\biggl|_{\partial \mathcal{V}} =0,
\end{equation}
es decir, la variación de la métrica se anula en la frontera  $\partial \mathcal{V}$. Usamos ahora los siguientes resultados \cite{Poisson},\cite{Carroll}
\begin{equation}\label{varmet1}
\delta g_{\alpha\beta} = -g_{\alpha\mu}g_{\beta\nu}\delta g^{\mu\nu}, \qquad \delta g^{\alpha\beta} = -g^{\alpha\mu}g^{\beta\nu}\delta g_{\mu\nu},
\end{equation}
\begin{equation}\label{deltag}
\delta \sqrt{-g} =  -\frac{1}{2}\sqrt{-g} g_{\alpha\beta}\delta g^{\alpha\beta},
\end{equation}
\begin{equation}
\delta R_{\beta\gamma\delta}^{\alpha} = \nabla_{\gamma}(\delta\Gamma_{\delta\beta}^{\alpha}) - \nabla_{\delta}(\delta\Gamma_{\gamma\beta}^{\alpha}),
\end{equation}
\begin{equation}\label{palatini}
\delta R_{\alpha\beta} = \nabla_{\gamma}(\delta\Gamma_{\beta\alpha}^{\gamma}) - \nabla_{\beta}(\delta\Gamma_{\gamma\alpha}^{\gamma}).
\end{equation}
Daremos a continuación un detallado repaso de los principios variacionales en RG siguiendo para esto \cite{Wald},\cite{Poisson}. La variación del término de Einstein-Hilbert es
\begin{equation}\label{varaccion1}
\delta S_{EH} = \int_{\mathcal{V}} d^4x \, \bigl(R\delta\sqrt{-g} + \sqrt{-g}\, \delta R\bigr).
\end{equation}
Ahora con $R = g^{\alpha\beta}R_{\alpha\beta}$, tendremos que la variación del escalar de Ricci es
\begin{equation}
\delta R = \delta g^{\alpha\beta}R_{\alpha\beta} + g^{\alpha\beta}\delta R_{\alpha\beta}.
\end{equation}
usando la identidad de Palatini (\ref{palatini}) podemos escribir \cite{Carroll}
\begin{align}\label{deltaR}
\delta R &= \delta g^{\alpha\beta}R_{\alpha\beta} + g^{\alpha\beta}\bigl(\nabla_{\gamma}(\delta\Gamma_{\beta\alpha}^{\gamma}) - \nabla_{\beta}(\delta\Gamma_{\alpha\gamma}^{\gamma})\bigr),\nonumber \\
&= \delta g^{\alpha\beta}R_{\alpha\beta} + \nabla_{\sigma
}\bigl(g^{\alpha\beta}(\delta\Gamma_{\beta\alpha}^{\sigma}) - g^{\alpha\sigma}
(\delta\Gamma_{\alpha\gamma}^{\gamma})\bigr),
\end{align}
donde hemos usado la compatibilidad métrica $\nabla_{\gamma}g_{\alpha\beta}\equiv 0$ y renombramos algunos índices mudos. Reemplazando estos resultados para las variaciones en la expresión (\ref{varaccion1}) tenemos
\begin{align}
\delta S_{EH} &= \int_{\mathcal{V}} d^4x \, \bigl(R\delta\sqrt{-g} + \sqrt{-g}\, \delta R\bigr),\nonumber \\
&= \int_{\mathcal{V}} d^4x \, \biggl(-\frac{1}{2}Rg_{\alpha\beta}\sqrt{-g}\, \delta g^{\alpha\beta} + R_{\alpha\beta}\sqrt{-g}\delta g^{\alpha\beta} + \sqrt{-g}\nabla_{\sigma}\bigl(g^{\alpha\beta}(\delta\Gamma_{\beta\alpha}^{\sigma}) - g^{\alpha\sigma}
(\delta\Gamma_{\alpha\gamma}^{\gamma})\bigr)\biggr),\nonumber \\
&= \int_{\mathcal{V}}  d^4x \, \sqrt{-g} \biggl(R_{\alpha\beta}-\frac{1}{2}Rg_{\alpha\beta}\biggr)\delta g^{\alpha\beta} + \int_{\mathcal{V}}  d^4x\sqrt{-g} \nabla_{\sigma}\bigl(g^{\alpha\beta}(\delta\Gamma_{\beta\alpha}^{\sigma}) - g^{\alpha\sigma}
(\delta\Gamma_{\alpha\gamma}^{\gamma})\bigr).
\end{align}
Denotando el término de divergencia por $\delta S_{B}$,
\begin{equation}
\delta S_B = \int_{\mathcal{V}} d^4x \, \sqrt{-g}\,  \nabla_{\sigma}\bigl(g^{\alpha\beta}(\delta\Gamma_{\beta\alpha}^{\sigma}) - g^{\alpha\sigma}
(\delta\Gamma_{\alpha\gamma}^{\gamma})\bigr),
\end{equation}
definimos
\begin{equation}\label{V}
V^{\sigma} = g^{\alpha\beta}(\delta\Gamma_{\beta\alpha}^{\sigma}) - g^{\alpha\sigma}
(\delta\Gamma_{\alpha\gamma}^{\gamma}),
\end{equation}
de modo que el término de frontera se puede escribir como
\begin{equation}\label{boundary}
\delta S_B = \int_{\mathcal{V}} d^4x \, \sqrt{-g}\,  \nabla_{\sigma}V^{\sigma}.
\end{equation}
Usando el teorema de Gauss-Stokes \cite{Poisson},\cite{Carroll}
\begin{equation}\label{gauss}
\int_{\mathcal{V}} d^{n}x\, \sqrt{|g|}\nabla_{\mu}A^{\mu} = \oint_{\partial \mathcal{V}}d^{n-1}y\, \varepsilon\sqrt{|h|}n_{\mu}A^{\mu},
\end{equation}
donde $n_{\mu}$ es un vector normal unitario a $\partial \mathcal{V}$. Usando esto podemos escribir (\ref{boundary}) como el siguiente término de frontera
\begin{equation}
\delta S_B = \oint_{\partial \mathcal{V}} d^{3}y\, \varepsilon \sqrt{|h|}n_{\sigma}V^{\sigma},
\end{equation}
con $V^{\sigma}$ dado en (\ref{V}). La variación $\delta \Gamma_{\beta\alpha}^{\sigma}$ se obtiene usando que $\Gamma_{\beta\alpha}^{\sigma}$ es el símbolo de Christoffel $\bigl\{_{\beta\alpha}^{\sigma}\bigr\}$
\begin{equation}\label{simbolo}
\Gamma_{\beta\gamma}^{\alpha} \equiv \Bigl\{_{\beta\gamma}^{\alpha}\Bigr\} = \frac{1}{2}g^{\alpha\sigma}\bigl[\partial_{\beta}g_{\sigma\gamma} + \partial_{\gamma}g_{\sigma\beta} - \partial_{\sigma}g_{\beta\gamma}\bigr],
\end{equation}
obteniendo
\begin{align}\label{varsim1}
\delta \Gamma_{\beta\alpha}^{\sigma} &= \delta \biggl(\frac{1}{2}g^{\sigma\gamma}\bigl[\partial_{\beta}g_{\gamma\alpha} + \partial_{\alpha}g_{\gamma\beta} - \partial_{\gamma}g_{\beta\alpha}\bigr]\biggr),\nonumber \\
&= \frac{1}{2}\delta g^{\sigma\gamma}\bigl[\partial_{\beta}g_{\gamma\alpha} + \partial_{\alpha}g_{\gamma\beta} - \partial_{\gamma}g_{\beta\alpha}\bigr] + \frac{1}{2}g^{\sigma\gamma}\bigl[\partial_{\beta}(\delta g_{\gamma\alpha}) + \partial_{\alpha}(\delta g_{\gamma\beta}) -  \partial_{\gamma}(\delta g_{\beta\alpha})\bigr].
\end{align}
De la condiciones de frontera $\delta g_{\alpha\beta} = \delta g^{\alpha\beta}=0$ la variación (\ref{varsim1}) es
\begin{equation}
\delta \Gamma_{\beta\alpha}^{\sigma}\Bigl|_{\partial \mathcal{V}} = \frac{1}{2}g^{\sigma\gamma}\bigl[\partial_{\beta}(\delta g_{\gamma\alpha}) + \partial_{\alpha}(\delta g_{\gamma\beta}) -  \partial_{\gamma}(\delta g_{\beta\alpha})\bigr],
\end{equation}
y
\begin{equation}
V^{\mu}\Bigl|_{\partial \mathcal{V}} = g^{\alpha\beta}\biggl[\frac{1}{2}g^{\mu\gamma}\bigl[\partial_{\beta}(\delta g_{\gamma\alpha}) + \partial_{\alpha}(\delta g_{\gamma\beta}) -  \partial_{\gamma}(\delta g_{\beta\alpha})\bigr]\biggr] - g^{\alpha\mu}\biggl[\frac{1}{2}g^{\nu\gamma}\partial_{\alpha}(\delta g_{\nu\gamma}) \biggr],
\end{equation}
podemos escribir
\begin{align}
V_{\sigma}\Bigl|_{\partial \mathcal{V}} = g_{\sigma\mu}V^{\mu}\Bigl|_{\partial \mathcal{V}} &= g_{\sigma\mu}g^{\alpha\beta}\biggl[\frac{1}{2}g^{\mu\gamma}\bigl[\partial_{\beta}(\delta g_{\gamma\alpha}) + \partial_{\alpha}(\delta g_{\gamma\beta}) -  \partial_{\gamma}(\delta g_{\beta\alpha})\bigr]\biggr]\\
& - g_{\sigma\mu}g^{\alpha\mu}\biggl[\frac{1}{2}g^{\nu\gamma}\partial_{\alpha}(\delta g_{\nu\gamma}) \biggr], \nonumber \\
&= \frac{1}{2}\delta_{\sigma}^{\gamma}g^{\alpha\beta}\bigl[\partial_{\beta}(\delta g_{\gamma\alpha}) + \partial_{\alpha}(\delta g_{\gamma\beta}) -  \partial_{\gamma}(\delta g_{\beta\alpha})\bigr] - \frac{1}{2}\delta_{\sigma}^{\alpha}g^{\nu\gamma}\bigl[\partial_{\alpha}(\delta g_{\nu\gamma}) \bigr],\nonumber \\
&= g^{\alpha\beta}\bigl[\partial_{\beta}(\delta g_{\sigma\alpha}) -  \partial_{\sigma}(\delta g_{\beta\alpha})\bigr].
\end{align}
Ahora evaluamos el término $n^{\sigma}V_{\sigma}\bigl|_{\partial \mathcal{V}}$ usando para esto que
\begin{equation}
g^{\alpha\beta} = h^{\alpha\beta} + \varepsilon n^{\alpha}n^{\beta},
\end{equation}
entonces
\begin{align}
n^{\sigma}V_{\sigma}\Bigl|_{\partial \mathcal{V}} &= n^{\sigma}(h^{\alpha\beta}+\varepsilon n^{\alpha}n^{\beta})[\partial_{\beta}(\delta g_{\sigma\alpha}) -  \partial_{\sigma}(\delta g_{\beta\alpha})], \nonumber \\
&= n^{\sigma}h^{\alpha\beta}[\partial_{\beta}(\delta g_{\sigma\alpha}) -  \partial_{\sigma}(\delta g_{\beta\alpha})],
\end{align}
donde usamos la parte antisimétrica de $\varepsilon n^{\alpha}n^{\beta}$ con $\varepsilon = n^{\mu}n_{\mu}=\pm 1$. Ahora, sabiendo que $\delta g_{\alpha\beta}=0$ en la frontera tendremos que $h^{\alpha\beta}\partial_{\beta}(\delta g_{\sigma\alpha})=0$ \cite{Poisson}. Obtenemos finalmente
\begin{equation}
n^{\sigma}V_{\sigma}\Bigl|_{\partial \mathcal{V}} = -n^{\sigma}h^{\alpha\beta}\partial_{\sigma}(\delta g_{\beta\alpha}).
\end{equation}
De modo que la variación del término de Einstein-Hilbert es
\begin{equation}
\delta S_{EH} = 2\int_{\mathcal{V}}  d^4x \, \sqrt{-g} \biggl(R_{\alpha\beta}-\frac{1}{2}Rg_{\alpha\beta}\biggr)\delta g^{\alpha\beta}-\oint_{\partial \mathcal{V}} d^{3}y\, \varepsilon \sqrt{|h|} h^{\alpha\beta}\partial_{\sigma}(\delta g_{\beta\alpha})n^{\sigma}.
\end{equation}
Como podemos observar, si fijamos $\delta g_{\alpha\beta}  = 0$ existe un termino de frontera adicional. En la mayoría de textos tal término se anula argumentando flujos nulos en el infinito; otro argumento es el de fijar tanto la variación de la métrica como su primer derivada a ser nulas sobre la frontera, es decir, $\delta g_{\alpha\beta} = 0$ y $\partial_{\gamma}\delta g_{\alpha\beta} = 0$. Aunque este último argumento nos conduce directamente a las ecuaciones de campo de Einstein (pues la contribución en la frontera es automáticamente cero), implica fijar dos condiciones en la variación. Para evitar esto Hawking, York y Gibbons introdujeron un término de frontera que permite tener un problema variacional bien definido con solo $\delta g_{\alpha\beta} = 0$. Es importante recalcar que aunque la acción total se vea modificada por términos adicionales sobre la frontera, las ecuaciones de campo obtenidas son las mismas.\\\\
Consideramos entonces la variación del término de frontera de Gibbons-York-Hawking
\begin{equation}
\delta S_{GYH} = \oint_{\partial \mathcal{V}} d^3y\, \varepsilon\sqrt{|h|}\delta K.
\end{equation}
Usando la definición de la traza de la curvatura extrínseca \cite{Poisson}
\begin{align}
K &= \nabla_{\alpha}n^{\alpha}, \nonumber \\
&= g^{\alpha\beta}\nabla_{\beta}n_{\alpha}, \nonumber \\
&= (h^{\alpha\beta}+\varepsilon n^{\alpha}n^{\beta})\nabla_{\beta}n_{\alpha}, \nonumber \\
&= h^{\alpha\beta}\nabla_{\beta}n_{\alpha}, \nonumber \\
&= h^{\alpha\beta}(\partial_{\beta}n_{\alpha}-\Gamma_{\beta\alpha}^{\gamma}n_{\gamma}),
\end{align}
la variación es
\begin{align}\label{deltaK}
\delta K &= -h^{\alpha\beta}\delta\Gamma_{\beta\alpha}^{\gamma}n_{\gamma}, \nonumber \\
&= -\frac{1}{2}h^{\alpha\beta}g^{\sigma\gamma}\bigl[\partial_{\beta}(\delta g_{\sigma\alpha}) + \partial_{\alpha}(\delta g_{\sigma\beta}) -  \partial_{\sigma}(\delta g_{\beta\alpha})\bigr]n_{\gamma}, \nonumber \\
&= -\frac{1}{2}h^{\alpha\beta}\bigl[\partial_{\beta}(\delta g_{\sigma\alpha}) + \partial_{\alpha}(\delta g_{\sigma\beta}) -  \partial_{\sigma}(\delta g_{\beta\alpha})\bigr]n^{\sigma}, \nonumber \\
&= \frac{1}{2}h^{\alpha\beta}\partial_{\sigma}(\delta g_{\beta\alpha})n^{\sigma}.
\end{align}
Esto viene de la variación $\delta\Gamma_{\beta\alpha}^{\gamma}$ evaluada en al frontera, y del hecho que $h^{\alpha\beta}\partial_{\beta}(\delta g_{\sigma\alpha})=0$, $h^{\alpha\beta}\partial_{\alpha}(\delta g_{\sigma\beta})=0$. Entonces tenemos para la variación del término de frontera
\begin{equation}
\delta S_{GYH} = \oint_{\partial \mathcal{V}} d^3y\, \varepsilon\sqrt{|h|}h^{\alpha\beta}\partial_{\sigma}(\delta g_{\beta\alpha})n^{\sigma}.
\end{equation}
Vemos que este término cancela exactamente la contribución en la frontera proveniente de la acción de Einstein-Hilbert. Ahora, si tenemos una acción de materia definida por
\begin{equation}\label{matteraction}
S_M = \int_{\mathcal{V}} d^4x\, \sqrt{-g} \mathcal{L}_M[g_{\alpha\beta},\psi],
\end{equation}
donde $\psi$ denota todos los campos de materia. La variación de esta acción toma la forma
\begin{align}
\delta S_M &= \int_{\mathcal{V}} d^4x\, \delta(\sqrt{-g} \mathcal{L}_M),\nonumber \\
&= \int_{\mathcal{V}} d^4x\, \biggl(\frac{\partial \mathcal{L}_M}{\partial g^{\alpha\beta}}\delta g^{\alpha\beta}\sqrt{-g} + \mathcal{L}_M\delta\sqrt{-g}\biggr),\nonumber \\
&= \int_{\mathcal{V}} d^4x\,\sqrt{-g} \biggl(\frac{\partial \mathcal{L}_M}{\partial g^{\alpha\beta}} -\frac{1}{2} \mathcal{L}_Mg_{\alpha\beta}\biggr)\delta g^{\alpha\beta},
\end{align}
como es usual, definimos el tensor de energía-momentum por
\begin{equation}\label{tensem}
T_{\alpha\beta} \equiv -2\frac{\partial \mathcal{L}_M}{\partial g^{\alpha\beta}} + \mathcal{L}_Mg_{\alpha\beta} = -\frac{2}{\sqrt{-g}}\frac{\delta S_M}{\delta g^{\alpha\beta}},
\end{equation}
entonces:
\begin{equation}\label{variacionener}
\delta S_M = -\frac{1}{2}\int_{\mathcal{V}} d^4x\,\sqrt{-g} T_{\alpha\beta}\delta g^{\alpha\beta},
\end{equation}
imponiendo que las variaciones totales permanezcan invariantes con respecto a $\delta g^{\alpha\beta}$, podemos escribir finalmente
\begin{equation}\label{campo1}
\frac{1}{\sqrt{-g}}\frac{\delta S}{\delta g^{\alpha\beta}} = 0,  \Longrightarrow R_{\alpha\beta} - \frac{1}{2}R g_{\alpha\beta} = \kappa T_{\alpha\beta} ,
\end{equation}
que corresponden a las ecuaciones de campo de Einstein (\ref{campo}).
\subsection{Relatividad General \textit{a la} Palatini}
\noindent Un método alternativo para encontrar las ecuaciones de campo fue introducido por Palatini en 1919. Su formulación consiste en tratar la métrica $g_{\alpha\beta}$ y la conexión $\Gamma_{\beta\gamma}^{\alpha}$ como dos campos independientes. Para esto definimos entonces un tensor de Ricci $\mathcal{R}_{\alpha\beta}$ definido por
\begin{equation}
\mathcal{R}_{\alpha\beta}=\mathcal{R}_{\alpha\eta\beta}^{\eta} = \partial_{\eta}\Gamma_{\alpha\beta}^{\eta} - \partial_{\beta}\Gamma_{\eta\alpha}^{\eta} + \Gamma_{\eta\sigma}^{\eta} \Gamma_{\beta\alpha}^{\sigma} - \Gamma_{\sigma\beta}^{\eta}\Gamma_{\eta\alpha}^{\sigma},
\end{equation}
y el escalar de Ricci como $\mathcal{R}= g^{\alpha}\mathcal{R}_{\alpha\beta}$. La acción de Palatini puede escribirse como
\begin{equation}
S_{P} = \frac{1}{2\kappa}\int_{\mathcal{V}} d^4x\, \sqrt{-g}\mathcal{R} + S_{M},
\end{equation}
con $S_{M}$ siendo nuevamente la acción asociada a los campos de materia (\ref{matteraction}). Usando los resultados (\ref{deltag}) y (\ref{deltaR}) asumiendo que la conexión es \textit{independiente} podemos escribir para la variación con respecto a $g_{\alpha\beta}$
\begin{align}
\delta S_{P} &= \frac{1}{2\kappa}\int_{\mathcal{V}} d^4x\,\biggl(\mathcal{R}_{\alpha\beta}\sqrt{-g}\delta g^{\alpha\beta}-\frac{1}{2}\mathcal{R}g_{\alpha\beta}\sqrt{-g}\, \delta g^{\alpha\beta}\biggr) -\frac{1}{2}\int_{\mathcal{V}} d^4x\,\sqrt{-g} T_{\alpha\beta}\delta g^{\alpha\beta},\nonumber \\
& = \int_{\mathcal{V}}d^4x\, \biggl(\mathcal{R}_{\alpha\beta}-\frac{1}{2}\mathcal{R}g_{\alpha\beta} -\kappa T_{\alpha\beta}\biggr)\sqrt{-g}\delta g^{\alpha\beta},
\end{align}
e imponiendo que la acción sea invariante con respecto a $\delta g_{\alpha\beta}$ obtenemos que
\begin{equation}
\frac{1}{\sqrt{-g}}\frac{\delta S_p}{\delta g^{\alpha\beta}} = 0,  \Longrightarrow \mathcal{R}_{\alpha\beta} - \frac{1}{2}\mathcal{R} g_{\alpha\beta} = \kappa T_{\alpha\beta} ,
\end{equation}
que corresponden a las ecuaciones de campo de Einstein. Consideramos ahora la variación con respecto a $\Gamma_{\beta\gamma}^{\alpha}$ usando para esto la identidad de Palatini (\ref{palatini})
\begin{equation}
\delta S_{P} = \frac{1}{2\kappa}\int_{\mathcal{V}} d^4x\,\sqrt{-g} g^{\alpha\beta}\bigl(\nabla_{\gamma}(\delta\Gamma_{\beta\alpha}^{\gamma}) - \nabla_{\beta}(\delta\Gamma_{\gamma\alpha}^{\gamma})\bigr),
\end{equation}
integrando por partes, y tomando en cuenta que $\delta \Gamma_{\beta\gamma}^{\alpha} = 0$ en al frontera, de modo que términos lineales en $\delta \Gamma_{\beta\gamma}^{\alpha}$ son cero, podemos escribir
\begin{align}
\delta S_{P} = -\frac{1}{2\kappa}\int_{\mathcal{V}} d^4x\,\bigl(\nabla_{\gamma}(\sqrt{-g} g^{\alpha\beta})\delta \Gamma_{\alpha\beta}^{\gamma} - \nabla_{\beta}(\sqrt{-g}g^{\alpha\beta})\delta\Gamma_{\gamma\alpha}^{\gamma}\bigr),\nonumber \\
\delta S_{P} = -\frac{1}{2\kappa}\int_{\mathcal{V}} d^4x\,\bigl(\nabla_{\gamma}(\sqrt{-g} g^{\alpha\beta}) - \nabla_{\sigma}(\sqrt{-g}g^{\alpha\sigma})\delta_{\gamma}^{\beta}\bigr)\delta \Gamma_{\alpha\beta}^{\gamma},
\end{align}
imponiendo de nuevo que esta variación sea invariante con respecto a $\delta \Gamma_{\beta\gamma}^{\alpha}$ tendremos
\begin{equation}
\frac{\delta S_p}{\delta \Gamma_{\beta\gamma}^{\alpha}} = 0,  \Longrightarrow (\nabla_{\gamma}(\sqrt{-g} g^{\alpha\beta})\delta \Gamma_{\alpha\beta}^{\gamma} - \nabla_{\sigma}(\sqrt{-g}g^{\alpha\sigma})\delta_{\gamma}^{\beta}\bigr) = 0 ,
\end{equation}
tomando la traza de esta expresión tenemos
\begin{equation}
\nabla_{\gamma}(\sqrt{-g} g^{\alpha\beta}) = 0 ,
\end{equation}
que es la condición de que la conexión sea de Levi-Civita. Con el método de Palatini se recuperan entonces las ecuaciones de campo y además se obtiene la relación entre la métrica y la conexión.\\\\
Las ecuaciones de campo de Einstein (\ref{campo}) son entonces el punto de partida para la evolución dinámica de las cantidades físicas en un espacio-tiempo.
\section{Efectos de Curvatura = Ecuación de Desvío Geodésico}\label{EDG}
\begin{figure}
\begin{center}
\includegraphics[scale=0.5]{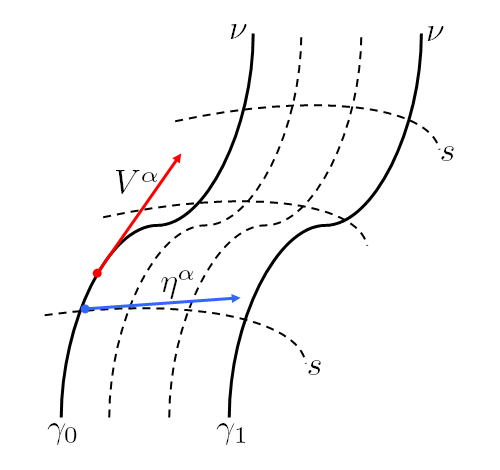}\\
\caption{\footnotesize{Desvío Geodésico.}}
\label{desviogeo}
\end{center}
\end{figure}

\noindent Una forma elegante de observar los efectos de la curvatura en una variedad es a través de la denominada \textit{Ecuación de Desvío Geodésico} \cite{Synge}. Vamos a continuación a hacer una descripción de tal ecuación siguiendo \cite{Wald}-\cite{Schutz}. Sean $\gamma_0$ y $\gamma_1$ dos geodésicas vecinas con parámetro afín $\nu$. Introducimos entre las geodésicas una familia de geodésicas que se interpolan con parámetro afín $s$, Figura \ref{desviogeo}. El campo vectorial $V^{\alpha} = \frac{dx^{\alpha}}{d\nu}$
es tangente a las geodésicas con parámetro afín $\nu$, y la familia $s$ tiene a $\eta^{\alpha}= \frac{dx^{\alpha}}{ds}$ como su campo vectorial tangente.
Vamos a derivar la expresión para la aceleración del vector desviación $\eta^{\alpha}$. Para esto consideremos las coordenadas $x_0^{\alpha}$ para la geodésica $\gamma_0$ y $x_1^{\alpha}$ para $\gamma_1$. La evolución de la geodésica $\gamma_0$ base viene descrita por la ecuación
\begin{equation}\label{geodésica1}
\frac{d^{2}x_{0}^{\alpha}}{d\nu^{2}}+\Gamma^{\alpha}_{\
\beta\gamma}\frac{dx^{\beta}_{0}}{d\nu}\frac{dx^{\gamma}_{0}}{d\nu}=0,
\end{equation}
y sean $x_{1}^{\alpha}=x_{0}^{\alpha}+\eta^{\alpha}$ las
coordenadas de la geodésica vecina, la cual satisface
\begin{equation}\label{geodésica2}
\frac{d^{2}x_{1}^{\alpha}}{d\nu^{2}}+\Gamma^{\alpha}_{\
\beta\gamma}\frac{dx_{1}^{\beta}}{d\nu}\frac{dx_{1}^{\gamma}}{d\nu}=0.
\end{equation}
La conexión afín a primer orden se puede escribir
\begin{equation}
\Gamma^{\alpha}_{\beta\gamma}(x_{1}^{\alpha})=\Gamma^{\alpha}_{\beta\gamma}(x_{0}^{\alpha}+\eta^{\alpha})\cong\Gamma^{\alpha}_{\beta\gamma}(x_{0}^{\alpha})
+\partial_{\sigma} \Gamma^{\alpha}_{\beta\gamma}(x_{0}^{\alpha})\eta^{\sigma},
\end{equation}
por otro lado
\begin{equation}
\frac{dx_{1}^{\alpha}}{d\nu}=\frac{dx_{0}^{\alpha}}{d\nu}+\frac{d\eta^{\alpha}}{d\nu},
\end{equation}
de modo, la ecuación resultante para la ecuación \eqref{geodésica2} es
\begin{equation}
\frac{d^{2}x_{0}^{\alpha}}{d\nu^{2}}+\frac{d^{2}\eta^{\alpha}}{d\nu^{2}}+
\left(\Gamma^{\alpha}_{\beta\gamma}(x_{0}^{\alpha})+\partial_{\sigma}\Gamma^{\alpha}_{\beta\gamma}(x_{0}^{\alpha})\eta^{\sigma}\right)\frac{dx_{1}^{\beta}}{d\nu}\frac{dx_{1}^{\gamma}}{d\nu}\cong
0.
\end{equation}
Reescribiendo $x_{0}^{\alpha}\equiv x^{\alpha}$,
$\Gamma^{\alpha}_{\beta\gamma}(x_{0}^{\alpha})\equiv\Gamma^{\alpha}_{\beta\gamma}$, tenemos
\begin{equation}
\frac{d^{2}\eta^{\alpha}}{d\nu^{2}}+\frac{d^{2}x^{\alpha}}{d\nu^{2}}+
\left(\Gamma^{\alpha}_{\
\beta\gamma}+\partial_{\sigma}\Gamma^{\alpha}_{\beta\gamma}\eta^{\sigma}\right)\left(\frac{dx^{\beta}}{d\nu}+\frac{d\eta^{\beta}}{d\nu}\right)\left(\frac{dx^{\gamma}}{d\nu}+\frac{d\eta^{\gamma}}{d\nu}\right)\cong 0,
\end{equation}
\begin{multline}
\frac{d^{2}\eta^{\alpha}}{d\nu^{2}}+\left(\frac{d^{2}x^{\alpha}}{d\nu^{2}}+\Gamma^{\alpha}_{\
\beta\gamma}\frac{dx^{\beta}}{d\nu}\frac{dx^{\gamma}}{d\nu}\right)+
\left(\Gamma^{\alpha}_{\
\beta\gamma}\frac{dx^{\beta}}{d\nu}\frac{d\eta^{\gamma}}{d\nu}+\Gamma^{\alpha}_{\
\beta\gamma}\frac{dx^{\gamma}}{d\nu}\frac{d\eta^{\beta}}{d\nu}\right)\\
+\partial_{\sigma}\Gamma^{\alpha}_{\beta\gamma}\frac{dx^{\beta}}{d\nu}\frac{dx^{\gamma}}{d\nu}\eta^{\sigma}\cong 0.
\end{multline}
El término del primer paréntesis se anula debido a la ecuación geodésica (\ref{geodésica1}). Los dos términos del segundo
paréntesis se pueden escribir en un solo término, debido a la simetría de la conexión, de modo que
\begin{equation}\label{separacióngeodésicas}
\frac{d^{2}\eta^{\alpha}}{d\nu^{2}}+\partial_{\sigma}\Gamma^{\alpha}_{\beta\gamma}\eta^{\sigma}\frac{dx^{\beta}}{d\nu}\frac{dx^{\gamma}}{d\nu}+2\Gamma^{\alpha}_{\beta\gamma}\frac{d\eta^{\beta}}{d\nu}\frac{dx^{\gamma}}{d\nu}=0.
\end{equation}
Esta ecuación es posible escribirla de manera más compacta acudiendo al concepto de derivada covariante a lo largo de una
curva.  Sea $A^{\alpha}$ un campo vectorial definido sobre una curva cuyo parámetro afín es $\nu$. Al igual que como se
define la derivada covariante sobre todo el espacio, la operación definida como
\begin{equation}
\frac{D
A^{\alpha}}{D\nu}:=\frac{dA^{\alpha}}{d\nu} + \Gamma_{\beta\gamma}^{\alpha}\frac{dx^{\beta}}{d\nu}A^{\gamma},
\end{equation}
define la derivada covariante a lo largo de la curva, con $V^{\beta}=\frac{dx^{\beta}}{d\nu}$ el vector tangente a la curva. Escrito de forma más compacta
\begin{equation}
\frac{D A^{\alpha}}{D\nu}=\nabla_{\gamma}A^{\alpha}V^{\gamma}.
\end{equation}
Como $\frac{D A^{\alpha}}{D\nu}$ es otro campo vectorial,
podemos tomar su derivada covariante a lo largo de esta curva
\begin{equation}
\frac{D^{2}A^{\alpha}}{D\nu^{2}}=\nabla_{\gamma}\left(\frac{D
A^{\alpha}}{D\nu}\right)V^{\gamma},
\end{equation}
utilizando la definición de derivada covariante tenemos
\begin{equation}
\nabla_{\gamma}\left(\frac{D
A^{\alpha}}{D\nu}\right)=\partial_{\gamma}\left(\frac{dA^{\alpha}}{d\nu}+\Gamma_{\sigma\delta}^{\alpha}V^{\sigma}A^{\delta}\right)+\Gamma_{
\gamma\beta}^{\alpha}\frac{D A^{\beta}}{D\nu},
\end{equation}
entonces
\begin{equation}
\nabla_{\gamma}\left(\frac{D
A^{\alpha}}{D\nu}\right)V^{\gamma}=\partial_{\gamma}\left(\frac{dA^{\alpha}}{d\nu}+\Gamma_{\sigma\delta}^{\alpha}V^{\sigma}A^{\delta}\right)V^{\gamma}+\Gamma_{
\gamma\beta}^{\alpha}\frac{D A^{\beta}}{D\nu}V^{\gamma}.
\end{equation}
Por otro lado tenemos que
\begin{equation}
\partial_{\gamma}\left(\frac{dA^{\alpha}}{d\nu}\right)V^{\gamma}=\frac{d}{dx^{\gamma}}\left(\frac{dA^{\alpha}}{d\nu}\right)\frac{dx^{\gamma}}{d\nu}=\frac{d}{d\nu}\left(\frac{dA^{\alpha}}{d\nu}\right)=\frac{d^{2}A^{\alpha}}{d\nu^{2}},
\end{equation}
de modo que
\begin{multline}
\frac{D^{2}A^{\alpha}}{D\nu^{2}}=\frac{d^{2}A^{\alpha}}{d\nu^{2}}+\partial_{\gamma}\Gamma_{\sigma\delta}^{\alpha}V^{\sigma}V^{\gamma}A^{\delta}+\Gamma_{
\sigma\delta}^{\alpha}\partial_{\gamma}V^{\sigma}V^{\gamma}A^{\delta} +\Gamma_{\sigma\delta}V^{\sigma}V^{\gamma}\partial_{\gamma}A^{\delta}\\+\Gamma_{
\gamma\beta}^{\alpha}\left(\frac{dA^{\beta}}{d\nu}+\Gamma^{\beta}_{\nu\rho}V^{\nu}A^{\rho}\right)V^{\gamma}.
\end{multline}
Debido a que $V^{\alpha}$ es el vector tangente a la curva
geodésica, esto es, la derivada covariante de $V^{\alpha}$ a lo
largo de la misma curva del vector tangente es nula, $\frac{D
V^{\alpha}}{D\nu}=0$, podemos utilizar
\begin{equation}
\partial_{\gamma}V^{\sigma}V^{\gamma}=-\Gamma^{\sigma}_{\nu\gamma}V^{\nu}V^{\gamma},
\end{equation}
de esta forma
\begin{multline}\label{segundaderivadacovariante}
\frac{D^{2}A^{\alpha}}{D\nu^{2}}=\frac{d^{2}A^{\alpha}}{d\nu^{2}} + \partial_{\gamma}\Gamma_{\sigma\delta}^{\alpha}V^{\sigma}V^{\gamma}A^{\delta}-\Gamma_{
\sigma\delta}^{\alpha}\Gamma_{\nu\gamma}^{\sigma}V^{\nu}V^{\gamma}A^{\delta} +\Gamma_{\sigma\delta}^{\alpha}V^{\sigma}\partial_{\gamma}A^{\delta}V^{\gamma}\\
+\Gamma_{\gamma\beta}^{\alpha}\left(\frac{dA^{\beta}}{d\nu}+\Gamma^{\beta}_{\nu\rho}V^{\nu}A^{\rho}\right)V^{\gamma},
\end{multline}
Ahora, volviendo a la ecuación \eqref{separacióngeodésicas},
ecuación para la separación de dos geodésicas próximas, despejando
$\frac{d^{2}\eta^{\mu}}{d\nu^{2}}$
\begin{equation}
\frac{d^{2}\eta^{\alpha}}{d\nu^{2}}=-\partial_{\sigma}\Gamma^{\alpha}_{\beta\gamma}\eta^{\sigma}V^{\beta}V^{\gamma}-2\Gamma^{\alpha}_{\beta\gamma}\frac{d\eta^{\beta}}{d\nu}V^{\gamma},
\end{equation}
De esta manera, si consideramos a $\eta^{\alpha}$ como el campo
vectorial $A^{\alpha}$, reemplazando el valor de la derivada parcial
segunda de la ecuación \eqref{segundaderivadacovariante} y
reescribiendo algunos índices mudos, tenemos
\begin{multline}
\frac{D^{2}\eta^{\alpha}}{D\nu^{2}}=-\partial_{\sigma}\Gamma^{\alpha}_{\beta\gamma}\eta^{\sigma}V^{\beta}V^{\gamma}-2\Gamma^{\alpha}_{\beta\gamma}\frac{d\eta^{\beta}}{d\nu}V^{\gamma} + \partial_{\sigma}\Gamma_{\beta\gamma}^{\alpha}V^{\beta}V^{\sigma}\eta^{\gamma}-\Gamma_{
\beta\gamma}^{\alpha}\Gamma_{\nu\rho}^{\beta}V^{\nu}V^{\rho}\eta^{\gamma}\\
+\Gamma_{\gamma\beta}^{\alpha}V^{\gamma}\frac{d \eta^{\beta}}{dx^{\delta}}V^{\delta}+\Gamma_{
\gamma\beta}^{\alpha}\left(\frac{d\eta^{\beta}}{d\nu}+\Gamma^{\beta}_{\nu\rho}V^{\nu}\eta^{\rho}\right)V^{\gamma},
\end{multline}
agrupando términos semejantes
\begin{multline}
\frac{D^{2}\eta^{\alpha}}{D\nu^{2}}=-\partial_{\sigma}\Gamma^{\alpha}_{\beta\gamma}\left[\eta^{\sigma}V^{\beta}V^{\gamma}-\eta^{\gamma}V^{\beta}V^{\sigma}\right]
+\left[-2\Gamma^{\alpha}_{\beta\gamma}\frac{d\eta^{\beta}}{d\nu}V^{\gamma}+\Gamma^{\alpha}_{\beta\gamma}V^{\gamma}\frac{d\eta^{\beta}}{d\nu}+\Gamma^{\alpha}_{\beta\gamma}\frac{d\eta^{\beta}}{d\nu}V^{\gamma}\right]\\
-\Gamma^{\alpha}_{\beta\gamma}\Gamma^{\beta}_{\nu\rho}\eta^{\gamma}V^{\nu}V^{\rho}+\Gamma^{\alpha}_{\gamma\beta}\Gamma^{\beta}_{\nu\rho}\eta^{\rho}V^{\nu}V^{\gamma},
\end{multline}
dado que $\Gamma^{\alpha}_{\beta\gamma}=\Gamma^{\alpha}_{\gamma\beta}$ (torsión nula), los términos del segundo paréntesis
se anulan y
\begin{equation}
\frac{D^{2}\eta^{\alpha}}{D\nu^{2}}=-\partial_{\sigma}\Gamma^{\alpha}_{\beta\gamma}\left[\eta^{\sigma}V^{\gamma}-\eta^{\gamma}V^{\sigma}\right]V^{\beta}
-\Gamma^{\alpha}_{\beta\gamma}\Gamma^{\beta}_{\nu\rho}[\eta^{\gamma}V^{\nu}V^{\rho}-\eta^{\rho}V^{\nu}V^{\gamma}],
\end{equation}
\noindent
\begin{equation}
\frac{D^{2}\eta^{\alpha}}{D\nu^{2}}=-\partial_{\delta}\Gamma^{\alpha}_{\beta\gamma}\eta^{\delta}V^{\gamma}V^{\beta}+\partial_{\delta}\Gamma^{\alpha}_{\beta\gamma}\eta^{\gamma}V^{\delta}V^{\beta}
-\Gamma^{\alpha}_{\sigma\delta}\Gamma^{\sigma}_{\gamma\beta}\eta^{\delta}V^{\gamma}V^{\beta}+\Gamma^{\alpha}_{\gamma\sigma}\Gamma^{\sigma}_{\delta\beta}\eta^{\beta}V^{\delta}V^{\gamma},
\end{equation}
donde de nuevo renombramos algunos índices mudos. Cambiando en el primer término después de la igualdad $\delta \longleftrightarrow \gamma$, en el tercer término $\delta \longleftrightarrow \gamma$, y en el cuarto término $\delta \longleftrightarrow \gamma$ y $\beta \longleftrightarrow \gamma$,  tendremos
\begin{align}
\frac{D^{2}\eta^{\alpha}}{D\nu^{2}}&=-\partial_{\gamma}\Gamma^{\alpha}_{\beta\delta}\eta^{\gamma}V^{\delta}V^{\beta}+\partial_{\delta}\Gamma^{\alpha}_{\beta\gamma}\eta^{\gamma}V^{\delta}V^{\beta}
-\Gamma^{\alpha}_{\sigma\gamma}\Gamma^{\sigma}_{\delta\beta}\eta^{\gamma}V^{\delta}V^{\beta}+\Gamma^{\alpha}_{\delta\sigma}\Gamma^{\sigma}_{\beta\gamma}\eta^{\gamma}V^{\beta}V^{\delta},\notag \\
&=-[\partial_{\gamma}\Gamma^{\alpha}_{\beta\delta}+\partial_{\delta}\Gamma^{\alpha}_{\beta\gamma}
-\Gamma^{\alpha}_{\sigma\gamma}\Gamma^{\sigma}_{\delta\beta}+\Gamma^{\alpha}_{\delta\sigma}\Gamma^{\sigma}_{\beta\gamma}]\eta^{\gamma}V^{\delta}V^{\beta},
\end{align}
El término entre paréntesis es el tensor de curvatura de Riemann (\ref{Riemanntens}),
por lo que la ecuación de desvío geodésico queda escrita como
\begin{equation}\label{GDE}
\boxed{\frac{D^2 \eta^{\alpha}}{D \nu^2} = - R_{\beta\gamma\delta}^{\alpha}V^{\beta}\eta^{\gamma}V^{\delta},}
\end{equation}
Esta importante ecuación nos muestra que la aceleración del vector desviación depende del tensor de Riemann, y por lo tanto da una directa prueba de la curvatura en una variedad. Como es de esperarse para un espacio-tiempo plano $R_{\beta\gamma\delta}^{\alpha} = 0$, y por lo tanto el vector desviación no sufre aceleración. Esta importante relación nos permitirá describir todas las propiedades geométricas de un espacio-tiempo, y nos dará de una forma elegante las cantidades observacionales (en particular las distancias) \cite{Ellis1}.

\newpage

{\color{white} .}

\chapter{El Modelo Estándar de la Cosmología}\label{Capitulo3}
\begin{flushright}
\textit{``An observer situated in a nebula and moving with the nebula\\
will observe the same properties of the universe as any other\\
similarly situated observer at any time. ''\\
Sir Hermann Bondi, 1948.}
\end{flushright}
\vspace{0.5cm}
\drop{L}a Relatividad General (RG) es la base para la construcción de modelos que describan la evolución del universo. Después de ser verificada experimentalmente, se empezaron a construir modelos a partir de las ecuaciones de campo de Einstein. Una vez dado el fracaso del modelo estático de Einstein motivado por observaciones en el corrimiento al rojo de galaxias, surgieron nuevos modelos basados en las propiedades estadísticas del universo a grandes escalas. El acuñado actualmente como ``Modelo Estándar de la Cosmología'' reposa sobre dos principios fundamentales: la homogeneidad e isotropía. \\\\
Este universo homogéneo e isotrópico descrito por la métrica de Robertson-Walker y denominado como el universo de Friedamnn-Lema\^{\i}tre-Robertson-Walker (FLRW) es la base actual de la cosmología moderna. Aunque es aceptado y varias pruebas experimentales lo soportan, existen varios modelos que guardan alguna de las propiedades del denominado principio cosmológico, entre ellos los universos de Bianchi (homogéneos pero no isotrópicos) y los de Lema\^{\i}tre-Tolman-Bondi (isotrópicos pero no homogénos).\\\\
En lo que sigue daremos los elementos básicos del modelo estándar de la cosmología centrándonos en sus problemas actuales y en el papel fundamental de la medición de distancias.

\section{La métrica de Robertson-Walker y la expansión del universo}
\begin{figure}
\begin{center}
\includegraphics[scale=0.7]{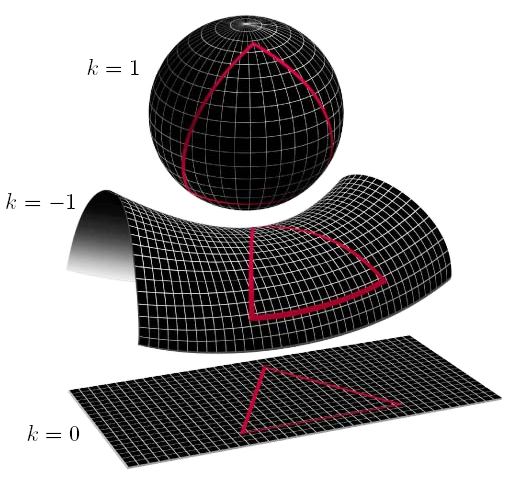}\\
\caption{Geometría espacial del Universo.}
\label{geometria}
\end{center}
\end{figure}

\noindent Como mencionamos anteriormente, el Modelo Estándar de la Cosmología esta basado en dos principios: La homogeneidad e isotropía del universo. Estos postulados permiten escribir el elemento de línea en un universo maximalmente simétrico por medio de la métrica de Robertson-Walker \cite{Weinberg2},\cite{Dodelson}, la cual esta dada por
\begin{equation}\label{robw}
\boxed{ds^2 = -dt^2 + a^2(t)\biggl[\frac{dr^2}{1-kr^2} + r^2d\theta^2 + r^2\sin^2\theta d\varphi^2\biggr],}
\end{equation}
siendo $a(t)$ el factor de escala y $k$ un parámetro que nos dice si la geometría del universo es plana ($k=0$), esférica ($k=1$) o hiperbólica ($k=-1$), Figura \ref{geometria}.
La variedad ahora esta representada por una foliación $\mathcal{M} = \Sigma\times\mathbb{R}$ en donde $\Sigma$ representa una tri-variedad maximalmente simétrica y $\mathbb{R}$ la coordenada temporal del espacio-tiempo. El principio cosmológico permite además escribir el tensor de energía-momentum en la forma de un fluido perfecto
\begin{equation}\label{e-m1}
\boxed{T_{\alpha\beta}= (\rho + p)u_{\alpha}u_{\beta} + p g_{\alpha\beta},}
\end{equation}
siendo $p$ la presión del fluido, $\rho$ la densidad de energía y $u^{\alpha}$ la cuadrivelocidad de los observadores fundamentales. Las ecuaciones de campo de Einstein (sin constante cosmológica) para la métrica (\ref{robw}) con el tensor de energía-momentum (\ref{e-m1}) son
\begin{equation}\label{fried1}
H^2 = \biggl(\frac{\dot{a}}{a}\biggr)^2 = \frac{\kappa}{3}\rho -\frac{k}{a^2} , \qquad \text{Componente temporal}
\end{equation}
y
\begin{equation}\label{fried2}
\frac{\ddot{a}}{a} = -\frac{\kappa}{6}\bigl(\rho + 3p\bigr), \qquad \text{Componentes espaciales}
\end{equation}
en donde definimos el parámetro de Hubble como $H \equiv \frac{\dot{a}}{a}$. La conservación de energía, expresada por la ecuación $\nabla_{\beta}T^{0\beta} =0$ se escribe para esta métrica como
\begin{equation}\label{fried3}
\frac{d}{dt}\bigl(\rho a^3\bigr) + p\frac{d}{dt}\bigl(a^3\bigr)=0.
\end{equation}
Para resolver estas ecuaciones es necesario conocer una relación entre la presión $p$ y la densidad de energía $\rho$. La forma más usual es considerar una ecuación de estado de la forma \cite{Dodelson}
\begin{equation}\label{ecestd}
p = w \rho,
\end{equation}
con $w$ un parámetro, que a lo más puede depender del tiempo $w=w(t)$. Al conjunto de ecuaciones (\ref{fried1})-(\ref{fried3}) se les conoce como \textit{ecuaciones de Friedmann-Lema\^{\i}tre}, y junto con la ecuación de estado (\ref{ecestd}) determinan el modelo cosmológico. Los modelos más trabajados son los de polvo ($w=0$) y radiación ($w=1/3$), los cuales dan cuenta de la evolución del universo. Sin embargo, hoy en día sabemos que nuestro universo se expande de forma acelerada. De esta manera surgen los modelos de constante cosmológica, que tratan de explicar esa expansión introduciendo formas exóticas de energía. Para que tengamos un universo en expansión se necesita que $\ddot{a} > 0 $, y a partir de la ecuación (\ref{fried2}), junto con la ecuación de estado (\ref{ecestd}), se llega a la condición $w < -1/3$. Los modelos con ecuaciones de estado de esta forma se denominan de \textit{Quintaesencia} \cite{Lobo}. El caso con $w=-1$ se obtiene suponiendo en la ecuación (\ref{fried3}) que la densidad es constante, lo que nos lleva a una ecuación de estado de la forma
\begin{equation}
p=-\rho,
\end{equation}
es decir, una ecuación de estado con presión negativa. A esta forma de energía que reproduce una presión negativa se le conoce como \textbf{energía oscura}. Podemos entonces asociar una densidad, que llamaremos densidad de energía del vacío relacionada con la constante cosmológica por la relación \cite{Carroll1}
\begin{equation}
\rho_{\Lambda} = \frac{\Lambda}{\kappa},
\end{equation}
que se puede obtener si introducimos el termino asociado a la constante cosmológica en la ecuación (\ref{campo})
\begin{equation}\label{campoconstante}
R_{\alpha\beta} - \frac{1}{2}R g_{\alpha\beta} = \kappa T_{\alpha\beta} + \Lambda g_{\alpha\beta}.
\end{equation}
Estas ecuaciones de campo nos conducen a la siguiente ecuación de Friedmann
\begin{equation}\label{fried4}
H^2  = \frac{\kappa}{3} \rho - \frac{k}{a^2} + \frac{\Lambda}{3},
\end{equation}
definimos los parámetros cosmológicos por las siguientes relaciones
\begin{equation}\label{params}
\Omega_{m} \equiv \frac{\kappa \rho}{3H^2}, \qquad  \Omega_{\Lambda} \equiv \frac{\Lambda}{3H^2}, \qquad \Omega_{k} \equiv -\frac{k}{a^2H^2},
\end{equation}
con lo cual la ecuación primera ecuación de Friedmann se escribe como
\begin{equation}
\Omega_{m} +\Omega_{\Lambda}+\Omega_{k} = 1,
\end{equation}
llamada como la regla de suma cósmica. Observaciones con supernovas de tipo Ia (Sn Ia) muestran que los valores actuales de estos parámetros son \cite{Perlmutter},\cite{Komatsu},\cite{Zaldarriaga}
\begin{equation}
\Omega_{m} \approx 0.28, \qquad \Omega_{\Lambda} \approx 0.72, \qquad \Omega_{k} \approx 0.
\end{equation}

\begin{figure}[htb]
\begin{center}
\includegraphics[height=11.74cm,width=9cm]{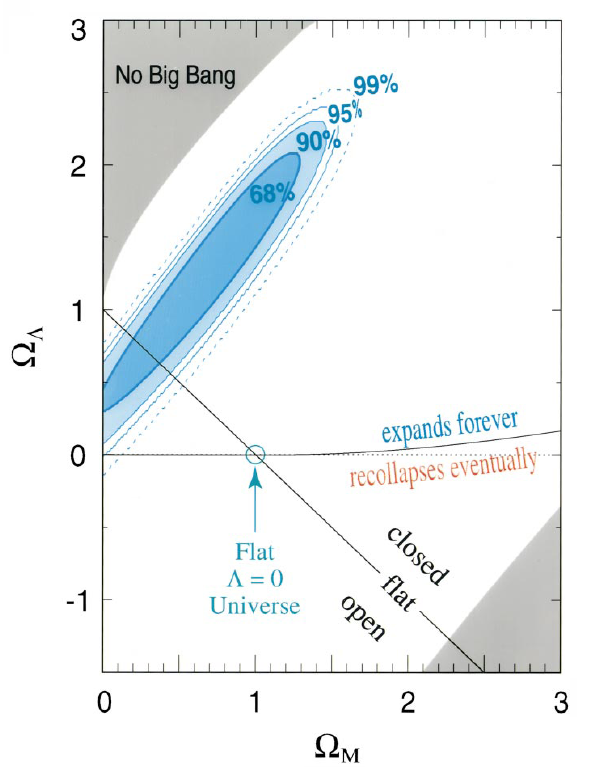}\\
\caption{Regiones de coincidencia para los parámetros $\Omega_{\Lambda}$ y $\Omega_m$ (Perlmutter
et al, (1999)) \cite{Perlmutter}.}
\label{planocoinc}
\end{center}
\end{figure}
\noindent La figura \ref{planocoinc} muestra las regiones de coincidencia en un plano de $\Omega_{\Lambda}$ y $\Omega_m$ a partir del cual se pueden estimar las cotas para los anteriores parámetros. La contribución al parámetro de curvatura $\Omega_k$ es nulo y $\Omega_{\Lambda} + \Omega_m = 1$. Por otro lado, gracias al estudio de la Radiación Cósmica de Fondo (CMB) por medio del proyecto Wilkinson Microwave Anisotropy Probe (WMAP), se demuestra que el contenido actual de materia en el universo esta compuesto en un 72.1 \% de energía oscura, un 23.3 \% de materia oscura y un restante 4.6\% de materia ordinaria (o mejor llamada bariónica)\footnote{Obtenidos también de \href{http://map.gsfc.nasa.gov/}{http://map.gsfc.nasa.gov/}.} \cite{Komatsu}.\\\\
Así pues nos encontramos actualmente en un universo que se expande aceleradamente y con abundancia de un contenido de materia y energía que aun desconocemos. Aunque hoy en día se maneja el modelo $\Lambda$CDM (constante cosmológica + materia oscura fría), se tienen sin embargo indicios que nos llevan a reconsiderar la introducción de la constante cosmológica. Dados los valores para los parámetros cosmológicos se puede dar una aproximación, al menos en órdenes de magnitud, para la densidad de energía relacionada del orden de $2 \times 10^{-10}$ erg/cm$^3$. Sin embargo a nivel de teorías cuánticas, la energía del punto cero de los campos, la cual se puede asociar a la constante cosmológica, es del orden de $2 \times 10^{110}$ erg/cm$^3$. Esta diferencia de 120 ordenes de magnitud para los valores experimentales y teóricos para la constante cosmológica es el conocido \textit{problema de la constante cosmológica}. Otro problema fundamental es el denominado \textit{problema de coincidencia}, y es el por que actualmente la densidad de energía de la constante cosmológica es comparable a la de la materia \cite{Carroll1}. \\\\
Es por estos problemas y la imposibilidad de dar una explicación razonable a la energía oscura que se han propuesto teorías alternativas para explicar la actual expansión acelerada del universo. Una de ellas es considerar teorías de gravedad de alto orden, en las cuales se pueda ver tal expansión como resultado de términos adicionales en las ecuaciones de campo de Einstein, debidos solamente a contribuciones geométricas.

\section{Distancias en el Modelo Estándar}
\noindent Un aspecto fundamental en el modelo estándar de la cosmología se refiere a la medición de distancias. Como veremos tal medición depende de la métrica usada y también del contenido de materia en el universo. Como vimos anteriormente, las ecuaciones de Friedmann dependen principalmente de la forma de las ecuaciones de campo, de modo que las expresiones para las distancias dependerán finalmente de la forma de la métrica y las ecuaciones de campo usadas. \\\\
Las observaciones astronómicas están basadas en la radiación que viaja a través del espacio-tiempo y que llegan a
nosotros a través de geodésicas nulas. En el caso del universo de FLRW es posible escoger geodésicas radiales, por lo que para estas
geodésicas $ds^{2}=d\theta=d\phi=0$, y por lo tanto $0=-dt^{2}+a^{2}(t)dr^{2}$. De aquí se sigue que la radiación
emitida en un punto $E$ y recibida en otro punto $O$ obedece las siguientes relaciones
\begin{equation}
r=\int_{E}^{O}dr=\int_{t_{E}}^{t_{O}}\frac{dt}{a(t)}=\int_{a_{E}}^{a_{O}}\frac{da}{a(t)\dot{a}(t)},
\end{equation}
con $\dot{a}(t) = \frac{d a(t)}{dt}$.
\subsection{Redshift}
\noindent Vamos a considerar un pulso de luz que se emite en n punto $E$ y llega a un punto $O$. Definimos $1+z_{c}$, siendo $z_{c}$ el redshift
cosmológico, como la razón entre la longitud de onda medida por el observador $O$, $\lambda_{O}$ y la longitud de onda emitida en
$E$, $\lambda_{E}$. Dado que la longitud de onda es inversamente proporcional al período entonces
\begin{equation}
(1+z)=\frac{\lambda_{0}}{\lambda_{E}}=\frac{\Delta
T_{0}}{\Delta T_{E}}.
\end{equation}
Podemos interpretar el período $\Delta T$ como el tiempo entre cresta y cresta de la onda de luz. Ahora los periodos $\Delta T_0$ y $\Delta T_E$ se relacionan con el factor de escala por \cite{Caceres1},\cite{Castaneda}
\begin{equation}
\frac{\Delta T_{0}}{a(t_{0})}=\frac{\Delta T_{E}}{a(t_{E})},
\end{equation}
de esta manera
\begin{equation}
(1+z)=\frac{\lambda_{0}}{\lambda_{E}}=\frac{a(t_{0})}{a(t_{E})},
\end{equation}
en donde $a(t_0)$ es el \textit{tamaño del universo} en el tiempo en el que la luz del objeto es observada, y
$a(t_e)$ es el tamaño en el tiempo en que la luz fue emitida. \\\\
Si consideramos que $a(t_0)$ corresponde al factor de escala evaluado actualmente de modo que podemos fijar $a(t_0) = 1$ y con $a(t)$ el redshift evaluado en algún tiempo cósmico correspondiente a un redshift $z$, la expresión anterior se reduce a la relación conocida
\begin{equation}\label{redshift}
\boxed{a(t) = \frac{1}{1+z}.}
\end{equation}
\subsection{Distancia Comóvil $\chi(z)$}
\noindent Para escribir las distancias de forma general vamos primero a escribir la ecuación de Friedmann (\ref{fried4}) considerando la densidad $\rho$ compuesta por una contribución de materia $\rho_m$ y otra de radiación $\rho_r$
\begin{equation}\label{fried5}
H^2  = \frac{\kappa}{3} (\rho_m + \rho_r) - \frac{k}{a^2} + \frac{\Lambda}{3},
\end{equation}
estas contribuciones satisfacen las ecuaciones de estado $p_m =0$ y $p_r = \frac{1}{3}\rho_r$, de modo que las densidades de materia y radiación satisfacen
\begin{equation}
\rho_m \propto a^{-3}, \qquad
\rho_r \propto a^{-4},
\end{equation}
relaciones que se deducen a partir de la ecuación de conservación (\ref{fried3}). Podemos escribir entonces el parámetro de Hubble de la siguiente forma
\begin{equation}\label{fried6}
H^2(a) = H_0^2 \bigl(\Omega_{m0} a^{-3} + \Omega_{r0} a^{-4} + \Omega_{k0} a^{-2} + \Omega_{\Lambda}\bigr),
\end{equation}
siendo $H_0$ el valor actual para el parámetro de Hubble y definimos además las siguientes cantidades (ver ecuación (\ref{params}))
\begin{equation}
\Omega_{m0} \equiv \frac{\kappa\rho_m}{3 H_0^2}, \qquad \Omega_{r0} \equiv  \frac{\kappa\rho_r}{3 H_0^2}, \qquad \Omega_{k0}=-\frac{k}{H^2 a^2}, \qquad \Omega_{\Lambda} \equiv \frac{\Lambda}{3H_0^2}.
\end{equation}
Podemos escribir el parámetro de Hubble en función del redshift usando (\ref{redshift})
\begin{equation}\label{fried7}
H^2(z) = H_0^2 \bigl(\Omega_{m0} (1+z)^{3} + \Omega_{r0} (1+z)^{4} + \Omega_{k0} (1+z)a^{2} + \Omega_{\Lambda}\bigr),
\end{equation}
con estas cantidades definimos la \textit{distancia comóvil} $\chi(z)$ como
\begin{equation}
\chi(z) = \int_0^z \frac{dz'}{H(z')}.
\end{equation}
Esta distancia es la más fundamental en cosmología, pues como veremos las otras relaciones están derivadas en función de ésta.
\subsection{Distancia Comóvil (transversal) $d_M(z)$}
\noindent Se define la distancia comóvil transversa $d_M$ por las relaciones
\begin{equation}\label{comoviltrans}
d_M (z) = \begin{cases}
\dfrac{1}{H_0 \sqrt{|\Omega_k|}}\sinh[\sqrt{\Omega_k}H_0 \chi(z)], & \Omega_k > 0, \\\\
\chi(z), & \Omega_k = 0, \\\\
\dfrac{1}{H_0 \sqrt{|\Omega_k|}}\sin[\sqrt{-\Omega_k}H_0 \chi(z)], & \Omega_k < 0, \\\\
\end{cases}
\end{equation}
esta distancia también es conocida como \textit{distancia de movimiento propio} definida como la razón entre la velocidad transversal y el movimiento propio \cite{Weinberg}.
\subsection{Distancia Diametral Angular $d_A(z)$}
\begin{figure}
\begin{center}
\includegraphics[scale=0.55]{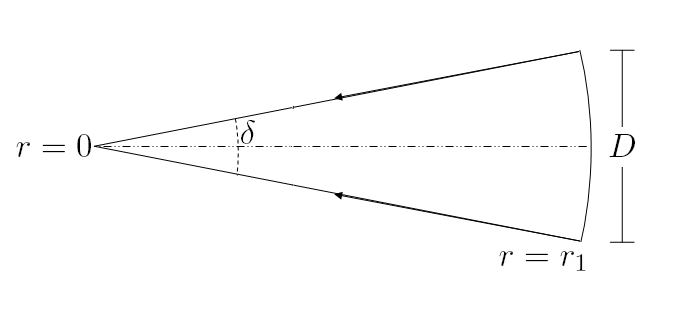}\\
\caption{Distancia diametral angular.}
\label{diametral}
\end{center}
\end{figure}

\noindent La distancia diametral angular $d_A$ es definida como la razón del diámetro propio de un objeto $D$ con su diámetro angular aparente $\delta$ (en radianes), Figura \ref{diametral}.
\begin{equation}
d_A = \frac{D}{\delta},
\end{equation}
de la métrica de Robertson-Walker ($r = \text{cte}$ y $\varphi = \text{cte}$), el diámetro propio de la fuente con $\Delta \theta = \delta$ es
\begin{equation}
D = a(t_1)r_1 \delta,
\end{equation}
de modo que la distancia diametral angular queda
\begin{equation}
d_A = a(t_1)r_1 .
\end{equation}
Esta distancia esta relacionada con la distancia comóvil transversa por
\begin{equation}
d_A = \frac{d_M}{(1+z)},
\end{equation}
de modo que usando la expresión (\ref{comoviltrans}) tendremos
\begin{equation}\label{disangular}
d_A (z) = \frac{1}{1+z}\begin{cases}
\dfrac{1}{H_0 \sqrt{|\Omega_k|}}\sinh[\sqrt{\Omega_k}H_0 \chi(z)], & \Omega_k > 0, \\\\
\chi(z), & \Omega_k = 0, \\\\
\dfrac{1}{H_0 \sqrt{|\Omega_k|}}\sin[\sqrt{-\Omega_k}H_0 \chi(z)], & \Omega_k < 0, \\\\
\end{cases}.
\end{equation}
\subsection{Distancia de Luminosidad $d_L(z)$}
\noindent Finalmente la distancia de luminosidad $d_L$ esta definida por la relación entre el flujo de una fuente $\mathcal{F}$ y su luminosidad $L$
\begin{equation}
d_L = \sqrt{\frac{L}{4\pi \mathcal{F}}}.
\end{equation}
La relación de la distancia de luminosidad con la distancia comóvil transversal y la distancia diametral angular por
\begin{equation}
d_L = (1+z)d_M = (1+z)^2 d_A,
\end{equation}
de modo que
\begin{equation}\label{dislumin}
d_L (z) = (1+z)\begin{cases}
\dfrac{1}{H_0 \sqrt{|\Omega_k|}}\sinh[\sqrt{\Omega_k}H_0 \chi(z)], & \Omega_k > 0, \\\\
\chi(z), & \Omega_k = 0, \\\\
\dfrac{1}{H_0 \sqrt{|\Omega_k|}}\sin[\sqrt{-\Omega_k}H_0 \chi(z)], & \Omega_k < 0, \\\\
\end{cases}.
\end{equation}
Las relaciones anteriores para las distancias nos permiten comparar entonces las observaciones con los modelos teóricos, y en especial determinar los valores de los parámetros cosmológicos.
\section{Ecuación de Desvío Geodésico en FLRW}\label{EDGFLRW}
\noindent Una forma muy elegante de tratar el problema de distancias en cosmología (y en general todas las propiedades del espacio-tiempo), es a través de la EDG. Daremos entonces a continuación un repaso de la EDG en el universo FLRW siguiendo \cite{Ellis1},\cite{Caceres1}.\\\\
El punto fundamental es relacionar las cantidades geométricas de las ecuaciones de campo (el tensor de Ricc $R_{\alpha\beta}$ y el escalar de curvatura $R$) con el tensor de energía-momentum $T_{\alpha\beta}$. Para esto vamos a considerar la siguiente descomposición del tensor de Riemann en sus partes libres de traza y con traza
\cite{Wald},\cite{Ellis}
\begin{equation}\label{RiemannC}
R_{\alpha\beta\gamma\delta} = C_{\alpha\beta\gamma\delta} + \frac{1}{2}\bigl(g_{\alpha\gamma}R_{\delta\beta} - g_{\alpha\delta}R_{\gamma\beta} + g_{\beta\delta}R_{\gamma\alpha}
- g_{\beta\gamma}R_{\delta\alpha} \bigr) - \frac{R}{6}\bigl(g_{\alpha\gamma}g_{\delta\beta} - g_{\alpha\delta}g_{\gamma\beta}\bigr),
\end{equation}
en donde $C_{\alpha\beta\gamma\delta}$ es el denominado tensor de Weyl. A partir del elemento de línea (\ref{robw}) se muestra que el tensor de Weyl es nulo para universos homogéneos e isotrópicos $C_{\alpha\beta\gamma\delta}=0$ \footnote{En universos con anisotropías, como por ejemplo los de Bianchi, el tensor de Weyl es en general no nulo.}.  A partir del tensor de energía-momentum (\ref{e-m1}) podemos obtener su traza
\begin{equation}
T = 3p-\rho.
\end{equation}
y de las ecuaciones de campo de Einstein (con constante cosmológica)
\begin{equation}
R_{\alpha\beta} - \frac{1}{2}R g_{\alpha\beta} + \Lambda g_{\alpha\beta} = \kappa T_{\alpha\beta},
\end{equation}
podemos escribir el escalar de curvatura $R$ y el tensor de Ricci $R_{\alpha\beta}$ como
\begin{equation}
R = \kappa(\rho - 3p) + 4\Lambda,
\end{equation}
\begin{equation}
R_{\alpha\beta} = \kappa(\rho + p)u_{\alpha}u_{\beta} + \frac{1}{2}\bigl[\kappa(\rho-p) + 2\Lambda\bigr]g_{\alpha\beta},
\end{equation}
de modo que reemplazando estos resultados en (\ref{RiemannC}) obtenemos
\begin{multline}
R_{\alpha\beta\gamma\delta}=\frac{1}{2}\kappa(\rho+p)[g_{\alpha\gamma}u_{\beta}u_{\delta}-g_{\alpha\delta}u_{\beta}u_{\gamma}+g_{\beta\delta}u_{\alpha}u_{\gamma}-g_{\beta\gamma}u_{\alpha}u_{\delta}]\\
+\frac{1}{3}(\kappa\rho+\Lambda)[g_{\alpha\gamma}g_{\beta\delta}-g_{\alpha\delta}g_{\beta\gamma}].
\end{multline}
Si el campo vectorial $V^{\alpha}$ está normalizado,
$V_{\alpha}V^{\alpha}=\epsilon$, tenemos
\begin{multline*}
R_{\alpha\beta\gamma\delta}V^{\beta}V^{\delta}=\frac{1}{2}\kappa(\rho+p)[g_{\alpha\gamma}u_{\beta}u_{\delta}V^{\beta}V^{\delta}-g_{\alpha\delta}V^{\delta}u_{\beta}u_{\gamma}V^{\beta}]
\\+\frac{1}{2}\kappa(\rho+p)[g_{\beta\delta}V^{\beta}V^{\delta}u_{\alpha}u_{\gamma}-g_{\beta\gamma}V^{\beta}V^{\delta}u_{\alpha}u_{\delta}]
+\frac{1}{3}(\kappa\rho+\Lambda)[g_{\alpha\gamma}g_{\beta\delta}V^{\beta}V^{\delta}-g_{\alpha\delta}g_{\beta\gamma}V^{\delta}V^{\beta}],
\end{multline*}
de donde obtenemos finalmente que
\begin{equation}
R_{\alpha\beta\gamma\delta}V^{\beta}V^{\delta}=\frac{1}{3}(\kappa\rho+\Lambda)[\epsilon
g_{\alpha\gamma}-V_{\alpha}V_{\gamma}]+\frac{1}{2}\kappa(\rho+p)[(u_{\beta}V^{\beta})^{2}g_{\alpha\gamma}-2(V_{\beta}u^{\beta})u_{(\alpha}V_{\gamma)}+\epsilon
u_{\alpha}u_{\gamma}],
\end{equation}
subiendo el primer índice del tensor de Riemann
\begin{equation}
R^{\alpha}_{\beta\gamma\delta}V^{\beta}V^{\delta}=\frac{1}{3}(\kappa\rho+\Lambda)(\epsilon\delta^{\alpha}_{\
\gamma}-V^{\alpha}V_{\gamma})+\frac{1}{2}\kappa(\rho+p)[(V_{\beta}u^{\beta})^{2}\delta^{\alpha}_{\
\gamma}-(V_{\beta}u^{\beta})u^{\alpha}V_{\gamma}-(V_{\beta}u^{\beta})V^{\alpha}u_{\gamma}],
\end{equation}
multiplicando por $\eta^{\gamma}$
\begin{multline}
R^{\alpha}_{\beta\gamma\delta}V^{\beta}\eta^{\gamma}V^{\delta}=\frac{1}{3}(\kappa\rho+\Lambda)[\epsilon\delta^{\alpha}_{\
\gamma}\eta^{\gamma}-V^{\alpha}(V_{\gamma}\eta^{\gamma})]+\frac{1}{2}\kappa(\rho+p)[(V_{\beta}u^{\beta})^{2}\delta^{\alpha}_{\
\gamma}\eta^{\gamma} \\
-(V_{\beta}u^{\beta})u^{\alpha}(V_{\gamma}\eta^{\gamma})-(V_{\beta}u^{\beta})V^{\alpha}(u_{\gamma}\eta^{\gamma})]
\end{multline}
usando ahora que $\eta_{\alpha}u^{\alpha}=0$ y $\eta_{\alpha}V^{\alpha}=0$ obtenemos finalmente \cite{Ellis1}
\begin{equation}\label{Pirani1}
\boxed{R_{\beta\gamma\delta}^{\alpha}V^{\beta}\eta^{\gamma}V^{\delta} = \biggl[\frac{1}{3}(\kappa \rho + \Lambda)\epsilon + \frac{1}{2}\kappa(\rho + p)E^2\biggr]\eta^{\alpha},}
\end{equation}
donde $E=-V_{\alpha}u^{\alpha}$. Esta es conocida en la
literatura como la \textit{ecuación de Pirani} \cite{Pirani}. Notemos que este término fuerza es proporcional al mismo
$\eta^{\alpha}$, i.e., de acuerdo con la EDG
\textit{solo la magnitud $\eta$ cambia a lo largo de la geodésica,
mientras que su orientación espacial permanece fija}.
Consecuentemente la EDG (\ref{GDE}) se reducirá simplemente a
una relación diferencial para la cantidad escalar $\eta$. Esto
refleja la isotropía espacial del tensor de curvatura de Riemann
alrededor de cada punto en la presente situación.\\\\
Consideraremos a continuación dos casos particulares de la GDE que nos permitirán determinar también las distancias cosmológicas en los universos de FLRW.
\subsection{EDG para Observadores Fundamentales}
\noindent En este caso, el campo vectorial geodésico $V^{\alpha}$
es el campo vectorial de la cuadrivelocidad del fluido
$u^{\alpha}$. El parámetro afín coincide con el tiempo propio de observadores fundamentales, i.e., $\nu=t$. Como estas son geodésicas
temporales, $\epsilon=-1$ y dado que los campos vectoriales están normalizados, $E=1$  de la ecuación (\ref{Pirani1}) se tiene
\begin{equation}\label{GDE1}
R^{\alpha}_{\beta\gamma\delta}u^{\beta}\eta^{\gamma}u^{\delta}=\left[\frac{\kappa}{6}(\rho + 3p)+\frac{1}{3}\Lambda\right]\eta^{\alpha}.
\end{equation}
Si el vector desviación es $\eta_{\alpha}= \ell e_{\alpha}$,
la isotropía implica que
\begin{equation}
\frac{D e^{\alpha}}{D t}=u^{\beta}\nabla_{\beta}e^{\alpha}=0,
\end{equation}
tenemos además que $e_{\alpha}e^{\alpha}=1$,
$e_{\alpha}u^{\alpha}=0$. Con base en esto, según (\ref{GDE1}) y la
EDG
\begin{equation}
\frac{D^{2}\eta^{\alpha}}{D t^{2}}=-R^{\alpha}_{\
\beta\gamma\delta}V^{\beta}\eta^{\gamma}V^{\delta}=-\left[\frac{\kappa}{6}(\rho+3p)-\frac{1}{3}\Lambda\right]\eta^{\alpha}.
\end{equation}
Dado que la derivada covariante cumple la regla de Leibnitz
\begin{align}
\frac{D\eta^{\alpha}}{D t}&=\frac{D}{D
t}(\ell e^{\alpha})=\frac{D \ell}{D
t}e^{\alpha}+\ell\left(\frac{D e^{\alpha}
}{D t}\right)=\frac{D \ell}{D t}e^{\alpha},\\
\frac{D^{2}\eta^{\alpha}}{D t^{2}}&=\frac{D}{D
t}\left(\frac{D\eta^{\alpha}}{D
t}\right)=\frac{D^{2}\ell}{D t^{2}}e^{\alpha}+\frac{D
\ell}{D t}\frac{D e^{\alpha}}{D t},\notag\\
\frac{D^{2}\eta^{\alpha}}{D
t^{2}}&=\frac{D^{2}\ell}{D t^{2}}e^{\alpha},
\end{align}
de modo que
\begin{equation}\label{53}
\boxed{ \frac{d^{2}\ell}{dt^{2}}=-\left[\frac{\kappa}{6}(\rho+3p)-\frac{1}{3}\Lambda\right]\ell,}
\end{equation}
Esta es la ecuación de Raychaudhuri \cite{Ellis1}. Escogiendo entonces $\ell=a$, siendo $a$ el factor de escala de modo que la ecuación de Raychaudhuri se reduce a
\begin{equation}\label{veintinueve}
\frac{\ddot{a}}{a}=-\frac{\kappa}{6}(\rho+3p)+\frac{\Lambda}{3},
\end{equation}
que corresponde a la ecuación de Friedmann (\ref{fried4}). Multiplicamos por $a \dot{a}$
\begin{equation}\label{54}
\ddot{a}\dot{a}+\frac{1}{6}\kappa(\rho+3p)a\dot{a}-\frac{1}{3}\Lambda
a\dot{a}=0,
\end{equation}
y usando la ecuación de conservación de la densidad de
materia (\ref{fried3}) podemos escribir \footnote{Escribimos esta ecuación multiplicando a ambos lados por $\kappa = 8\pi G$, la cual no quita generalidad a la expresión y se usa solo por estar de acuerdo con nuestras unidades.}
\begin{equation}
\kappa\dot{\rho}=-3\frac{\dot{a}}{a}\kappa(\rho+p),
\end{equation}
de esta manera, al considerar la derivada de $\kappa \rho a^{2}$
\begin{align}
\frac{d(\kappa \rho a^{2})}{dt}&=a^{2}\kappa\dot{\rho}+2a\kappa\rho\dot{a},\notag\\
&=a^{2}\left(-3\frac{\dot{a}}{a}\kappa(\rho+p)\right)+2a\kappa\rho\dot{a},
\end{align}
reorganizando términos
\begin{equation}\label{55b}
\frac{d(\kappa \rho a^{2})}{dt}=-\kappa(\rho+3p)a\dot{a}.
\end{equation}
Reemplazando \eqref{55b} en \eqref{54}
\begin{equation}
\ddot{a}\dot{a}-\frac{1}{6}\frac{d(\kappa\rho
a^{2})}{dt}-\frac{1}{3}\Lambda a\dot{a}=0,
\end{equation}
el primer término se puede escribir como la derivada del cuadrado de la derivada de $a(t)$
\begin{equation}
\frac{1}{2}\frac{d}{dt}\dot{a}^2-\frac{1}{6}\frac{d(\kappa\rho
a^{2})}{dt}-\frac{\Lambda}{6}\frac{d}{dt}a^{2}=0,
\end{equation}
escribiendo todo como una sola derivada
\begin{equation}
\frac{d}{dt}\left[\dot{a}^2-\frac{\kappa}{3}\rho
a^{2}-\frac{1}{3}\Lambda a^{2}\right]=0,
\end{equation}
por lo tanto, lo que está dentro del paréntesis debe ser igual a una constante $-k$
\begin{equation}
\dot{a}^{2}-\frac{1}{3}\kappa\rho
a^{2}-\frac{1}{3}\Lambda a^{2}=-k,
\end{equation}
dividiendo por $a^{2}$ tenemos
\begin{equation}
H^2= \frac{\kappa}{3}\rho + \frac{\Lambda}{3}-\frac{k}{a^{2}},
\end{equation}
en donde hemos usado de nuevo la definición del parámetro de Hubble $H = \frac{\dot{a}}{a}$. Vemos que podemos recobrar
las ecuaciones dinámicas estándares de los modelos de FLRW a partir de la EDG.
\subsection{EDG para Campos Vectoriales Nulos}\label{Seccion3.4}
\noindent La EDG para campos vectoriales nulos nos dará de forma elegante las ecuaciones diferenciales para las distancias, pues considera la separación de geodésicas nulas (rayos de luz o fotones). \noindent Debemos considerar la EDG para campos vectoriales nulos
dirigidos al pasado. En este caso $V^{\alpha}=k^{\alpha}$,
$k_{\alpha}k^{\alpha}=0$, $k^{0}<0$. La ecuación de Pirani (\ref{Pirani1}) ahora es
\begin{equation}
R_{\beta\gamma\delta}^{\alpha}k^{\beta}\eta^{\gamma}k^{\delta}=\frac{1}{2}\kappa(\rho+p)E^{2}\eta^{\alpha},
\end{equation}
que se denomina \textit{Enfocamiento de Ricci}. Así, si escribimos $\eta^{\alpha}=\eta e^{\alpha}$,
$e_{\alpha}e^{\alpha}=1$,
$e_{\alpha}u^{\alpha}=e_{\alpha}k^{\alpha}=0$ y usando una base
alineada y propagada paralelamente, $\frac{\delta
e^{\alpha}}{\delta\nu}=k^{\beta}\nabla_{\beta}e^{\alpha}=0$, por
la EDG encontramos
\begin{equation}\label{setentaynueve}
\frac{d^{2}\eta}{d\nu^{2}}=-\frac{1}{2}\kappa(\rho+p)E^{2}\eta.
\end{equation}
En este caso, todas las geodésicas nulas dirigidas al
pasado o al futuro experimentan enfocamiento si $\kappa(\rho+p)>0$
(mientras que el signo de $\Lambda$ no tiene importancia).\\\\
Se pretende obtener la ecuación \eqref{setentaynueve}
escrita en términos del parámetro no afín redshift $z$. Veamos
como se transforman los operadores diferenciales
\begin{equation}
\frac{d}{d\nu} = \frac{dz}{d\nu}\frac{d}{dz},
\end{equation}
\begin{align}
\frac{d^2}{d\nu^2} &= \frac{dz}{d\nu}\frac{d}{dz}\biggl(\frac{d}{d\nu}\biggr), \notag \\
& = \biggl(\frac{d\nu}{dz}\biggr)^{-2}\biggl[-\biggl(\frac{d\nu}{dz}\biggr)^{-1}\frac{d^2\nu}{dz^2}\frac{d}{dz}+\frac{d^2}{dz^2}\biggr].
\end{align}
Para geodésicas nulas tenemos que
\begin{equation}
(1+z) = \frac{a_0}{a}=\frac{E}{E_0} \quad \longrightarrow \quad \frac{dz}{1+z}=- \frac{da}{a},
\end{equation}
con $a$ el factor de escala, y $a_0=1$ el valor presente ($a_0=1$. Entonces para el caso dirigido al pasado
\begin{equation}
dz = (1+z) \frac{1}{a}\frac{da}{d\nu}\, d\nu = (1+z) \frac{\dot{a}}{a} E\, d\nu = E_0 H (1+z)^2\, d\nu,
\end{equation}
donde de nuevo usamos la definición para el parámetro de Hubble. Así obtenemos
\begin{equation}
\frac{d\nu}{dz} = \frac{1}{E_0 H (1+z)^2},
\end{equation}
y
\begin{equation}
\frac{d^2\nu}{dz^2} =- \frac{1}{E_0 H (1+z)^3}\biggl[\frac{1}{H}(1+z)\frac{dH}{dz}+2\biggr],
\end{equation}
expresamos $\frac{dH}{dz}$ como
\begin{equation}
\frac{dH}{dz} = \frac{d\nu}{dz}\frac{dt}{d\nu}\frac{dH}{dt} = - \frac{1}{H(1+z)} \frac{dH}{dt},
\end{equation}
el signo menos viene de la condición de geodésicas nulas dirigidas al pasado, cuando $z$ aumenta, $\nu$ disminuye. Usamos también que $\frac{dt}{d\nu} = E_0 (1+z)$.
Ahora, de la definición del parámetro de Hubble $H$ tenemos
\begin{equation}
\dot{H} \equiv \frac{dH}{dt} = \frac{d}{dt}\frac{\dot{a}}{a} = \frac{\ddot{a}}{a} - H^2,
\end{equation}
de la ecuación de Raychaudhuri \eqref{veintinueve} se obtiene
\begin{equation}
\frac{d^{2}\nu}{dz^{2}}=-\frac{3}{E_{0}H(1+z)^{3}}\left[1+\frac{\kappa}{18H^{2}}(\rho+3p)-\frac{\Lambda}{9H^{2}}\right],
\end{equation}
y
\begin{equation}
\frac{d^{2}\eta}{d\nu^{2}}=\left[EH(1+z)\right]^{2}\left[\frac{d^{2}\eta}{dz^{2}}+\frac{3}{1+z}\left(1+\frac{\kappa}{18H^{2}}(\rho+3p)-\frac{\Lambda}{9H^{2}}\right)\frac{d\eta}{dz}\right].
\end{equation}
De esta manera, la ecuación \eqref{setentaynueve} se
convierte en
\begin{equation}
(EH(1+z))^2\left[\frac{d^{2}\eta}{dz^{2}}+\frac{3}{1+z}\left(1+\frac{\kappa}{18H^{2}}(\rho+3p)-\frac{\Lambda}{9H^{2}}\right)\frac{d\eta}{dz}\right]=-\frac{\kappa}{2}(\rho+p)E^{2}\eta,
\end{equation}
dividiendo por $E^2$ y organizando términos
\begin{equation}\label{ochentayseis}
\frac{d^{2}\eta}{dz^{2}}+\frac{3}{1+z}\left[1+\frac{\kappa}{18H^{2}}(\rho+3p)-\frac{\Lambda}{9H^{2}}\right]\frac{d\eta}{dz}+\frac{1}{2(1+z)^{2}}\frac{\kappa}{H^{2}}(\rho+p)\eta=0.
\end{equation}
Si consideramos de nuevo una mezcla no interactuante de materia y de radiación
\begin{equation}
\kappa\rho = 3H_0^2 \Omega_{m0}(1 + z)^3 + 3H_0^2\Omega_{r0}(1 + z)^4, \qquad \kappa p = H_0^2\Omega_{r0}(1 + z)^4,
\end{equation}
usemos además la ecuación de Friedmann
\begin{equation}
H^2 = H_0^2\bigl[\Omega_{m0}(1 + z)^3 + \Omega_{r0}(1 + z)^4 + \Omega_{\Lambda}  + \Omega_{k}(1+z)^2\bigr],
\end{equation}
de modo que la EDG para campos vectoriales nulos es
\begin{multline}\label{GDEnull1}
\frac{d^2\eta}{dz^2} + \frac{4\Omega_{r0}(1 + z)^4 + (7/2)
\Omega_{m0}(1 + z)^3 + 3\Omega_{k0}(1 + z)^2 + 2\Omega_{\Lambda}
}{(1+z)\bigl(\Omega_{r0}(1 + z)^4 + \Omega_{m0}(1 + z)^3 + \Omega_{k0}(1 + z)^2 + \Omega_{\Lambda}\bigr)}\, \frac{d\eta}{dz}\\ + \frac{2\Omega_{r0}(1 + z)^2 + (3/2)
\Omega_{m0}(1 + z)}{\Omega_{r0}(1 + z)^4 + \Omega_{m0}(1 + z)^3 + \Omega_{k0}(1 + z)^2 + \Omega_{\Lambda}}\, \eta = 0.
\end{multline}
La relación de Mattig es obtenida en el caso $\Omega_{\Lambda} = 0$ y escribiendo $\Omega_{k0}=1-\Omega_{m0} - \Omega_{r0}$ con lo cual obtenemos \cite{Ellis1}
\begin{multline}
\frac{d^2\eta}{dz^2} + \frac{6 +
\Omega_{m0}(1 + 7z) + \Omega_{r0}(1 + 8z + 4z^2)}{2(1 + z)(1 + \Omega_{m0}z +
\Omega_{r0}z(2 + z))}\, \frac{d\eta}{dz}\\ + \frac{3\Omega_{m0} + 4\Omega_{r0}(1 + z)}{2(1 + z)(1 + \Omega_{m0}z +
\Omega_{r0}z(2 + z))}\, \eta = 0.
\end{multline}
Vamos a considerar ahora la relación entre la magnitud del vector desviación $\eta$ con la distancia diametral angular. En un espacio-tiempo esféricamente simétrico, como el modelo FLRW, la magnitud del vector desviación $\eta$ se relaciona con el área propia $dA$ de una fuente a redshift $z$ por $d\eta \propto \sqrt{dA}$, y de esta expresión, la definición de la distancia diametral angular $d_A$ puede ser escrita como
\begin{equation}
d_A = \sqrt{\frac{dA}{d\Omega}},
\end{equation}
con $d\Omega$ el ángulo sólido. De modo que la forma de la relación de Mattig es la misma cambiando $\eta$ por $d_A$ (el factor de proporcionalidad se puede descartar de la ecuación diferencial). Así tendremos la siguiente ecuación diferencial para la distancia diametral angular
\begin{multline}
\frac{d^2 d_A}{dz^2} + \frac{6 +
\Omega_{m0}(1 + 7z) + \Omega_{r0}(1 + 8z + 4z^2)}{2(1 + z)(1 + \Omega_{m0}z +
\Omega_{r0}z(2 + z))}\, \frac{d d_A}{dz}\\ + \frac{3\Omega_{m0} + 4\Omega_{r0}(1 + z)}{2(1 + z)(1 + \Omega_{m0}z +
\Omega_{r0}z(2 + z))}\, d_A = 0.
\end{multline}
Fijando entonces los parámetros cosmológicos y resolviendo esta ecuación diferencial, podemos encontrar todas las expresiones para las distancias cosmológicas, usando para esto las relaciones existentes entre ellas (\ref{disangular}), (\ref{dislumin}).
\newpage

{\color{white} . } 
\chapter{Ecuaciones de Campo en Teorías $f(R)$}\label{Capitulo4}
\begin{flushright}
\textit{``The next case in simplicity includes those manifoldnesses in\\
which the line-element may be expressed as the fourth-root\\
of a quartic differential expression''\\
Bernard Riemann, 1854}
\end{flushright}

\drop{A}unque la Relatividad General se supone válida, y se han probado varias soluciones experimentalmente, desde su génesis se han desarrollado teorías alternativas. En particular, Weyl y Eddington en 1919 propusieron teorías de gravedad que consistían en lagrangianos cuadráticos (proporcionales al cuadrado del escalar de Ricci), y se han estudiado las soluciones para estos espacio-tiempos, en particular la de Schwarzschild \cite{Pechlaner}. \\\\
El punto de partida fundamental de estas teorías de gravedad modificada consisten entonces en los lagrangianos y hamiltonianos que describen el campo gravitacional. La primera acción como vimos es la de Einstein-Hilbert
\begin{equation}
S_{EH} = \int_{\mathcal{V}} d^4x\, \sqrt{-g}R,
\end{equation}
como mostramos en el Capítulo \ref{Capitulo2} las ecuaciones de campo de Einstein se encuentran a partir de un principio variacional de esta acción. La acción para una teoría de gravedad cuadrática puede entonces escribirse como \cite{Wands}
\begin{equation}
S_{R^2} = \int_{\mathcal{V}} d^4x\, \sqrt{-g}(R + \alpha R^2),
\end{equation}
en donde $\alpha$ es una constante con que permite tener el lagrangiano en la acción con las unidades correctas. Esta acción con un lagrangiano cuadrático conduce a unas ecuaciones de campo que son de cuarto orden en la métrica, lo que conlleva a que surjan nuevos grados de libertad en la teoría. Puede además pensarse en teorías con lagrangianos compuestos con todos los posibles escalares que pueden obtenerse a partir del tensor de Riemann y sus derivadas, sin embargo tales teorías poseen el mismo problema de grados adicionales.\\\\
Después de que Einstein propusiera la RG y Hilbert encontrara el lagrangiano que la describe, Kretschmann \cite{Kretschmann} propuso una acción construida con el escalar $R_{\alpha\eta\gamma\delta}R^{\alpha\eta\gamma\delta}$ en vez del escalar de Ricci, el cual fue denominado tiempo después como el \textit{escalar de Kretschmann}
\begin{equation}
S_{K} = \int_{\mathcal{V}} d^4x\, \sqrt{-g}R_{\alpha\eta\gamma\delta}R^{\alpha\eta\gamma\delta}.
\end{equation}
Puesto que el tensor de Riemann $R_{\alpha\eta\gamma\delta}$ es el tensor fundamental para la gravitación y el escalar de Kretschamn es su cuadrado, este lagrangiano parece ser una buena opción; además, esta teoría preserva las identidades de Bianchi. Sin embargo, la aparición de términos de la forma $\nabla_{\alpha}\nabla_{\beta}R_{\, \, (\gamma\delta)}^{\alpha \, \, \, \, \beta}$, conduce a fijar los elementos de la métrica, sus primeras, segundas y terceras derivadas, de modo que la teoría introduce más grados de libertad en comparación con RG. La aparición de estos términos de cuarto orden en la métrica se puede evitar considerando el escalar de Gauss-Bonnet \cite{Stelle}
\begin{equation}
\mathcal{G} = R^2 - 4R_{\alpha\beta}R^{\alpha\beta} + R_{\alpha\beta\gamma\delta}R^{\alpha\beta\gamma\delta},
\end{equation}
de modo que la acción pueda escribirse como
\begin{equation}
S_{\mathcal{G}} = \int_{\mathcal{V}} d^4x\, \sqrt{-g}\mathcal{G}.
\end{equation}
Existe sin embargo una teoría más general de gravedades modificadas denominadas \textit{Teorías de Lovelock} en la cual se da una generalización del escalar de Gauss-Bonnet \cite{Lovelock1},\cite{Lovelock2}. Tales teorías se basan en un lagrangiano general de la forma
\begin{equation}
\mathcal{L}_{L} = \sqrt{-g} \sum_{n} \frac{1}{2^n}\alpha_n \delta_{\gamma_1\, \gamma_2 \cdots \gamma_{2n}}^{\beta_1\, \beta_2 \cdots \beta_{2n}}\, R_{\gamma_1\, \gamma_2}^{\beta_1 \, \beta_2} \cdots R_{\gamma_{2n-1}\, \gamma_{2n}}^{\beta_{2n-1} \, \beta_{2n}},
\end{equation}
en donde $\delta_{\gamma_1\, \gamma_2 \cdots \gamma_{2n}}^{\beta_1\, \beta_2 \cdots \beta_{2n}}$ es la delta de Kronecker generalizada de orden $2n$, y $\alpha_n$ son constantes. \\\\
Otra posibilidad interesante, pero que después de su aparición fue en cierto modo olvidada, son las teorías $C^2$ siendo $C$ la traza del tensor de Weyl, y denominadas \textit{teorías de gravedad conforme}, que tienen la particularidad de ser una teoría localmente invariante, y una de las candidatas para la formulación cuántica de la gravedad. Las ecuaciones de campo en esta gravedad conforme fueron publicadas primeramente en \cite{Bach} y su estudio se justifica en gran medida por ser una alternativa físicamente viable a problemas que la cosmología estándar posee, en particular la hipótesis de la materia oscura.\\\\
Algunas extensiones de la acción de Einstein-Hilbert a modelos con escalares de orden superior son consideradas en \cite{Schmidt}-\cite{Querella}. Como una extensión  natural de RG y de las  teorías de orden superior surgen las \textit{teorías $f(R)$} al considerar una función arbitraria del escalar de Ricci en vez de solamente el escalar como en el caso de la acción de Einstein-Hilbert \cite{Felice}. En las teorías de gravedad modificada $f(R)$ debemos distinguir 3 casos diferentes, los cuales nos van a  conducir a diferentes ecuaciones de campo. Estas 3 versiones se pueden resumir en
\begin{enumerate}
\item Formalismo Métrico: Para este caso escribimos la acción como
\begin{equation}\label{accion1}
S_{mod} = S_{met} + S_M(g_{\alpha\beta},\psi), \quad S_{met} = \frac{1}{2\kappa}\int_{\mathcal{V}} d^4x\, \sqrt{-g} f(R),
\end{equation}
 con $\kappa = 8\pi G$. Aquí la variación se hace con respecto a $g^{\alpha\beta}$ y $\psi$ denota todos los campos de materia.
\item Formalismo de Palatini: En este caso la métrica $g^{\alpha\beta}$ y la conexión $\Gamma_{\beta\gamma}^{\alpha}$ son independientes. Sea entonces $\mathcal{R}_{\alpha\beta}$ el tensor de Ricci construido con la conexión independiente y $\mathcal{R}=g^{\alpha\beta}\mathcal{R}_{\alpha\beta}$. La acción puede escribirse como
\begin{equation}\label{accion2}
S_{Pal} = \frac{1}{2\kappa}\int_{\mathcal{V}} d^4x\, \sqrt{-g} f(\mathcal{R}) .
\end{equation}
La acción total incluye la acción asociada a todos los campos de materia, la cual es \textit{independiente} de la conexión.
\item Formalismo Métrico-Afín: Esta forma es similar a la de Palatini, pero en este caso la acción de materia \textit{depende} de la conexión
\begin{equation}\label{accion3}
S_{MA} = \frac{1}{2\kappa}\int_{\mathcal{V}} d^4x\, \sqrt{-g} f(\mathcal{R}) + S_M(g_{\alpha\beta},\Gamma_{\beta\gamma}^{\alpha},\psi).
\end{equation}
\end{enumerate}

\noindent A continuación mostraremos la obtención de las ecuaciones de campo en los 3 formalismos, dando especial atención a las ecuaciones de campo en el formalismo métrico, incluyendo términos de frontera.
\section{Ecuaciones de Campo en el Formalismo Métrico}
\noindent  Las ecuaciones de campo en el formalismo métrico fueron deducidas en \cite{Buchdahl}, e incluyendo términos de frontera en gravedades de cuarto orden en \cite{Barth}. El termino similar al de Gibbons-York-Hawking en teorías $f(R)$ fue explorado en \cite{Madsen}, con un principio variacional aumentado en \cite{Francaviglia}, y usando un marco tensor-escalar en \cite{Casadio}-\cite{Dyer}.\\\\
Consideramos de nuevo el espacio-tiempo como un par $(\mathcal{M},g)$ con $\mathcal{M}$ una variedad cuadri-dimensional y  $g_{\alpha\beta}$ una métrica sobre $\mathcal{M}$. El lagrangiano ahora es una función arbitraria del escalar de Ricci $\mathcal{L}[g_{\alpha\beta}] = f(R)$, de nuevo asumimos que la variedad posee una conexión de Levi-Civita, es decir, un símbolo de Christoffel. La acción general que incluye un termino de frontera se puede escribir como \cite{Dyer}
\begin{equation}\label{accionF}
S_{mod} = \frac{1}{2\kappa}\bigl(S_{met} + S'_{GYH}\bigr) + S_{M},
\end{equation}
con
\begin{equation} \label{accionmet}
S_{met} = \int_{\mathcal{V}} d^4x\, \sqrt{-g}f(R),
\end{equation}\\
y el termino del tipo Gibbons-York-Hawking \cite{Dyer}
\begin{equation}
S'_{GYH} = 2\oint_{\partial \mathcal{V}} d^3y\, \varepsilon\sqrt{|h|} f'(R)K,
\end{equation}
con $f'(R) = df(R)/dR$. De nuevo, $S_M$ representa la acción asociada con todos los campos de materia
(\ref{matteraction}). Vamos entonces a encontrar las ecuaciones de campo en el formalismo métrico de gravedad $f(R)$ siguiendo \cite{Guarnizo}. Realizamos la variación con respecto al tensor métrico sujeto a la condición
\begin{equation}\label{frontera2}
\delta g_{\alpha\beta}\biggl|_{\partial \mathcal{V}} =0.
\end{equation}
Primero realizamos la variación del término $S_{met}$
\begin{equation}\label{varaccion2}
\delta S_{met} = \int_{\mathcal{V}} d^4x \, \bigl(f(R)\delta\sqrt{-g} + \sqrt{-g}\, \delta f(R)\bigr),
\end{equation}
la derivada funcional de $f(R)$ se puede escribir como
\begin{equation}
\delta f(R) = f'(R) \delta R.
\end{equation}
Usando la expresión para la variación del escalar de Ricci
\begin{equation}
\delta R = \delta g^{\alpha\beta}R_{\alpha\beta} + \nabla_{\sigma}\bigl(g^{\alpha\beta}(\delta\Gamma_{\beta\alpha}^{\sigma}) - g^{\alpha\sigma}
(\delta\Gamma_{\alpha\gamma}^{\gamma})\bigr),
\end{equation}
en donde la variación del termino $g^{\alpha\beta}(\delta\Gamma_{\beta\alpha}^{\sigma}) - g^{\alpha\sigma}
(\delta\Gamma_{\alpha\gamma}^{\gamma})$ esta dada en Apéndice \ref{Appendix A}.
Con este resultado tendremos
\begin{align}
\delta R &= \delta g^{\alpha\beta}R_{\alpha\beta} + \nabla_{\sigma
}\bigl(g^{\alpha\beta}(\delta\Gamma_{\beta\alpha}^{\sigma}) - g^{\alpha\sigma}
(\delta\Gamma_{\alpha\gamma}^{\gamma})\bigr),\nonumber \\
&= \delta g^{\alpha\beta}R_{\alpha\beta} + g_{\mu\nu}\nabla_{\sigma}\nabla^{\sigma}(\delta g^{\mu\nu}) - \nabla_{\sigma}\nabla_{\gamma}(\delta g^{\sigma\gamma}),\nonumber \\
&= \delta g^{\alpha\beta}R_{\alpha\beta} + g_{\alpha\beta}\square(\delta g^{\alpha\beta}) - \nabla_{\alpha}\nabla_{\beta}(\delta g^{\alpha\beta}).
\end{align}
Con $\square \equiv\nabla_{\sigma}\nabla^{\sigma}$ y renombramos algunos índices mudos. Reemplazando estos resultados en la variación de la acción (\ref{varaccion2})
\begin{align}\label{varaccionf}
\delta S_{met} &= \int_{\mathcal{V}} d^4x \, \bigl(f(R)\delta\sqrt{-g} + \sqrt{-g}\, f'(R) \delta R\bigr), \nonumber \\
&= \int_{\mathcal{V}} d^4x \, \biggl(-\frac{1}{2}f(R)\sqrt{-g}\, g_{\alpha\beta}\delta g^{\alpha\beta} + f'(R)\sqrt{-g}\Bigl(\delta g^{\alpha\beta}R_{\alpha\beta} + g_{\alpha\beta}\square(\delta g^{\alpha\beta}) - \nabla_{\alpha}\nabla_{\beta}(\delta g^{\alpha\beta})\Bigr)\biggr),\nonumber \\\
&= \int_{\mathcal{V}} d^4x \, \sqrt{-g} \biggl(f'(R)\Bigl(\delta g^{\alpha\beta}R_{\alpha\beta} + g_{\alpha\beta}\square(\delta g^{\alpha\beta}) - \nabla_{\alpha}\nabla_{\beta}(\delta g^{\alpha\beta})\Bigr)-\frac{1}{2}f(R)\, g_{\alpha\beta}\delta g^{\alpha\beta}\biggr).
\end{align}
Consideramos ahora las siguientes integrales
\begin{equation}\label{integrals}
\int_{\mathcal{V}} d^4x \, \sqrt{-g} f'(R)g_{\alpha\beta}\square(\delta g^{\alpha\beta}), \qquad \int_{\mathcal{V}} d^4x \, \sqrt{-g} f'(R)\nabla_{\alpha}\nabla_{\beta}(\delta g^{\alpha\beta}).
\end{equation}
Vamos a escribir estas integrales de una forma diferente realizando para esto una integración por partes. Definimos las siguientes cantidades
\begin{equation}\label{M}
M_{\tau} = f'(R)g_{\alpha\beta}\nabla_{\tau}(\delta g^{\alpha\beta}) - \delta g^{\alpha\beta} g_{\alpha\beta}\nabla_{\tau}(f'(R)),
\end{equation}
y
\begin{equation}\label{N}
N^{\sigma} = f'(R)\nabla_{\gamma}(\delta g^{\sigma\gamma}) - \delta g^{\sigma\gamma}\nabla_{\gamma}(f'(R)).
\end{equation}
La combinación $g^{\sigma\tau}M_{\tau} + N^{\sigma}$ es
\begin{equation}
g^{\sigma\tau}M_{\tau} + N^{\sigma} = f'(R)g_{\alpha\beta}\nabla^{\sigma}(\delta g^{\alpha\beta}) - \delta g^{\alpha\beta} g_{\alpha\beta}\nabla^{\sigma}(f'(R)) +
f'(R)\nabla_{\gamma}(\delta g^{\sigma\gamma}) - \delta g^{\sigma\gamma}\nabla_{\gamma}(f'(R)),
\end{equation}
en el caso particular $f(R) = R$, la anterior combinación se reduce a la expresión (\ref{V}) con la ecuación (\ref{vargamma}). Las cantidades $M_{\tau}$ y $N^{\sigma}$ nos permiten escribir la variación del termino (\ref{varaccionf}) de la siguiente forma (ver Apéndice \ref{Appendix B})
\begin{multline}\label{varf1}
\delta S_{met} = \frac{1}{2\kappa}\int_{\mathcal{V}} d^4x \, \sqrt{-g} \biggl(f'(R)R_{\alpha\beta} + g_{\alpha\beta}\square f'(R) - \nabla_{\alpha}\nabla_{\beta}f'(R)-\frac{1}{2}f(R)\, g_{\alpha\beta}\biggr)\delta g^{\alpha\beta}\\ + \oint_{\partial \mathcal{V}} d^{3}y\, \varepsilon\sqrt{|h|}n^{\tau}M_{\tau} + \oint_{\partial \mathcal{V}} d^{3}y\, \varepsilon\sqrt{|h|}n_{\sigma}N^{\sigma}.
\end{multline}
A continuación estudiaremos la contribución en la frontera (\ref{varf1}), y mostraremos que este término se cancela con las variaciones de la acción $S'_{GYH}$.
\subsection{Término de Frontera en el Formalismo Métrico $f(R)$}
\noindent Expresamos las cantidades $M_{\tau}$ y $N^{\sigma}$ calculadas en la frontera $\partial \mathcal{V}$. Es conveniente primero expresar estas cantidades en función de las variaciones $\delta g_{\alpha\beta}$. Usando la ecuación (\ref{varmet1}) en (\ref{M}) y (\ref{N}) obtenemos
\begin{equation}\label{M1}
M_{\tau} = -f'(R)g^{\alpha\beta}\nabla_{\tau}(\delta g_{\alpha\beta}) + g^{\alpha\beta}\delta g_{\alpha\beta} \nabla_{\tau}(f'(R)),
\end{equation}
y
\begin{equation}\label{N1}
N^{\sigma} = -f'(R)g^{\sigma\mu}g^{\gamma\nu}\nabla_{\gamma}(\delta g_{\mu\nu}) + g^{\sigma\mu}g^{\gamma\nu}\delta g_{\mu\nu}\nabla_{\gamma}(f'(R)).
\end{equation}
Para evaluar estas cantidades en la frontera usamos el hecho que  $\delta g_{\alpha\beta}|_{\partial \mathcal{V}}=\delta g^{\alpha\beta}|_{\partial \mathcal{V}}=0$, de modo que los únicos términos no nulos son las derivadas de $\delta g_{\alpha\beta}$ en las derivadas covariantes. Así tendremos
\begin{equation}\label{M2}
M_{\tau}\biggl|_{\partial \mathcal{V}} =-f'(R)g^{\alpha\beta}\partial_{\tau}(\delta g_{\alpha\beta}),
\end{equation}
y
\begin{equation}\label{M1}
N^{\sigma}\biggl|_{\partial \mathcal{V}} = -f'(R)g^{\sigma\mu}g^{\gamma\nu}\partial_{\gamma}(\delta g_{\mu\nu}),
\end{equation}
Calculamos ahora $n^{\tau}M_{\tau}\bigl|_{\partial \mathcal{V}}$ y $n_{\sigma}N^{\sigma}\bigl|_{\partial \mathcal{V}}$ que son los términos en las integrales sobre la frontera (\ref{varf1})
\begin{align}
n^{\tau}M_{\tau}\biggl|_{\partial \mathcal{V}} &= -f'(R)n^{\tau}(\varepsilon n^{\alpha}n^{\beta}+h^{\alpha\beta})\partial_{\tau}(\delta g_{\alpha\beta}), \nonumber\\
&= -f'(R)n^{\sigma}h^{\alpha\beta}\partial_{\sigma}(\delta g_{\alpha\beta}),
\end{align}
donde renombramos el índice mudo $\tau$. Por otro lado
\begin{align}
n_{\sigma}N^{\sigma}\biggl|_{\partial \mathcal{V}} &= -f'(R)n_{\sigma}(h^{\sigma\mu}+\varepsilon n^{\sigma}n^{\mu})(h^{\gamma\nu}+\varepsilon n^{\gamma}n^{\nu})\partial_{\gamma}(\delta g_{\mu\nu}), \nonumber\\
&= -f'(R)n^{\mu}(h^{\gamma\nu}+\varepsilon n^{\gamma}n^{\nu})\partial_{\gamma}(\delta g_{\mu\nu}), \nonumber\\
&= -f'(R)n^{\mu}h^{\gamma\nu}\partial_{\gamma}(\delta g_{\mu\nu}), \nonumber\\
&= 0,
\end{align}
donde usamos que $n_{\sigma}h^{\sigma\mu}=0$, $\varepsilon^2=1$ y el hecho de que la derivada tangencial $h^{\gamma\nu}\partial_{\gamma}(\delta g_{\mu\nu})$ se anula. Con estos resultados la variación de la acción $S_{met}$ es
\begin{multline}\label{varf2}
\delta S_{met} = \frac{1}{2\kappa}\int_{\mathcal{V}} d^4x \, \sqrt{-g} \biggl(f'(R)R_{\alpha\beta} + g_{\alpha\beta}\square f'(R) - \nabla_{\alpha}\nabla_{\beta}f'(R)-f(R)\frac{1}{2}\, g_{\alpha\beta}\biggr)\delta g^{\alpha\beta}\\ -\oint_{\partial \mathcal{V}} d^{3}y\, \varepsilon\sqrt{|h|}f'(R)n^{\sigma}h^{\alpha\beta}\partial_{\sigma}(\delta g_{\alpha\beta}).
\end{multline}
Procedemos ahora con el termino de frontera $S'_{GYH}$ en la acción total. La variación de este término es
\begin{align}
\delta S'_{GYH} &= 2\oint_{\partial \mathcal{V}} d^3y \, \varepsilon\sqrt{|h|}\bigl(\delta f'(R)K + f'(R)\delta K\bigr), \nonumber \\
&= 2\oint_{\partial \mathcal{V}} d^3y \, \varepsilon\sqrt{|h|}\bigl(f''(R)\delta R\, K + f'(R)\delta K\bigr).
\end{align}
Usando la expresión para la variación de $K$, ecuación (\ref{deltaK}), podemos escribir
\begin{align}\label{varfront}
\delta S'_{GYH} &= 2\oint_{\partial \mathcal{V}} d^3y \, \varepsilon\sqrt{|h|}\biggl(f''(R)\delta R\, K + \frac{1}{2}f'(R)h^{\alpha\beta}\partial_{\sigma}(\delta g_{\beta\alpha})n^{\sigma}\biggr), \nonumber \\
&= 2\oint_{\partial \mathcal{V}} d^3y \, \varepsilon\sqrt{|h|}f''(R)\delta R\, K + \oint_{\partial \mathcal{V}} d^3y \, \varepsilon\sqrt{|h|}f'(R)h^{\alpha\beta}\partial_{\sigma}(\delta g_{\beta\alpha})n^{\sigma}.
\end{align}
Observamos que el segundo término en (\ref{varfront}) cancela el término de frontera en la variación (\ref{varf2}), y adicionalmente necesitamos imponer $\delta R = 0$ en la frontera. Un argumento similar esta dado en \cite{Dyer}.\\\\
Finalmente, con la variación de la acción de materia, dada en (\ref{variacionener}), la variación total de la acción en el formalismo métrico de gravedad $f(R)$ es
\begin{multline}
\delta S_{mod} = \frac{1}{2\kappa}\int_{\mathcal{V}} d^4x \, \sqrt{-g} \biggl(f'(R)R_{\alpha\beta} + g_{\alpha\beta}\square f'(R) - \nabla_{\alpha}\nabla_{\beta}f'(R)-\frac{1}{2}f(R)\, g_{\alpha\beta}\biggr)\delta g^{\alpha\beta}\\ -\frac{1}{2}\int_{\mathcal{V}} d^4x\,\sqrt{-g} T_{\alpha\beta}\delta g^{\alpha\beta}.
\end{multline}
Imponiendo que esta variación sea estacionaria tendremos
\begin{equation}\label{variacionfinal}
\frac{1}{\sqrt{-g}}\frac{\delta S_{mod}}{\delta g^{\alpha\beta}} = 0  \Longrightarrow f'(R)R_{\alpha\beta} + g_{\alpha\beta}\square f'(R) - \nabla_{\alpha}\nabla_{\beta}f'(R)-\frac{1}{2}f(R)\, g_{\alpha\beta} = \kappa T_{\alpha\beta},
\end{equation}
de modo que las ecuaciones de campo en el formalismo métrico $f(R)$ son
\begin{equation}\label{ecuacionescampo}
\boxed{f'(R)R_{\alpha\beta}-\frac{1}{2}f(R)\, g_{\alpha\beta} + g_{\alpha\beta}\square f'(R) - \nabla_{\alpha}\nabla_{\beta}f'(R) = \kappa T_{\alpha\beta}.}
\end{equation}
Estas ecuaciones son claramente ecuaciones de cuarto orden en la métrica (por la contribución de segundas derivadas en $R$ de los operadores $\square$ y $\nabla_{\alpha}\nabla_{\beta}$). La traza de estas ecuaciones de campo es
\begin{equation}\label{tracefield}
\boxed{f'(R)R - 2f(R) + 3\square f'(R) = \kappa T,}
\end{equation}
en donde podemos observar que a diferencia de la RG, la relación entre $R$ y $T$ pasa de ser una simple relación algebraica a una relación diferencial para $R$. Este hecho conlleva a resultados interesantes. El teorema de Birkhoff que asegura que la solución de Schwarzschild es la solución más general a las ecuaciones de campo en el vacío, deja de ser valido en teorías $f(R)$. En RG la relación entre $R$ y $T$ viene dada por $R= -\kappa T$, lo que implica que en el vacío las ecuaciones se reducen a $R=0$; en este caso sin embargo si $T=0$ no necesariamente se obtiene un escalar nulo. \\\\
Es posible escribir las ecuaciones de campo en la forma de las ecuaciones de Einstein con un tensor de energía-momentum efectivo compuesto de términos de curvatura. Para esto escribamos la expresión (\ref{ecuacionescampo}) en la forma
\begin{align}
G_{\alpha\beta} &\equiv R_{\alpha\beta} - \frac{1}{2}R\, g_{\alpha\beta},\notag\\
& = \frac{\kappa T_{\alpha\beta}}{f'(R)} + g_{\alpha\beta}\frac{[f(R) - Rf'(R)]}{2f'(R)} + \frac{[\nabla_{\alpha}\nabla_{\beta}f'(R)-g_{\alpha\beta}\square f'(R)]}{f'(R)},
\end{align}
o
\begin{equation}
G_{\alpha\beta} = \frac{\kappa}{f'(R)}\bigl(T_{\alpha\beta} + T_{\alpha\beta}^{eff}\bigr),
\end{equation}
donde
\begin{equation}\label{tenseff}
\boxed{T_{\alpha\beta}^{eff} \equiv \frac{1}{\kappa}\biggl[\frac{[f(R) - Rf'(R)]}{2}g_{\alpha\beta} + [\nabla_{\alpha}\nabla_{\beta}-g_{\alpha\beta}\square]f'(R)\biggr].}
\end{equation}
De esta manera los efectos de la modificación en las ecuaciones de campo se pueden interpretar como una contribución de un \textit{fluido de curvatura} correspondiente a términos netamente geométricos. Se puede mostrar que este tensor de energía-momentum efectivo satisface también una ley de conservación $\nabla_{\beta}T^{eff\, \alpha\beta}=0$ \cite{Carloni}. \\\\
A continuación mostraremos muy resumidamente las ecuaciones de campo en los formalismos de Palatini y métrico-afín siguiendo \cite{Sotiriou3},\cite{Sotiriou4}.
\section{Ecuaciones de Campo en Formalismo de Palatini}
\noindent La idea básica en este formalismo es que el tensor métrico $g_{\alpha\beta}$ y la conexión $\Gamma_{\beta\gamma}^{\alpha}$ son independientes. Consideremos entonces $\mathcal{R}_{\alpha\beta}$ el tensor de Ricci construido a partir de esta conexión independiente y sea $\mathcal{R} = g^{\alpha\beta}\mathcal{R}_{\alpha\beta}$ el respectivo escalar de Ricci. Partiendo de la acción dada en la ecuación (\ref{accion2}), y haciendo las variaciones con respecto a $g^{\alpha\beta}$ y $\Gamma_{\beta\gamma}^{\alpha}$, respectivamente, encontramos las ecuaciones
\begin{equation}
f'(\mathcal{R})\mathcal{R}_{(\alpha\beta)} -\frac{1}{2}f(\mathcal{R})\, g_{\alpha\beta} =\kappa \, T_{\alpha\beta},
\end{equation}
\begin{equation}
-\bar{\nabla}_{\gamma}\bigl(\sqrt{-g} f'(\mathcal{R}) g^{\alpha\beta}\bigr) + \bar{\nabla}_{\sigma}\bigl(\sqrt{-g} f'(\mathcal{R}) g^{\sigma(\alpha}\bigr)\delta_{\gamma}^{\beta)} =0,
\end{equation}
con $\bar{\nabla}_{\gamma}$ la derivada covariante definida con la conexión independiente, y en donde los índices entre paréntesis están simetrizados. Tomando la traza de la ecuación anterior se pueden reducir las ecuaciones a
\begin{equation}\label{varpal1}
f'(\mathcal{R})\mathcal{R}_{(\alpha\beta)} -\frac{1}{2}f(\mathcal{R})\, g_{\alpha\beta} =\kappa \, T_{\alpha\beta},
\end{equation}
\begin{equation}\label{varpal2}
\bar{\nabla}_{\gamma}\bigl(\sqrt{-g} f'(\mathcal{R}) g^{\alpha\beta}\bigr) =0.
\end{equation}
Consideremos la ecuación (\ref{varpal2}), definamos una métrica conforme a $g_{\alpha\beta}$ de la siguiente forma
\begin{equation}
h_{\alpha\beta} = f'(\mathcal{R})g_{\alpha\beta},
\end{equation}
por lo tanto la ecuación (\ref{varpal2}) se escribe como
\begin{equation}
\bar{\nabla}_{\gamma}\bigl(\sqrt{-h}\, h^{\alpha\beta}\bigr) =0,
\end{equation}
que se puede interpretar como la conexión de Levi-Civita para $h_{\alpha\beta}$
\begin{align}
\Gamma_{\beta\gamma}^{\alpha} &= \frac{1}{2}h^{\alpha\sigma}\bigl(\partial_{\alpha}h_{\sigma\beta} + \partial_{\beta}h_{\sigma\alpha} - \partial_{\sigma}h_{\alpha\beta}\bigr)\\
&= \frac{1}{2}\frac{1}{f'(\mathcal{R})} g^{\alpha\sigma}\bigl(\partial_{\alpha}(f'(\mathcal{R})g_{\sigma\beta}) + \partial_{\beta}(f'(\mathcal{R})g_{\sigma\alpha}) - \partial_{\sigma}(f'(\mathcal{R})g_{\alpha\beta})\bigr).
\end{align}
Puesto que la traza relaciona $\mathcal{R}$ algebraicamente con $T$, y como tenemos una expresión explicita para $\Gamma_{\beta\gamma}^{\alpha}$ en términos de $\mathcal{R}$ y $g^{\alpha\beta}$, podemos en principio eliminar la conexión independiente de las ecuaciones de campo y expresar esta solo en función de la métrica y los campos de materia. Veamos ahora como se escribe el tensor de Ricci bajo una transformación conforme
\begin{equation}\label{riccipal}
\mathcal{R}_{\alpha\beta} = R_{\alpha\beta} + \frac{3}{2}\frac{1}{\bigl(f'(\mathcal{R})\bigr)^2}\bigl(\nabla_{\alpha}(f'(\mathcal{R}))\bigr)(\nabla_{\beta}(f'(\mathcal{R}))\bigr)-\frac{1}{f'(\mathcal{R})}
\bigl(\nabla_{\alpha}\nabla_{\beta}-g_{\alpha\beta}\square\bigr)f'(\mathcal{R}),
\end{equation}
contrayendo esta expresión con $g^{\alpha\beta}$ nos da
\begin{equation}\label{rpal}
\mathcal{R} = R + \frac{3}{2}\frac{1}{\bigl(f'(\mathcal{R})\bigr)^2}\bigl(\nabla_{\alpha}(f'(\mathcal{R}))\bigr)(\nabla^{\alpha}(f'(\mathcal{R}))\bigr)+\frac{3}{f'(\mathcal{R})}
\square f'(\mathcal{R}),
\end{equation}
con lo cual podemos obtener
\begin{multline}
G_{\alpha\beta} = \frac{k}{f'(\mathcal{R})}T_{\alpha\beta} - \frac{1}{2}\biggl(\mathcal{R}-\frac{f(\mathcal{R})}{f'(\mathcal{R})}\biggr)g_{\alpha\beta} + \frac{1}{f'(\mathcal{R})}\biggl(\nabla_{\alpha}\nabla_{\beta}-\frac{1}{2}g_{\alpha\beta}\square\biggr)f'(\mathcal{R})\\\\ - \frac{3}{2}\frac{1}{(f'(\mathcal{R}))^2}\biggl((\nabla_{\alpha} f'(\mathcal{R}))(\nabla_{\beta} f'(\mathcal{R}))-\frac{1}{2}g_{\alpha\beta}(\nabla f'(\mathcal{R}))^2\biggr),
\end{multline}
en donde escribimos
\begin{equation}
G_{\alpha\beta} \equiv R_{\alpha\beta} - \frac{1}{2}R g_{\alpha\beta}.
\end{equation}
Como es de esperar, las ecuaciones de campos en los formalismos métrico y de Palatini se reducen a las ecuaciones de campo de Einstein en el caso $f(R)=R$, o con la contribución de la constante cosmológica cuando $f(R) = R - 2\Lambda$.
\section{Ecuaciones de Campo en el Formalismo Métrico-Afín}
\noindent Este formalismo es similar al de Palatini, excepto que ahora los campos de materia dependen de la conexión $\Gamma_{\beta\gamma}^{\alpha}$. Además se esto se da la libertad para que la variedad $\mathcal{M}$ en este formalismo tenga torsión ($\Gamma_{\beta\gamma}^{\alpha} \neq \Gamma_{\gamma\beta}^{\alpha}$). Esta libertad permite que en algunos regímenes de energía el espín de las partículas pueda interactuar con la geometría, lo cual directamente nos da la existencia de la torsión en la teoría. Considerando entonces la acción (\ref{accion3}), y usando los siguiente resultados para la variación del escalar $\mathcal{R}$ (asumiendo torsión diferente de cero)
\begin{equation}
\delta \mathcal{R}_{\alpha\beta} = \bar{\nabla}_{\gamma}\delta\Gamma_{\alpha\beta}^{\gamma} - \bar{\nabla}_{\beta}\delta\Gamma_{\alpha\gamma}^{\gamma} +
2\Gamma_{[\beta\gamma]}^{\sigma}\delta \Gamma_{\alpha\sigma}^{\gamma},
\end{equation}
y la variación de la acción de materia $S_M=S_M(g_{\alpha\beta},\Gamma_{\alpha\beta}^{\gamma},\psi)$
\begin{equation}
\delta S_M = \frac{\delta S_M}{\delta g^{\alpha\beta}}\delta g^{\alpha\beta}+\frac{\delta S_M}{\delta \Gamma_{\alpha\beta}^{\gamma}}\delta \Gamma_{\alpha\beta}^{\gamma},
\end{equation}
donde definimos ahora un nuevo tensor llamado el \textit{Hipermomentum} como
\begin{equation}
\Delta_{\gamma}^{\alpha\beta} \equiv-\frac{2}{\sqrt{-g}}\frac{\delta S_M}{\delta \Gamma_{\alpha\beta}^{\gamma}},
\end{equation}
llegamos a las ecuaciones de campo
\begin{equation}
f'(\mathcal{R})\mathcal{R}_{(\alpha\beta)} -\frac{1}{2}f(\mathcal{R})\, g_{\alpha\beta} = \kappa T_{\alpha\beta},
\end{equation}
y
\begin{multline}
\frac{1}{\sqrt{-g}}\biggl[-\bar{\nabla}_{\gamma}\bigl(\sqrt{-g} f'(\mathcal{R}) g^{\alpha\beta}\bigr) + \bar{\nabla}_{\sigma}\bigl(\sqrt{-g} f'(\mathcal{R}) g^{\alpha\sigma}\bigr)\delta_{\gamma}^{\beta}\biggr]\\\\ + 2\sqrt{-g}\,f'(\mathcal{R})\bigl(g^{\alpha\beta}S_{\gamma\sigma}^{\sigma}-g^{\alpha\rho}S_{\rho\sigma}^{\sigma} + g^{\alpha\sigma}S_{\sigma\gamma}^{\beta}\bigr) = \kappa \Delta_{\gamma}^{\alpha\beta},
\end{multline}
donde escribimos la última expresión en función del tensor de torsión de Cartan (\ref{cartan}). Un ejemplo de campos que tengan hipermomentum diferente de cero, son los campos de Dirac, así que el estudio de tales campos en este formalismo nos podrían dar evidencia de la existencia de torsión en la teoría \cite{Sotiriou4}.
\section{Equivalencia con Teorías de Brans-Dicke}\label{Escalares}
\noindent Una formulación diferente a la RG que además del tensor métrico (y de éste los tensores de curvatura) incluye un campo escalar $\varphi$, fueron trabajadas por Brans y Dicke en 1961 \cite{Brans}. La motivación principal de estas teorías era la de cambiar el valor de la constante de gravitación universal $G$ de lugar a lugar, con el fin de incorporar el principio de Mach\footnote{Que en resumen nos dice que la inercia de un cuerpo depende en principio de todo el contenido de materia del universo.} en la RG.\\\\
La acción general para las teorías de Brans-Dicke se puede escribir como
\begin{equation}\label{accionbrans}
S = \int_{\mathcal{V}} d^4x\, \sqrt{-g} \biggl[\frac{1}{2}\phi R - \frac{\omega_{\text{BD}}}{2\phi}(\nabla \phi)^2 - U(\phi)\biggr] + S_M(g_{\alpha\beta},\psi),
\end{equation}
en donde $\omega_{\text{BD}}$ es el denominado parámetro de Brans-Dicke y $(\nabla \phi)^2 \equiv g^{\alpha\beta}\partial_{\alpha}\phi \partial_{\beta}\phi$. \\\\
Las teorías $f(R)$ en el formalismo métrico se pueden escribir en la forma de las teorías de Brans-Dicke \cite{Felice} con un potencial para el campo escalar como grado de libertad (\textit{escalaron}). Para esto consideramos la siguiente acción (sin términos en la frontera) para un nuevo campo $\chi$
\begin{equation}\label{accionescalar}
S = \frac{1}{2\kappa}\int_{\mathcal{V}} d^4x\, \sqrt{-g} \bigl[f(\chi) + f_{,\chi}(R-\chi)\bigr] + S_M(g_{\alpha\beta},\psi),
\end{equation}
en donde $f_{,\chi}$ corresponde a la derivada de la función $f$ con respecto a $\chi$, cabe aclarar que el termino $f_{,\chi}(R-\chi)$ es un producto entre $f_{,\chi}$ y $(R-\chi)$. Variando esta acción con respecto a $\chi$ obtenemos
\begin{equation}
f_{,\chi\chi}(R-\chi) = 0,
\end{equation}
fijando $f_{,\chi\chi} \neq 0$ se sigue que $R = \chi$. De este modo la acción (\ref{accionescalar}) recupera la acción  (\ref{accion1}) en teorías $f(R)$. Si definimos ahora
\begin{equation}
\phi \equiv f_{,\chi}(\chi),
\end{equation}
la acción (\ref{accionescalar}) puede escribirse como
\begin{equation}\label{accionfescalar}
S = \int_{\mathcal{V}} d^4x\, \sqrt{-g} \biggl[\frac{1}{2\kappa} \phi R - U(\phi)\biggr]  + S_M(g_{\alpha\beta},\psi),
\end{equation}
donde $U(\phi)$ esta definido como
\begin{equation}
U(\phi) = \frac{\chi(\phi)\phi- f\bigl(\chi(\phi)\bigr)}{2\kappa}.
\end{equation}
De esta forma si comparamos la expresión (\ref{accionbrans}) con (\ref{accionfescalar}) observamos que las teorías $f(R)$ con equivalentes a las de Brans-Dicke con $\omega_{\text{BD}} = 0$ (y en unidades de $\kappa = 1$). En el formalismo de Palatini donde la métrica $g_{\alpha\beta}$ y la conexión $\Gamma_{\beta\gamma}^{\alpha}$ con considerados como variables independientes, el escalar de Ricci es diferente que el dado en el formalismo métrico. Se puede mostrar que para el formalismo de Palatini en teorías $f(R)$ es equivalente a las teorías de Brans-Dicke con el parámetro $\omega_{\text{BD}}=-3/2$ \cite{Felice}. \\\\
Hemos entonces discutido los principios variacionales en teorías $f(R)$, haciendo especial énfasis en el formalismo métrico. Nuestro punto de partida entonces para construir modelos cosmológicos son las ecuaciones de campo modificadas (\ref{ecuacionescampo}).

\newpage

{\color{white} . } 
\chapter{Modelos Cosmológicos en Teorías $f(R)$}\label{Capitulo5}
\drop{E}n este capítulo aplicaremos las ecuaciones de campo en el formalismo métrico de las teorías $f(R)$ para un universo homogéneo e isotrópico descrito por la métrica de Robertson-Walker. Varias formas de la función $f(R)$ se han propuesto con el fin de obtener la actual expansión acelerada del universo, sin embargo, algunos poseen problemas (en particular inestabilidades) en el régimen de bajas energías (sistema solar).\\\\
El estudio de de las ecuaciones de Friedmann modificadas, modeladas como un sistema dinámico autónomo, permiten restringir la forma de las funciones y dar criterios de viabilidad para teorías que reproduzcan una época de radiación, una de materia y una de expansión acelerada.
\section{Ecuaciones de Friedmann modificadas en gravedad $f(R)$}\label{Seccion5.1}
\noindent Debemos entonces hallar las expresiones para las ecuaciones de Friedmann modificadas asumiendo para esto de nuevo la validez del Modelo Estándar de la Cosmología. De modo que nuestro punto de partida es la métrica de Robertson-Walker, ecuación (\ref{robw})
\begin{equation}
ds^2 = -dt^2 + a^2(t)\biggl[\frac{dr^2}{1-kr^2} + r^2d\theta^2 + r^2\sin^2\theta d\varphi^2\biggr],
\end{equation}
siendo $a(t)$ el factor de escala y $k$ la curvatura espacial del universo. Para esta métrica el escalar de Ricci es
\begin{equation}\label{RWscalar}
R = 6\biggl[\frac{\ddot{a}}{a}+ \biggl(\frac{\dot{a}}{a}\biggr)^2 + \frac{k}{a^2}\biggr] = 6\biggl[\dot{H} + 2H^2 + \frac{k}{a^2}\biggr],
\end{equation}
consideraremos también el tensor de energía-momentum  de un fluido perfecto, ecuación (\ref{e-m1})
\begin{equation}
T_{\alpha\beta}= (\rho + p)u_{\alpha}u_{\beta} + p g_{\alpha\beta},
\end{equation}
siendo $p$ la presión del fluido, $\rho$ la densidad de energía y $u^{\alpha}$ la cuadrivelocidad de los observadores fundamentales.  A partir de las ecuaciones de campo (\ref{ecuacionescampo}), y con la métrica (\ref{robw}), obtenemos las siguientes ecuación de Friedmann modificadas (ver Apéndice \ref{AppendixC})
\begin{equation}\label{Friedmod1}
\boxed{ H^2 + \frac{k}{a^2}= \frac{1}{3 f'(R)}\biggl[\kappa\rho + \frac{(R f'(R)-f(R))}{2} - 3Hf''(R)\dot{R}\biggr], }
\end{equation}
\begin{equation}\label{Friedmod2}
\boxed{2\dot{H} + 3H^2 + \frac{k}{a^2} = -\frac{1}{f'(R)}\biggl[\kappa p+2H\dot{R}f''(R) +\frac{(f(R) - Rf'(R))}{2}  + \ddot{R}f''(R)  + (\dot{R})^2f'''(R)\biggl],}
\end{equation}
en donde usamos que $H = \frac{\dot{a}}{a}$. A partir de la traza de la ecuación de campo (\ref{tracefield}), podemos encontrar otra relación para $f(R)$ y sus derivadas con la presión y la densidad de energía
\begin{equation}
f'(R)R - 2 f(R) + 3 \square f'(R) = \kappa (3p-\rho).
\end{equation}
A partir de las ecuaciones (\ref{Friedmod1}) y (\ref{Friedmod2}), podemos definir una densidad y presión asociadas a los términos adicionales de curvatura de la siguiente forma
\begin{equation}\label{rhoDE}
\rho_{DE}= \frac{1}{f'(R)}\biggl[\frac{(R f'(R)-f(R))}{2} - 3Hf''(R)\dot{R}\biggr],
\end{equation}
\begin{equation}\label{pDE}
p_{DE} = -\frac{1}{f'(R)}\biggl[2H\dot{R}f''(R) +\frac{(f(R) - Rf'(R))}{2}  + \ddot{R}f''(R)  + (\dot{R})^2f'''(R)\biggl],
\end{equation}
de modo que la ecuación de estado puede escribir se como
\begin{align}\label{wde}
w_{DE} = \frac{p_{DE}}{\rho_{DE}} &= \frac{2H\dot{R}f''(R) +\frac{(f(R) - Rf'(R))}{2}  + \ddot{R}f''(R)  + (\dot{R})^2f'''(R)}{\frac{(R f'(R)-f(R))}{2} - 3Hf''(R)\dot{R}}, \notag \\
&= -1 + \frac{\ddot{R}f''(R) - H\dot{R}f''(R) + (\dot{R})^2f'''(R)}{\frac{(R f'(R)-f(R))}{2} - 3Hf''(R)\dot{R}}.
\end{align}
De forma similar a la ecuación de continuidad para las contribuciones de densidad y materia
\begin{equation}\label{contmateria}
\dot{\rho}_m + 3H\rho_{m} =0,
\end{equation}
\begin{equation}\label{contmateria}
\dot{\rho}_r + 4H\rho_{m} =0,
\end{equation}
podemos obtener una expresión para la ecuación de estado efectiva para las componentes $\rho_{DE}$ y $p_{DE}$ (ecuaciones (\ref{rhoDE}) y (\ref{pDE}) respectivamente) \cite{Capozziello}
\begin{equation}
\dot{\rho}_{DE} + 3H(1+w_{DE})\rho_{DE} = 3 H_0^2 \Omega_m \frac{\dot{R} f''(R)}{\bigl(f'(R)\bigr)^2}.
\end{equation}
Si consideramos ahora las ecuaciones de campo modificadas constituidas por un fluido efectivo (\ref{tenseff}), la ecuación de Friedmann (\ref{fried2}) se puede escribir como
\begin{equation}
\frac{\ddot{a}}{a} = -\frac{\kappa}{6}\bigl(\rho_{\text{tot}} + 3p_{\text{tot}}\bigr),
\end{equation}
en donde el subíndice  tot denota la suma del fluido de curvatura y la contribución de materia a la densidad de energía y presión. De esta relación la condición de aceleración, para un modelo en el dominio de materia es
\begin{equation}
\rho_{DE} + \rho_m + 3p_{DE}< 0, \qquad \Longrightarrow \qquad w_{DE} < - \frac{\rho_{\text{tot}}}{\rho_{DE}},
\end{equation}
esta relación nos permite entonces determinar si dada alguna forma específica de la función $f(R)$ y por lo tanto unas componentes para $\rho_{DE}$ y $p_{DE}$, obtener universos con expansión acelerada. A los modelos que tienen una ecuación de estado con $w < -1/3$ los denominaremos \textit{modelos phantom} o de Quintaesencia.\\\\
Aunque algunos modelos se han utilizado para explicar la expansión del universo, en particular el modelo \cite{Capozziello2},\cite{Carroll4}
\begin{equation}
f(R) = R - \frac{\alpha}{R^n}, \quad (\alpha > 0, n > 0),
\end{equation}
presenta sin embargo un numero de problemas, en particular la inestabilidad en la materia \cite{Faraoni2},\cite{Dolgov}, inestabilidad en perturbaciones cosmológicas
\cite{Carroll5}-\cite{Sawicki2}, y la incapacidad de satisfacer restricciones locales (sistema solar) \cite{Olmo}-\cite{Capoziello5}. Este tipo de pruebas han llevado a considerar algunas restricciones en las funciones $f(R)$ \cite{Starobinsky} en particular
\begin{equation}\label{restricciones}
\boxed{f'(R) > 0, \qquad f''(R) > 0 \qquad \text{para } R \geq R_0 (> 0),}
\end{equation}
siendo $R_0$ el valor actual para el escalar de Ricci. Consideraremos a continuación la aproximación de sistema dinámico en teorías $f(R)$ que nos permitirá obtener más criterios de viabilidad para las funciones $f(R)$.
\section{Aproximación de Sistema Dinámico en ecuaciones de Friedmann modificadas}
\noindent
El análisis de las ecuaciones de campos modificadas modeladas como un sistema dinámico, nos permitirá encontrar criterios para modelos cosmológicamente viables. Realizaremos un análisis de tal enfoque siguiendo \cite{Felice},\cite{Amendola1}-\cite{Amendola3}. Vamos primero a escribir las ecuaciones de Friedmann modificadas (\ref{Friedmod1}) y (\ref{Friedmod2}) en el caso de un universo plano ($k=0$) y considerando las contribuciones en la densidad de energía y la presión para los dominios de radiación y materia. Para esto usamos que la ecuación de estado para el caso de materia es $p_m =0$ y en el caso de radiación $p_r = 1/3 \rho_r$, con esto las ecuaciones pueden escribirse como
\begin{equation}\label{Friedmod4}
3 f' H^2 = \kappa (\rho_m + \rho_r) + \frac{f'R-f}{2}-3H\dot{f}',
\end{equation}
\begin{equation}\label{Friedmod5}
-2f'\dot{H} = \kappa\biggl(\rho_m + \frac{4}{3}\rho_r\biggr) + \ddot{f}' - H\dot{f}',
\end{equation}
en donde usamos la siguiente notación $\dot{f}' = \partial_{t}(f') = f'' \dot{R}$, y eliminamos la dependencia de $R$ en las funciones $f$ y $f'$.  Con $\mathcal{N} \equiv \ln a$, definimos un operador diferencial de la siguiente manera
\begin{equation}
\frac{d}{d \mathcal{N}} = \frac{d}{d \ln a} = \frac{1}{H}\frac{d}{dt},
\end{equation}
así que la ecuación (\ref{Friedmod4}) se puede escribir como
\begin{equation}\label{Friedmod4}
1 = \frac{\kappa\rho_m}{3 f' H^2} + \frac{\kappa\rho_r}{3 f' H^2} + \frac{R}{6H^2} - \frac{f}{3 f' H^2}-\frac{\dot{f}'}{H f'}.
\end{equation}
Definamos las siguientes cantidades adimensionales
\begin{equation}\label{xi}
x_1 = -\frac{\dot{f}'}{Hf'}, \qquad x_2 = -\frac{f}{6 f'H^2}, \qquad
x_3 = 2+\frac{\dot{H}}{H^2}, \qquad x_4 = \frac{\kappa \rho_r}{3 f' H^2},
\end{equation}
\begin{equation}
\Omega_{m} = \frac{\kappa\rho_m}{3 f' H^2} = 1-x_1-x_2-x_3-x_4,
\end{equation}
y también
\begin{equation}\label{OmegaDE}
\boxed{\Omega_{DE} = x_1 + x_2 + x_3,}
\end{equation}
el cual lo asociaremos con la contribución adicional a las ecuaciones de campo, o equivalentemente la contribución a la energía oscura (razón por la cual usamos los subíndices DE: Dark Energy). Con estas definiciones se pueden obtener las siguientes ecuaciones de movimiento
\begin{subequations}\label{sistemadinamico}
\begin{equation}
\frac{dx_1}{d\mathcal{N}} = -1-x_3-3x_2+x_1^2-x_1 x_3 + x_4,
\end{equation}
\begin{equation}
\frac{dx_2}{d\mathcal{N}} = \frac{x_1x_3}{m}-x_2(2x_3-4-x_1),
\end{equation}
\begin{equation}
\frac{dx_3}{d\mathcal{N}} = -\frac{x_1 x_3}{m}-2x_3(x_3-2),
\end{equation}
\begin{equation}
\frac{dx_4}{d\mathcal{N}} = -2x_3x_4 + x_1x_4,
\end{equation}
\end{subequations}
donde hemos definido
\begin{equation}\label{mr}
m \equiv \frac{d\ln f'}{d\ln R} = \frac{R f''}{f'} = \frac{R\dot{f}'}{\dot{f}}, \qquad r \equiv -\frac{d\ln f}{d\ln R} = - \frac{Rf'}{f}=\frac{x_3}{x_2}.
\end{equation}
De la expresión para $r$, el escalar $R$ se puede escribir como $x_3/x_2$, y puesto que $m$ depende de $R$, esto significa
que $m$ es una función de $r$, $m = m(r)$. El modelo $\Lambda$CDM, $f(R) = R - 2\Lambda$, corresponde a $m = 0$. \textit{Por lo tanto la cantidad $m$ caracteriza la desviación en la dinámica de un modelo $\Lambda$CDM}. A partir de las ecuaciones (\ref{Friedmod4}) y (\ref{Friedmod5}) podemos definir una ecuación de estado efectiva por la siguiente relación
\begin{equation}\label{weff}
w_{eff}  = -1 - \frac{2\dot{H}}{3H^2} = - \frac{(2x_3-1)}{3}.
\end{equation}
la cual se puede interpretar como la ecuación total para una contribución de densidad de energía efectiva $\rho_{eff}= \rho_m + \rho_r + \rho_{DE}$ y una relación similar para la presión efectiva.\\\\
La idea ahora es determinar la estabilidad del sistema dinámico. Vamos a realizar el siguiente análisis en el caso en que no hay radiación, es decir, para $x_4=0$. Notamos primero que las ecuaciones dinámicas se pueden escribir en la forma
\begin{equation}
\frac{dx_i}{d\mathcal{N}} = g_i(x_1,x_2,x_3), \qquad i=1,2,3,
\end{equation}
de modo que los puntos fijos del sistema son los puntos ($x'_1,x'_2,x'_3$) que satisfacen de forma simultanea $g_i(x'_1,x'_2,x'_3)=g_j(x'_1,x'_2,x'_3)=0, i,j=1,2,3, \, i\neq j$. En la tabla \ref{tabla1} se muestran tales puntos, el respectivo valor para ecuación de estado efectiva (\ref{weff})y el valor para el parámetro $\Omega_m$.
\begin{table}
\begin{center}
\begin{tabular}{||l||c||c||} \hline
$P:(x_1,x_2,x_3)$  & $w_{eff}$  & $\Omega_{m}$ \\ \hline
$P_1 = (0,-1,2)$  & $-1$  & $0$ \\
& & \\
$P_2 = (-1,0,0)$  & $1/3$  & $2$ \\
& & \\
$P_3 = (1,0,0)$  & $1/3$   & $0$ \\
& & \\
$P_4 = (-4,5,0)$  & $1/3$  & $0$ \\
& & \\
$P_5 = \biggl(\dfrac{3m}{1 + m},-\dfrac{1 + 4m}{2(1 + m)^2},\dfrac{1 + 4m}{2(1 + m)}\biggr)$  & $-\dfrac{m}{1 + m}$  & $ 1 -\dfrac{m(7 + 10m)}{2(1 + m)^2}$ \\
& & \\
$P_6 = \biggl(\dfrac{2(1 - m)}{1 + 2m},\dfrac{1 - 4m}{m(1 + 2m)},\dfrac{-(1 -4 m)(1+m)}{m(1 + 2m)}\biggr)$  & $\dfrac{2 - 5m- 6m^2}{3m(1 + 2m)}$  & $0$ \\
& & \\  \hline\hline
\end{tabular}
\caption{Valores para los puntos críticos del sistema dinámico (\ref{sistemadinamico}) y valores para la ecuación de estado efectiva $w_{eff}$ (\ref{weff}).}
\label{tabla1}
\end{center}
\end{table}
Los puntos  $P_5$ y  $P_6$ satisfacen la ecuación
\begin{equation}
x_3 = -\bigl(m(r)+1\bigr)x_2,
\end{equation}
es decir
\begin{equation}\label{mP51}
\boxed{m(r) = - r-1.}
\end{equation}
Cuando $m(R)$ no es constante, se debe resolver tal ecuación, para cada raíz $r_i$ uno obtiene un punto del tipo $P_5$ o $P_6$ con $m = m(r_i)$.
\subsection{Estabilidad de los puntos críticos}
\begin{figure}
\begin{center}
\includegraphics[scale=0.58]{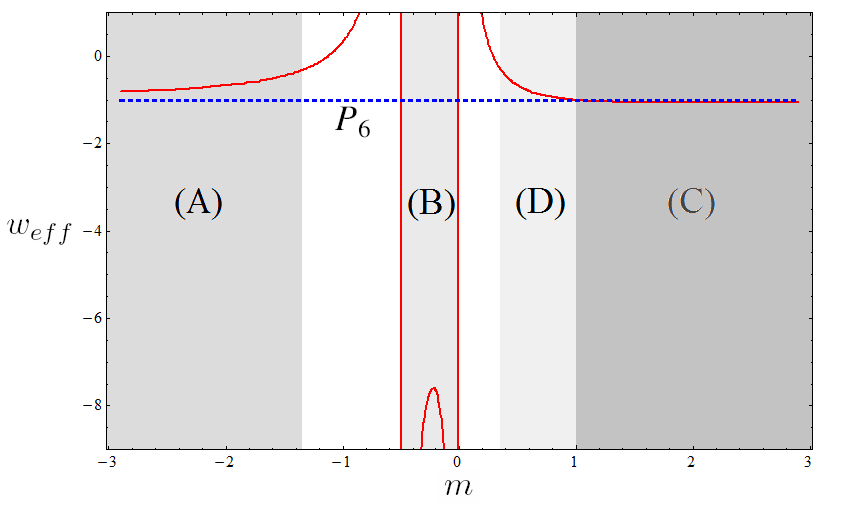}\\
\caption{Comportamiento de $w_{eff}$ en función de $m$ para el punto $P_6$, obtenido de \cite{Amendola2}.}
\label{wefffigura}
\end{center}
\end{figure}

\noindent Para determinar la estabilidad de los puntos críticos vamos a seguir \cite{Copeland}. Consideremos por el momento el siguiente sistema de ecuaciones diferenciales acopladas para dos variables $x(t)$ y $y(t)$
\begin{equation}
\dot{x}  = f(x,y,t), \qquad \dot{y} = g(x,y,t),
\end{equation}
donde $f$ y $g$ son funciones de $x$, $y$ y $t$. Como vimos anteriormente, un punto crítico $(x_c, y_c)$ es un punto tal que
\begin{equation}
\bigl(f,g\bigr)\bigl|_{x_c,y_c} =0.
\end{equation}
Para estudiar la estabilidad vamos a considerar pequeñas perturbaciones $\delta x$ y $\delta y$ alrededor de los puntos críticos $(x_c, y_c)$, es decir
\begin{equation}
x = x_c + \delta x,\qquad  y = y_c + \delta_y.
\end{equation}
de modo que reemplazando esto en el sistema dinámico obtenemos que
\begin{equation}
\frac{d}{d \mathcal{N}}\begin{pmatrix}
\delta_x \\
\delta_y
\end{pmatrix} = \mathcal{J}\begin{pmatrix}
\delta_x \\
\delta_y
\end{pmatrix},
\end{equation}
donde nuevo $\mathcal{N} \equiv \ln a$ y con $\mathcal{J}$ dado por
\begin{equation}
\mathcal{J} =
\begin{pmatrix}
\dfrac{\partial g}{\partial x} &   & \dfrac{\partial g}{\partial y}\\\\
\dfrac{\partial f}{\partial x} &   & \dfrac{\partial f}{\partial y}
\end{pmatrix}_{\bigl|_{x_c,y_c}.}
\end{equation}
Esta matriz posee dos autovalores $j_1$ y $j_2$, de modo que la solución general a la evolución de las perturbaciones lineales es
\begin{subequations}
\begin{equation}
\delta x = C_1 e^{j_1 \mathcal{N}} + C_2 e^{j_2 \mathcal{N}},
\end{equation}
\begin{equation}
\delta y = C_3 e^{j_1 \mathcal{N}} + C_4 e^{j_2 \mathcal{N}},
\end{equation}
\end{subequations}
siendo $C_1,C_2,C_3,C_4$ constantes de integración. Entonces, la estabilidad de los puntos fijos depende de la naturaleza de los autovalores
\begin{itemize}
\item \textit{(i)} Punto Estable: $j_1 < 0$ y $j_2 < 0$.
\item \textit{(ii)} Punto Inestable: $j_1 > 0$ y $j_2 > 0$.
\item \textit{(iii)} Punto de silla: $j_1 < 0$ y $j_2 > 0$ (o $j_1 > 0$ y $j_2 < 0$) .
\item \textit{(iv)} Espiral estable: El determinante de la matriz $\mathcal{J}$  es negativo y las partes reales de $j_1$ y $j_2$ son negativas.
\end{itemize}
Un punto fijo es un atractor en los casos \textit{(i)} y \textit{(iv)},
pero no lo es en los casos \textit{(ii)} y \textit{(iii)}. \\\\\
De esta manera construimos entonces la matriz $\mathcal{J}$ a partir de las derivadas parciales de las funciones $g_{i}$ con respecto a cada una de las variables $x_i$.
\begin{equation}\label{matriz}
\mathcal{J}(x_1,x_2,x_3) = \begin{pmatrix}
2x_1-x_3 & -3 & -1-x_1  \\\\
\dfrac{x_3}{m}+x_2 & x_1 x_3 \partial_{x_2}\biggl(\dfrac{1}{m}\biggr)-2x_3 + 4 + x_1 & x_1\partial_{x_3}\biggl(\dfrac{x_3}{m}\biggr)-2x_2 \\\\
-\dfrac{x_3}{m} & -x_1 x_3 \partial_{x_2}\biggl(\dfrac{1}{m}\biggr) & -x_1\partial_{x_3}\biggl(\dfrac{x_3}{m}\biggr)-4x_3+4
\end{pmatrix}
\end{equation}
determinamos los valores propios del jacobiano a partir de la relación
\begin{equation}
\det (\mathcal{J}\bigl|_{x_{1c},x_{2c},x_{3c}}-\lambda \mathbb{I}) = 0,
\end{equation}
con $\mathbb{I}$ la matriz identidad. La estabilidad de los puntos críticos dependerá entonces de los signos de los autovalores de la matriz (\ref{matriz}) evaluados en los puntos críticos. \\\\
$\bullet$ \quad $P_1$: Punto de de Sitter.\\\\
Puesto que $w_{eff} =-1$ el punto $P_1$ corresponde a las soluciones de de Sitter que satisfacen
\begin{equation}
F(R) R -2f(R) = 0,
\end{equation}
que corresponde a la traza de las ecuaciones de campo (\ref{tracefield}) en el vacío ($T=0$). Para este punto los autovalores de (\ref{matriz}) son
\begin{equation}
-3, \qquad -\dfrac{3}{2} \pm \dfrac{\sqrt{25-16/m_1}}{2},
\end{equation}
donde $m_1 = m(r_1), r_1 =-2$. Así, $P_1$ es estable cuando $0 < m_1 \leq 1$ y un punto de silla en los demás valores. La condición entonces para la estabilidad del punto de de Sitter es
\begin{equation}
0 < m(r=-2) \leq 1.
\end{equation}\\\\\\
$\bullet$ \quad $P_2$: $\phi$-EDM. \\\\
Este punto es caracterizado por una época ``cinética'' en la cual la materia y un campo $\phi$ coexisten con fracciones de energía constantes. Denotamos entonces este punto por $\phi$-EDM ($\phi$-Época de Domino de Materia) siguiendo \cite{Amendola4}. Los autovalores son
\begin{equation}
-2,\qquad \frac{1}{2}\biggl[7+\frac{1}{m_2} - \frac{m'_2}{m_2^2}r(1+r) \pm \sqrt{\biggl(7+\frac{1}{m_2} - \frac{m'_2}{m_2^2}r(1+r)\biggr)^2 - 4 \biggl(12 + \frac{3}{m_2}-\frac{m'_2}{m_2^2}r(3+4r)\biggr)}\biggr],
\end{equation}
en donde $m'_2 \equiv \frac{dm}{dr}(r_2)$. Si  $m(r)$ es constante, los autovalores se reducen a
\begin{equation}
3,\qquad  -2, \qquad  4 + \dfrac{1}{m_2},
\end{equation}
de modo que $P_2$ es un punto de silla.  \\\\
$\bullet$ \quad $P_3$: Punto puramente cinético.\\\\
Este punto también corresponde a una época cinética, pero a diferencia del punto $P_2$ la fracción de energía en la materia se hace cero. Este punto además se puede ver como un caso particular del punto $P_6$ escogiendo $m = 1/4$. Los autovalores son
\begin{equation}
2,\qquad \frac{1}{2}\biggl[9+\frac{1}{m_3} + \frac{m'_3}{m_3^2}r(1+r) \pm \sqrt{\biggl(9+\frac{1}{m_3} - \frac{m'_3}{m_3^2}r(1+r)\biggr)^2 - 4 \biggl(20 - \frac{5}{m_3}+\frac{m'_3}{m_3^2}r(5+4r)\biggr)}\biggr].
\end{equation}
Si $m(r)$ es constante los autovalores se reducen a
\begin{equation}
5,\qquad 2, \qquad 4 + \dfrac{1}{m_3},
\end{equation}
en este caso $P_3$ es inestable para  $m_3 < 0$ and $m_3 > 1/4$, y un punto de silla para los demás casos.\\\\
$\bullet$ \quad $P_4$. \\\\
Este punto tiene una propiedad similar al punto $P_3$ pues ambos tienen los mismos valores para $w_{eff}$ y $\Omega_m$. También es un caso especial del punto $P_6$ con $m=-1$. Los autovalores son
\begin{equation}
-5, \qquad -3 \qquad 4 + \dfrac{4}{m_4},
\end{equation}
de modo que es estable para  $-1< m_ 4 < 0$ y un punto de silla para los demás valores. Ninguno de los puntos $P_3$ o $P_4$ puede ser usado para la época de dominio de materia, ni para la época de expansión acelerada.\\\\
$\bullet$ \quad $P_5$.\\\\
El punto $P_5$ corresponde a \textit{scaling solutions} las cuales dan una razón constante $\Omega_m/\Omega_{DE}$. En el límite $m_5 \longrightarrow 0$ el punto representa una era de materia estándar con $a \propto t^{2/3}$ y $\Omega_m =1$. De modo que la condición necesaria para que $P_5$ sea una era exacta de materia está dada por
\begin{equation}
m(r=-1) =0.
\end{equation}
Los autovalores para $P_5$ son
\begin{equation}\label{eigen5}
3(1+m'_5),\qquad  \frac{-3m_5 \pm \sqrt{m_5(256m_5^3 + 160m_5^2 - 31m_5 - 16)}}{4m_5(1+m_5)},
\end{equation}
en el límite $|m_5| \ll 1$ los autovalores se pueden escribir aproximadamente por
\begin{equation}
3(1+m'_5),\qquad  -\frac{3}{4} \pm \sqrt{-\frac{1}{m_5}},
\end{equation}
para modelos con $m_5 < 0$, las soluciones no pueden permanecer por un periodo largo de tiempo alrededor del punto $P_5$ debido al comportamiento divergente de los autovalores cuando $m \longrightarrow 0^{-}$ ($m$ tendiendo a cero por la izquierda). El radical de los dos últimos autovalores se hace negativo para $0 < m_5 < \text{0.327}$, de modo que los autovalores son complejos con partes reales negativas, así que exigiendo que $m'_5 > -1$, el punto $P_5$  corresponde a un punto de silla.  La condición entonces para que este sea un punto de silla es
\begin{equation}
m(r \leq - 1) > 0 , m'( r \leq - 1) > -1.
\end{equation}
$\bullet$ \quad $P_6$: Punto de dominio de curvatura.\\\\
Este corresponde a un punto de dominio de curvatura \cite{Amendola2} cuya ecuación de estado efectiva depende de $m$. La condición para un universo con expansión acelerada
$ (w_{eff} < -1/3)$ se obtiene cuando $m_6 < -(1+\sqrt{3})/2$, $-1/2 < m_6 < 0$, y $m_6 > (\sqrt{3}-1)/2$. En la Figura \ref{wefffigura}  se muestra el comportamiento de $w_{eff}$ en función de $m$
Los autovalores para este punto son
\begin{equation}
\frac{1-4m_6}{m_6},\qquad  \frac{2-3m_6 - 8m_6^2}{m_6(1+2m_6)}, \qquad \dfrac{2(1-m_6)(1+m'_6)}{m_6(1+2m_6)},
\end{equation}
de modo que la estabilidad de $P_6$ depende tanto de $m_6$ como de $m'_6$. En el límite  $m_6 \longrightarrow \pm \infty$ tenemos $P_6 \rightarrow  (-1,0,2)$ con una ecuación de estado $w_{eff} =-1$. Este punto es estable con
$m'_6 > -1$. $P_6$ es también un punto de de Sitter para  $m_6 = 1$, que coincide con $P_1$, y puesto que en este caso $r=-2$, este punto se caracteriza por
\begin{equation}
m(r=-2) \longrightarrow 1.
\end{equation}
El punto $P_6$ es estable y acelerado para 4 diferentes rangos:
\begin{enumerate}[(I.)]
\item $m'_6 > - 1$\\
Cuando $m'_6 > -1$, $P_6$ es estable y acelerado en las tres siguientes regiones:\\
(A) $m_6 < -(1 + \sqrt{3})/2$: $P_6$ es acelerado pero no con condición de phantom cosmology, es decir, $w_{eff} > -1$. Se tiene el límite $w_{eff} =- 1$ cuando $m_6 \rightarrow -\infty$.\\
(B) $-1/2 < m_6 < 0$: $P_6$ obedece una ecuación de estado phantom con $w_{eff} < -7.6$.\\
(C) $m_6 \geq 1$: $P_6$ tiene una ecuación de estado phantom con $-1.07 < w_{eff} < -1$. Se tiene $w_{eff} = -1$ en el límite $m_6 \rightarrow + \infty$ y $m_6 \rightarrow 1$.
\item $m'_6 < - 1$\\
Cuando $m'_6 < -1$, $P_6$ es estable y acelerado en la siguiente región:\\
(D) $(\sqrt{3}+1)/2 < m_6 < 1$: en este caso $P_6$ no tiene una ecuación de estado phantom $w_{eff} > -1$.
\end{enumerate}
De esto se deriva entonces una primer conclusión general  con respecto a las funciones $f(R)$ \cite{Amendola2}: \textit{La aceleración asintótica NO puede tener una ecuación de estado en el rango $-7.6 < w_{eff} < -1.07$.}\\\\
Si consideramos ahora el caso con radiación los puntos $P_{1 - 6}$ se mantienen iguales (con $x_4 =0$) y además obtenemos dos nuevos puntos, Tabla \ref{tabla2}.
\begin{table}
\begin{center}
\begin{tabular}{||l||c||c||} \hline
$P:(x_1,x_2,x_3)$  & $w_{eff}$  & $\Omega_{m}$ \\ \hline
$P_7 = (0,0,0,1)$  & $1/3$  & $0$ \\
& & \\
$P_8 = \biggl(\dfrac{4m}{1 + m},-\dfrac{2m}{(1 + m)^2},\dfrac{2m}{(1 + m)},\dfrac{1-2m-5m^2}{(1 + m)^2}\biggr)$  & $\dfrac{1-3m}{3(1 + m)}$  & $0$ \\
& & \\
\hline\hline
\end{tabular}
\caption{Valores adicionales para los puntos críticos del sistema dinámico (\ref{sistemadinamico}) considerando contribuciones en radiación.}
\label{tabla2}
\end{center}
\end{table}
Observamos que el punto $P_7$ es un punto estándar de radiación. Cuando $m(r)$ es constante los autovalores de $P_7$ son
\begin{equation}
1,  \qquad 4, \qquad 4, \qquad -1,
\end{equation}
lo que implica que $P_7$ es un punto de silla.  El punto $P_8$ es un punto nuevo de radiación que contiene contribución a la energía oscura diferente de cero. Puesto que la ecuación de estado efectiva esta restringida por nucleosínteis a ser cercana a $1/3$, $P_8$ es un punto aceptable para la época de radiación con $m_8$ cercano a 0. Los autovalores para el punto $P_8$ son
\begin{equation}
1, \qquad 4(1+m'_8), \qquad \frac{m_8 - 1 \pm \sqrt{81m_8^2 + 30 m_8 - 15}}{2(m_8+1)},
\end{equation}
En el límite $m_8 \longrightarrow 0$ los últimos dos autovalores son complejos con partes reales negativas, lo que muestra que el punto $P_8$ es un punto de silla alrededor del punto de radiación.  A diferencia del punto de dominio de materia $P_5$ no existen singularidades en los autovalores para el punto $P_8$ en el límite
$m_8 \longrightarrow 0$. También notamos que el punto $P_8$ esta sobre la línea $m=-r-1$ como en el caso del punto $P_5$ . Si la condición para la existencia del punto de dominio de materia $P_5$ se satisface ($m(r) \simeq 0^{+}, r=-1$), entonces existe un punto de radiación en la misma región \cite{Amendola2}.\\\\
De este modo existen dos clases de modelos que son cosmológicamente viables:
\begin{itemize}
\item (A) Modelos que conectan $P_5 (r\simeq -1, m \simeq 0^{+})$ con $P_1 (r = -2, 0 < m \leq 1))$
\item (B) Modelos que conectan $P_5 (r\simeq -1, m \simeq 0^{-})$ con $P_6 (m=-r-1, -(\sqrt{3}+1)/2 < m < 0)$.
\end{itemize}
De la expresión (\ref{restricciones}) los modelos viables de $f(R)$ para explicar energía oscura deben satisfacer $m > 0$ que esta en acuerdo con los argumentos anteriores.\\\\
\section{Algunos Modelos Específicos}
\noindent En esta sección consideraremos algunos modelos $f(R)$ para los cuales se puede escribir explícitamente $m$ como función de $r$ y se estudiara la posibilidad de obtener una época de dominio de materia seguida por una de expansión acelerada, todo esto siguiendo \cite{Amendola2}
\subsection{$f(R) = \alpha R^{-n}$}
\begin{figure}
\begin{center}
\includegraphics[scale=0.42]{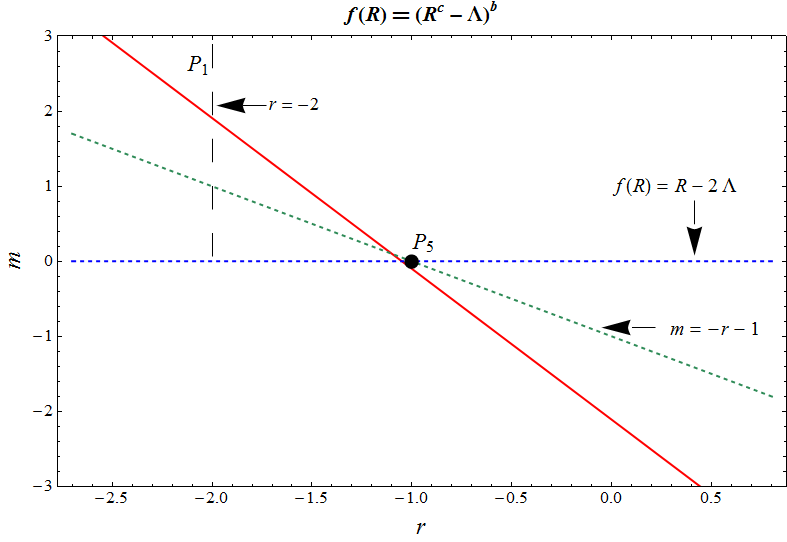}
\caption{Plano $(r,m)$ para $f(R)=(R^{b}-\Lambda)^{c}$.}
\label{plano1}
\end{center}
\end{figure}

\noindent Para este modelo obtenemos a partir de (\ref{mr}) la siguiente expresión
\begin{equation}\label{m1}
m = -n -1,
\end{equation}
con $r= n$. La curva $m(r)$ es degenerada, de modo que este caso se reduce a un sistema bidimensional (en el caso en que no hay radiación), debido a la relación  $x_3 = r x_2 = nx_2$. La condición $m(r=-1) =0$ se satisface entonces para $n=-1$, es decir, en el caso de la Relatividad General. Ahora para que el termino $f'(R) =-n\alpha R^{-n-1}$ sea mayor que cero, se necesita que  $\alpha <0$ para $n>0$ y $\alpha >0$ para $n<0$. De la expresión (\ref{m1}) se tiene $m > 0$ para $n < - 1$ y $m<0$ para $n>-1$. Ahora a partir de los autovalores para el punto $P_5$ (\ref{eigen5}), el punto para la época de materia es una espiral estable para $n < - 1$, de modo que las soluciones no abandonan la era de materia para entrar en la expansión acelerada. \\\\
Por otro lado, $P_5$ es un punto de silla para $-1 < n < -0.713$ mientras que el punto $P_2$ es estable en la región $-1<n<-3/4$. Sin embargo, uno de los autovalores de $P_5$ muestra un divergencia positiva en el límite $m \rightarrow 0^{-}$, lo que significa que el punto de materia se vuelve repulsivo para $m$ cercano a $0^{-}$. Como vimos anteriormente la ecuación de estado efectiva para $P_6$ tiene un valor phantom de $< -7.6$ para $m=0^{-}$, lo que implica que el punto de Silla $P_5$ se conecta al punto $P_2$ o al punto $P_6$.\\\\
El modelo $\Lambda$CDM $f(R) = R - 2\Lambda$ corresponde a la línea horizontal $m=0$, que conecta la época de materia $(r,m)=(-1,0)$ con el punto de de Sitter $P_1$ $(r,m)=(-2,0)$. Una posible generalización del modelo $\Lambda$CDM
\begin{equation}
f(R) = (R^{b}-\Lambda)^{c},
\end{equation}
con lo cual $r=-c(R^{b}-\Lambda)^{-1}bR^{b}$ y
\begin{equation}
m(r)=\frac{(1-c)}{c}r + b -1.
\end{equation}
Si la intersección $m= -1 +bc$ con la línea crítica es en  $0 <m \ll 1$ y la pendiente esta dada por $-1 < (1-c)/c < 0$, entonces la época de materia esta conectada co $P_1$ y el modelo es aceptable. En la Figura \ref{plano1} se muestra el plano $(r,m)$ para $f(R) =(R^{b}-\Lambda)^{c}$.
\subsection{$f(R) = R + \alpha R^{-n}$}
\begin{figure}[htb]
\begin{center}
\includegraphics[scale=0.42]{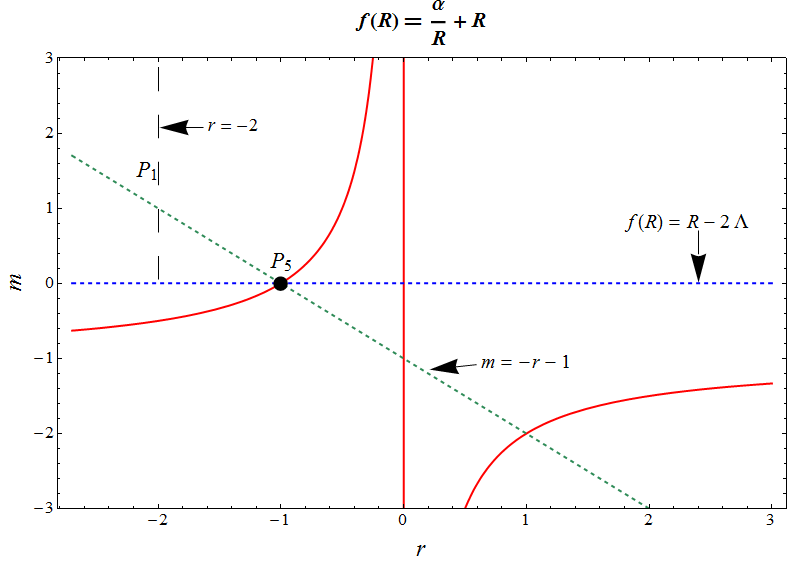}
\caption{Plano $(r,m)$ para $f(R)=R+\alpha/R$.}
\label{plano2}
\end{center}
\end{figure}

\begin{figure}[htb]
\begin{center}
\includegraphics[scale=0.42]{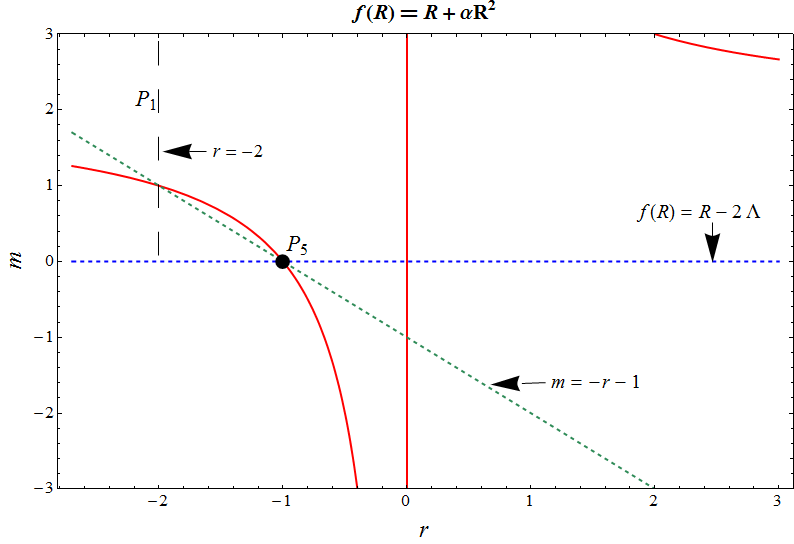}
\caption{Plano $(r,m)$ para $f(R)=R+\alpha R^2$, modelo de Starobinsky.}
\label{plano3}
\end{center}
\end{figure}

\noindent Este modelo esta propuesto en \cite{Carroll4} para explicar la actual expansión acelerada. Para este modelo tendremos
\begin{equation}\label{mP52}
m(r) = - \frac{n(1+r)}{r},
\end{equation}
vemos que tal modelo es independiente de $\alpha$, y puesto que $m(r=-1)=0$ el modelo satisface la condición necesaria para la existencia del punto de materia $P_5$.
Reemplazando (\ref{mP51}) para (\ref{mP52}), obtenemos la solución $m_a = 0$ o  $m_b =-(n+1)$, que es cierto para los puntos $P_5$ y $P_6$. En este caso tales puntos están dados por
\begin{equation}
P_{5a} = \biggl(0, -\frac{1}{2}, \frac{1}{2}\biggr), \qquad \Omega_{m} = 1, \qquad w_{eff} =0,
\end{equation}
\begin{equation}
P_{5b} = \biggl(-\frac{3(n+1)}{n}, \frac{4n+3}{2n^2}, \frac{4n+3}{2n}\biggr), \qquad \Omega_{m} = -\frac{8n^2 +13 n +3}{2n^2}, \qquad w_{eff} =-1-\frac{1}{n},
\end{equation}
\begin{multline}
P_{6b} = \biggl(-\frac{2(n+2)}{2n+1}, \frac{4n+5}{(n+1)(2n+1)}, \frac{n(4n+5)}{(n+1)(2n+1)}\biggr), \qquad \Omega_{m} = 0, \\
w_{eff} =--\frac{6n^2+7n-1}{3(n+1)(2n+1)}.
\end{multline}
Esta familia de modelos se puede dividir en tres casos: $(1) n < -1$, $(2) -1 < n < 1$ y $(3) n > 1$. Para el caso $(1)$ tendremos que
$m'=n/r^2$, implica que $m'(r=-1)< -11$ y por lo tanto la época de materia alrededor de $m \approx 0^{+}$ es estable ($P_1$ es estable también para $2<n<0$). El caso con $n=-2$ corresponde al modelo de inflación de Starobinsky \cite{Starobinsky}.\\\\
Para el caso $(2)$ la condición $r=-1$ se satisface con $R \rightarrow \infty$ y vemos que
\begin{equation}
m = \frac{n(n+1)\alpha R^{-n-1}}{(1-n\alpha R^{-n-1})},
\end{equation}
se aproxima a cero cuando si $\alpha = 0$. Para este caso existen oscilaciones alrededor de la época de materia y un punto final de de Sitter $P_1$. Para el caso $(3)$ el punto $P_6$ es estable y acelerado en la región (A) (para $w_{eff}$ diferente de un valor phantom) debido a la condición $m=-n-1 < -1$. Si $\alpha > 0$, $m$ se aproxima a cero por el lado derecho, y entonces existen oscilaciones alrededor de la época de materia, pero el punto de aceleración $P_1$ es inestable. Cuando $\alpha <0$, $m$ se aproxima a cero por el lado izquierdo y de nuevo el punto de materia es inestable, pues uno de los autovalores muestra una divergencia. En la figura \ref{plano2} se muestra el plano $(r,m)$ para el modelo $f(R) = R + \alpha/R$,  $(n=1)$ donde se aprecia claramente que NO es un modelo viable, pues no conecta el punto de la época de materia con el de expansión acelerada. En la figura \ref{plano3} se muestra el plano $(r,m)$ para el modelo de Starobinsky.
\subsection{$f(R) = R^{p}e^{q R}$}
\begin{figure}[htb]
\begin{center}
\includegraphics[scale=0.42]{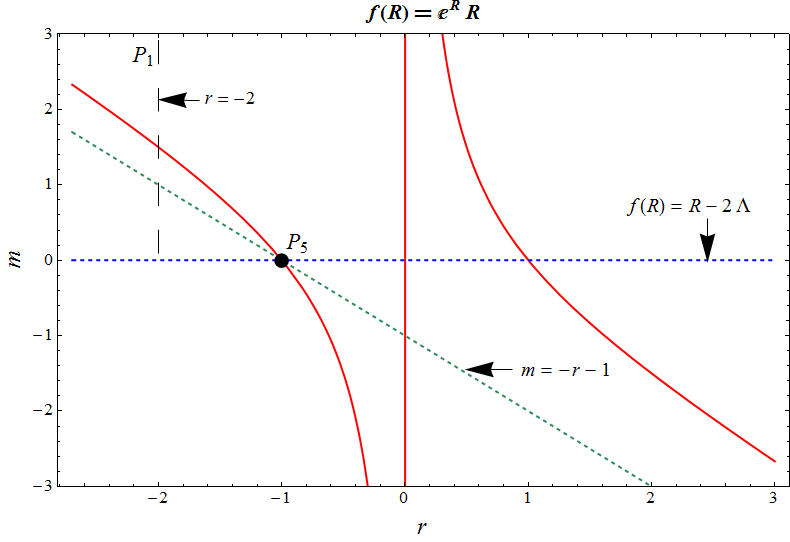}
\caption{Plano $(r,m)$ para $f(R)=R e^{R}$.}
\label{plano4}
\end{center}
\end{figure}

\noindent En este modelo $m$ esta dado por
\begin{equation}
m(r) = -r + \frac{p}{r},
\end{equation}
En el caso exponencial puro ($p = 0$) se tiene $m=-1$ y $x_3/m \rightarrow x_2 \rightarrow 0$ de modo que el punto $P_2$ existe mientras que $P_5$ no. Por otro lado vemos que la función $m$ se hace cero para $r= \pm\sqrt{p}$, lo que implica que la condición $m(r=-1)=0$ para la existencia de un punto de materia se cumple para $p=1$. Sin embargo, puesto que en este caso $m'(r=-1)=-2 < - 1$, el punto $P_5$ es una espiral estable para $m>0$.\\\\
En el límite $m \rightarrow 0^{+}$, el punto $P_6$ no puede ser usado para la expansión acelerada, además el punto $P_5$ es estable.  Por otra parte, cuando $m(r=-2)=3/2$ para $p=1$, el punto de de Sitter no es estable. Estos modelos entonces no tienen una secuencia de época de materia y expansión acelerada para $p=1$. En la figura \ref{plano4} se muestra el plano $(r,m)$ para $f(R) = R e^{R}$ ($p=q=1$).
\subsection{$f(R) = R^{p}(\ln \alpha R)^{q} $}
\begin{figure}[htb]
\begin{center}
\includegraphics[scale=0.42]{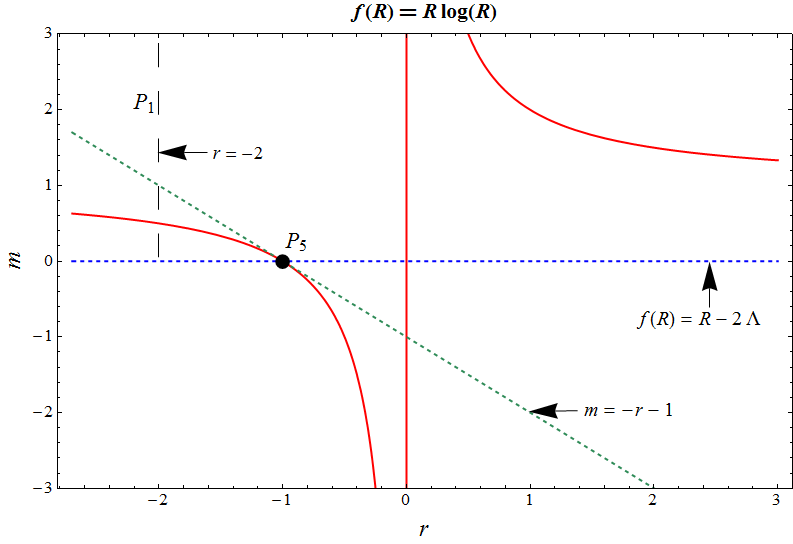}
\caption{Plano $(r,m)$ para $f(R)=R\ln R$. En la figura se debe entender el nombre $\log$ como el logaritmo natural $\ln$.}
\label{plano5}
\end{center}
\end{figure}

\noindent Para este modelo tenemos la relación
\begin{equation}
m(r) = \frac{p^2 + 2pr - r(q-r+qr)}{qr},
\end{equation}
puesto que $m(r=-1)=-(p-1)^2/q$, la época de materia existe solo para $p = 1$. Cuando $p=1$ se tiene $m(r=-2)=1-1/2q$, lo que significa que $P_1$ es estable para $q > 0$ pero no lo es para $q <0$.  La derivada de $m(r)$ es
\begin{equation}
m'(r)  = -1 + \frac{r^2-1}{qr^2},
\end{equation}
como $m'(r=-1)=-1$, el punto $P_5$ es estable. Similarmente el punto de aceleración $P_6$ es estable en la región (C) para $ q > 0$, pero no lo es para $q <0$.
Esto implica que la curva $m(r)$ no atraviesa el punto $P_6$ en la región (C), de modo que las únicas trayectorias posibles son desde el punto de materia $P_5$ al punto de de Sitter $P_1$, $(r=-2) $. En la figura \ref{plano5} se muestra el plano $(r,m)$ para $f(R) = R \ln R$, ($p=\alpha=q=1$).
\begin{figure}[htb]
\begin{center}
\includegraphics[scale=0.42]{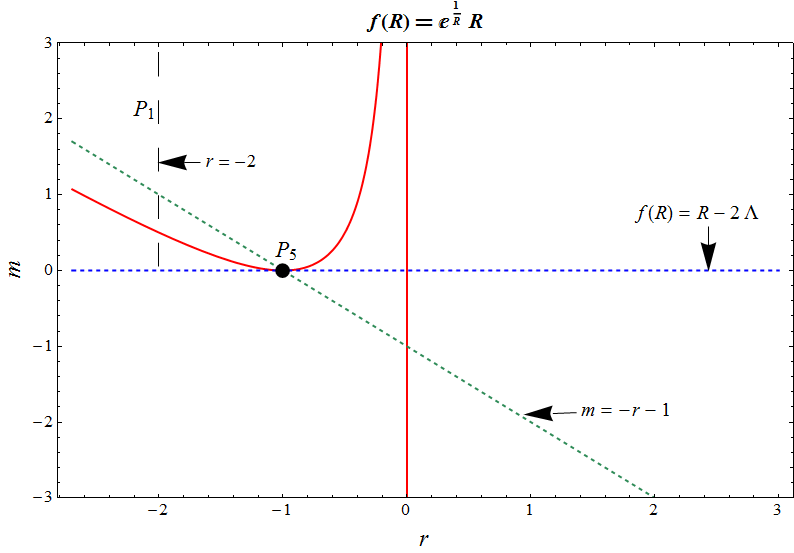}
\caption{Plano $(r,m)$ para $f(R)=Re^{1/R}$.}
\label{plano6}
\end{center}
\end{figure}

\subsection{$f(R) = R^{p}e^{q/R} $}
\noindent Para este modelo se tiene que
\begin{equation}
m(r) = - \frac{p+r(2+r)}{r},
\end{equation}
la cual es independiente de $q$. Tenemos que  $m(r=-1)=p-1$ de modo que la época de materia existe para $p=1$. En este caso se tiene $m(r)=-(r+1)^2/r >0$ para $r<0$, y puesto que $m(r=-2) = 1/2$ para $p=1$, el punto $P_1$ es una espiral estable. La derivada de $m(r)$ es $m'(r)=-1+1/r^2$, que implica entonces que $m'(r=-1)=0$ y $m'(r<-1)>-1$. Esto muestra que el punto $P_5$ es un punto de silla mientras que $P_6$ en la región (C) es estable. La curva $m(r)$ satisface $m(r) < - r-1$ en la región $r<-1$ para $p=1$ y también un comportamiento asintótico $m(r) \rightarrow -r$ en el límite $r \rightarrow -\infty$. Entonces, en principio, es posible tener una secuencia $P_5 \rightarrow P_6$ ($r \rightarrow -\infty$), pero la trayectoria desde el punto $P_5$ es \textit{atrapada} por el punto estable de de Sitter $P_1$ que existe en $(r,m)=(-2,1/2)$. En la figura \ref{plano6} se muestra el plano $(r,m)$ para $f(R) = Re^{1/R}$.\\

\noindent Si observamos ahora las figuras \ref{plano1} a \ref{plano6} encontramos que las únicas que nos pueden dar una época de materia seguida por una expansión acelerada son los modelos $f(R) = R \ln R$ y $f(R) = R e^{1/R}$. La razón de estos es que los demás modelos \textit{atraviesan} la región delimitada por $m = -r -1$ para $m > -1$ hasta alcanzar el punto de de Sitter ubicado en la línea $r=-2$.\\\\
Podemos entonces resumir las condiciones que debe cumplir una función $f(R)$ para ser cosmológicamente viable:\\
\begin{enumerate}
\item Un modelo $f(R)$ debe tener una época estándar de dominio de materia solo si satisface las condiciones
\begin{equation}
m(r) \approx 0^{+}, \qquad m'(r) > -1 \, \text{en } r \approx -1,
\end{equation}
\item La época de materia es seguida por una aceleración tipo de Sitter ($w_{eff}=-1$) solo si
\begin{equation}
0 < m(r) \leq 1 \, \text{en } r=-2, \, \text{o } \, m(r) = -r-1 \rightarrow \pm \infty,
\end{equation}
\item La época de materia es seguida por un atractor con $w_{eff} > -1$ solo si $m = -r-1$ y se cumplen simultáneamente
\begin{equation}
(\sqrt{3}-1)/2 < m(r) \leq 1 \, \text{ y } \, m'(r) < - 1.
\end{equation}
\end{enumerate}
En la siguiente sección estudiaremos el problema de las distancias cosmológicas en teorías $f(R)$ estudiando para esto la Ecuación de Desvío Geodésico en estas teorías.

\newpage

{\color{white} . }

\chapter{Distancias Cosmológicas en Teorías $f(R)$}\label{Capitulo6}
\drop{C}onsideramos ahora el problema de las distancias cosmológicas en teorías $f(R)$. Este estudio es de gran importancia, pues permite restringir los modelos $f(R)$ mediante observaciones, en particular por ejemplo con la medición de distancias en supernovas de tipo IA \cite{Perlmutter}. Si bien, las ecuaciones obtenidas en el Capítulo \ref{Capitulo5} nos permiten en principio determinar las distancias, usando también las expresiones dadas en el Capítulo \ref{Capitulo3}, debemos notar sin embargo que tal proceso conlleva una complejidad inherente en cuanto la misma forma de la función $f(R)$.\\\\
La expresión (\ref{Friedmod1}) nos da entonces una relación para el parámetro de Hubble $H$ en función de la densidad de energía $\rho$, la presión $p$ y fuertemente en el escalar de Ricci $R$, por otro lado, sin embargo, de la métrica de Robertson-Walker se tiene una relación para $R$ en función de $H$ y $\dot{H}$, y además una relación diferencial para $R$ en función de la traza del tensor energía-momentum, a través de la traza de las ecuaciones de campo (\ref{tracefield}). De esta forma es necesario o bien fijar la forma para $H(z)$, puede ser por observaciones o suponiendo un ansatz, o bien fijar la forma de la función $f(R)$ para poder determinar $H(z)$ partiendo de las ecuaciones modificadas. \\\\
En el formalismo de Palatini, por ejemplo, las distancias cosmológicas pueden ser determinadas y comparadas con datos observacionales, fijando la forma de la función $f(R)$ \cite{Mortonson}. Cabe decir que la forma de las ecuaciones de campo en el formalismo de Palatini permiten en principio determinar la forma para $H(z)$ de una manera más \textit{simple} que en el caso del formalismo métrico, debido a que en estas ecuaciones los operadores $\square$ y $\nabla_{\alpha}\nabla_{\beta}$ no aparecen de forma explícita en las ecuaciones. \\\\
Un método interesante para atacar el problema de las distancias cosmológicas en teorías $f(R)$ es presentado en \cite{Capozziello4} a través de la denominada \textit{cosmografía}. El punto clave en la cosmografía es la definición de nuevas parámetros cosmológicos que contienen derivadas del factor de escala hasta de quinto orden. Ya conocemos uno de ellos, el parámetro de Hubble dado por $H(t) \equiv \frac{\dot{a}}{a}$, además definimos \cite{Visser}
\begin{subequations}
\begin{equation}
q(t) = - \frac{\ddot{a}}{a H^2}, \qquad \text{Parámetro de desaceleración}
\end{equation}
\begin{equation}
j(t) = +\frac{\dddot{a}}{a H^3}, \qquad \text{Parámetro de \textit{jerk}}
\end{equation}
\begin{equation}
s(t) = +\frac{a^{(4)}}{a H^4}, \qquad \text{Parámetro de \textit{snap}}
\end{equation}
\begin{equation}
l(t) = +\frac{a^{(5)}}{a H^5}, \qquad \text{Parámetro de \textit{lerk}}
\end{equation}
\end{subequations}
con lo cual el factor de escala $a(t)$ se puede escribir como una serie en términos de $t$. Estos parámetros nos permiten además obtener las expresiones para las distancias (diametral angular y de luminosidad) como series en función del redshift \cite{Visser}. Aunque esta aproximación tiene algunas ventajas (es cosmología sin ecuaciones de campo) conlleva al problema de que nos da tan solo una aproximación a los valores para las distancias, dependiendo los ordenes usados en la expansión. El método seguido entonces en \cite{Capozziello4}, extiende tal método para la determinación de las distancias en teorías $f(R)$ fijando la forma de $f(R)$ como una expansión en serie de Taylor.\\\\
Nuestro objetivo en este capitulo es utilizar al EDG extendida a teorías $f(R)$ para la determinación de distancias de una forma alternativa.
\section{Ecuación de Desvío Geodésico en Teorías $f(R)$}
\noindent Como vimos en las secciones \ref{EDG} y \ref{EDGFLRW} el estudio de la Ecuación de Desvío Geodésico (EDG) nos permite caracterizar las propiedades más importantes de un espacio-tiempo, y lo más importante tratar el problema de las distancias cosmológicas de una manera muy general. Nuestro objetivo es entonces extender los resultados de la EDG para las teorías $f(R)$ en el formalismo métrico siguiendo para esto \cite{Guarnizo2}; en el formalismo de Palatini la EDG se ha estudiado en \cite{Shojai}.\\\\
Nuestro punto de partida es entonces son las expresiones para el tensor de Ricci y el escalar de Ricci escritos a partir de las ecuaciones de campo y su traza respectivamente (\ref{ecuacionescampo}) y (\ref{tracefield})
\begin{equation}\label{ricci}
R_{\alpha\beta}  = \frac{1}{f'(R)}\biggl[\kappa T_{\alpha\beta}+\frac{f(R)}{2}\, g_{\alpha\beta} - g_{\alpha\beta}\square f'(R) + \nabla_{\alpha}\nabla_{\beta}f'(R)\biggr],
\end{equation}
\begin{equation}\label{scalar}
R  = \frac{1}{f'(R)}\biggl[\kappa T+2f(R) - 3\square f'(R)\biggr].
\end{equation}
Usando estas expresiones podemos escribir el tensor de Riemann (\ref{RiemannC}) como
\begin{multline}
R_{\alpha\beta\gamma\delta} = C_{\alpha\beta\gamma\delta} + \frac{1}{2f'(R)}\Biggl[\kappa(T_{\delta\beta}g_{\alpha\gamma}-T_{\gamma\beta}g_{\alpha\delta} + T_{\gamma\alpha}g_{\beta\delta}-T_{\delta\alpha}g_{\beta\gamma}) + f(R)\bigl(g_{\alpha\gamma}g_{\delta\beta} - g_{\alpha \delta}g_{\gamma\beta}\bigr)\\
+  \bigl(g_{\alpha\gamma}\mathcal{D}_{\delta\beta}
 - g_{\alpha\delta}\mathcal{D}_{\gamma\beta} + g_{\beta\delta}\mathcal{D}_{\gamma\alpha}- g_{\beta\gamma}\mathcal{D}_{\delta\alpha}\bigr)f'(R)\Biggr] - \frac{1}{6f'(R)}\biggl(\kappa T + 2f(R) + 3\square f'(R)\biggr)\bigl(g_{\alpha\gamma}g_{\delta\beta} - g_{\alpha \delta}g_{\gamma\beta}\bigr),
\end{multline}
donde definimos el operador
\begin{equation}
\boxed{\mathcal{D}_{\alpha\beta} \equiv \nabla_{\alpha}\nabla_{\beta} - g_{\alpha\beta}\square.}
\end{equation}
Entonces subiendo el primer índice en el tensor de Riemann
\begin{multline}
R_{\beta\gamma\delta}^{\alpha} = C_{\beta\gamma\delta}^{\alpha} + \frac{1}{2f'(R)}\Biggl[\kappa(T_{\delta\beta}\delta_{\gamma}^{\alpha}-T_{\gamma\beta}\delta_{\delta}^{\alpha} + T_{\gamma}^{\, \, \alpha}g_{\beta\delta}-T_{\delta}^{\, \, \alpha}g_{\beta\gamma}) + f(R)\bigl(\delta_{\gamma}^{\alpha}g_{\delta\beta} - \delta_{\delta}^{\alpha}g_{\gamma\beta}\bigr)\\
+  \bigl(\delta_{\gamma}^{\alpha}\mathcal{D}_{\delta\beta}
 - \delta_{\delta}^{\alpha}\mathcal{D}_{\gamma\beta} + g_{\beta\delta}\mathcal{D}_{\gamma}^{\, \, \alpha}- g_{\beta\gamma}\mathcal{D}_{\delta}^{\, \, \alpha}\bigr)f'(R)\Biggr] - \frac{1}{6f'(R)}\biggl(\kappa T + 2f(R) + 3\square f'(R)\biggr)\bigl(\delta_{\gamma}^{\alpha}g_{\delta\beta} - \delta_{\delta}^{\alpha}g_{\gamma\beta}\bigr),
\end{multline}
y contrayendo con $V^{\beta}\eta^{\gamma}V^{\delta}$ el lado derecho de la EDG (\ref{GDE}) nos queda
\begin{multline}
R_{\beta\gamma\delta}^{\alpha}V^{\beta}\eta^{\gamma}V^{\delta} = C_{\beta\gamma\delta}^{\alpha}V^{\beta}\eta^{\gamma}V^{\delta} + \frac{1}{2f'(R)}\Biggl[\kappa(T_{\delta\beta}\delta_{\gamma}^{\alpha}-T_{\gamma\beta}\delta_{\delta}^{\alpha} + T_{\gamma}^{\, \, \alpha}g_{\beta\delta}-T_{\delta}^{\, \, \alpha}g_{\beta\gamma}) + f(R)\bigl(\delta_{\gamma}^{\alpha}g_{\delta\beta} - \delta_{\delta}^{\alpha}g_{\gamma\beta}\bigr)\\
+  \bigl(\delta_{\gamma}^{\alpha}\mathcal{D}_{\delta\beta}
 - \delta_{\delta}^{\alpha}\mathcal{D}_{\gamma\beta} + g_{\beta\delta}\mathcal{D}_{\gamma}^{\, \, \alpha}- g_{\beta\gamma}\mathcal{D}_{\delta}^{\, \, \alpha}\bigr)f'(R)\Biggr]V^{\beta}\eta^{\gamma}V^{\delta}\\ - \frac{1}{6f'(R)}\biggl(\kappa T + 2f(R) + 3\square f'(R)\biggr)\bigl(\delta_{\gamma}^{\alpha}g_{\delta\beta} - \delta_{\delta}^{\alpha}g_{\gamma\beta}\bigr)V^{\beta}\eta^{\gamma}V^{\delta},
\end{multline}
Esta ecuación nos permitirá entonces encontrar la EDG para cualquier espacio-tiempo y para cualquier forma del tensor de energía-momentum en teorías $f(R)$.\\\\
En la siguiente sección mostraremos en detalle los pasos para encontrar la EDG en teorías $f(R)$ usando la métrica de Robertson-Walker, nuestro propósito también es el de comparar los resultados con los de RG para el caso límite  $f(R) = R - 2\Lambda$.
\subsection{Ecuación de Desvío Geodésico para el Universo FLRW}
\noindent Usando entonces la métrica de Robertson-Walker (\ref{robw}) y el tensor de energía-momentum para un fluido perfecto (\ref{e-m1}) tendremos
\begin{equation}\label{RicciRWF}
R_{\alpha\beta}  = \frac{1}{f'(R)}\biggl[\kappa(\rho + p)u_{\alpha}u_{\beta}+ \biggl(\kappa p + \frac{f(R)}{2}\biggr)g_{\alpha\beta} +
\mathcal{D}_{\alpha\beta}f'(R)\biggr],
\end{equation}
\begin{equation}\label{ScalarRWF}
R  = \frac{1}{f'(R)}\biggl[\kappa (3p-\rho) + 2f(R) - 3\square f'(R)\biggr],
\end{equation}
con estas expresiones podemos escribir el tensor de Riemann como
\begin{multline}
R_{\alpha\beta\gamma\delta} = \frac{1}{2f'(R)}\Biggl[\kappa(\rho+p)\bigl(u_{\delta}u_{\beta}g_{\alpha\gamma}-u_{\gamma}u_{\beta}g_{\alpha\delta} + u_{\gamma}u_{\alpha}g_{\beta\delta}-u_{\delta}u_{\alpha}g_{\beta\gamma}\bigr)\\
+ \biggl(\kappa p + \frac{\kappa\rho}{3} + \frac{f(R)}{3}+ \square f'(R)\biggr)\bigl(g_{\alpha\gamma}g_{\delta\beta} - g_{\alpha \delta}g_{\gamma\beta}\bigr)+ (g_{\alpha\gamma}\mathcal{D}_{\delta\beta}
 - g_{\alpha\delta}\mathcal{D}_{\gamma\beta} + g_{\beta\delta}\mathcal{D}_{\gamma\alpha}- g_{\beta\gamma}\mathcal{D}_{\delta\alpha})f'(R)\Biggr],
\end{multline}
donde de nuevo hemos usado que para este espacio-tiempo el tensor de Weyl $C_{\alpha\beta\gamma\delta}$ es nulo. Ahora, si el campo vectorial $V^{\alpha}$ esta normalizado, tendremos que $V^{\alpha}V_{\alpha} = \epsilon$, y por lo tanto
\begin{multline}
R_{\alpha\beta\gamma\delta}V^{\beta}V^{\delta} = \frac{1}{2f'(R)}\Biggl[\kappa(\rho+p)\bigl(g_{\alpha\gamma}(u_{\beta}V^{\beta})^2-2(u_{\beta}V^{\beta})V_{(\alpha}u_{\gamma)}
 + \epsilon u_{\alpha}u_{\gamma}\bigr)\\ + \biggl(\kappa p + \frac{\kappa\rho}{3} + \frac{f(R)}{3}+ \square f'(R)\biggr)\bigl(\epsilon g_{\alpha\gamma}-V_{\alpha}V_{\gamma}\bigr)
 + (g_{\alpha\gamma}\mathcal{D}_{\delta\beta} - g_{\alpha\delta}\mathcal{D}_{\gamma\beta} + g_{\beta\delta}\mathcal{D}_{\gamma\alpha}- g_{\beta\gamma}\mathcal{D}_{\delta\alpha})f'(R)V^{\beta}V^{\delta}\Biggr],
\end{multline}
subiendo el primer índice en el tensor de Riemann y multiplicando por $\eta^{\gamma}$
\begin{multline}
R_{\beta\gamma\delta}^{\alpha}V^{\beta}\eta^{\gamma}V^{\delta} = \frac{1}{2f'(R)}\Biggr[\kappa(\rho+p)\bigl((u_{\beta}V^{\beta})^2\eta^{\alpha}-(u_{\beta}V^{\beta})V^{\alpha}(u_{\gamma}\eta^{\gamma})-(u_{\beta}V^{\beta})u^{\alpha}(V_{\gamma}\eta^{\gamma}) + \epsilon u^{\alpha}u_{\gamma}\eta^{\gamma}\bigr)\\
+ \biggl(\kappa p + \frac{\kappa\rho}{3} + \frac{f(R)}{3}+ \square f'(R)\biggr)\bigl(\epsilon \eta^{\alpha}-V^{\alpha}(V_{\gamma}\eta^{\gamma})\bigr) + \bigl[(\delta_{\gamma}^{\alpha}\mathcal{D}_{\delta\beta} - \delta_{\delta}^{\alpha}\mathcal{D}_{\gamma\beta} +
g_{\beta\delta}\mathcal{D}_{\gamma}^{\, \, \alpha}- g_{\beta\gamma}\mathcal{D}_{\delta}^{\, \,  \alpha})f'(R)\bigr]V^{\beta}V^{\delta}\eta^{\gamma}\Biggr],
\end{multline}
con $E = - V_{\alpha}u^{\alpha}, \eta_{\alpha}u^{\alpha}=\eta_{\alpha}V^{\alpha}=0$ \cite{Ellis1} esta expresión se reduce a
\begin{multline}\label{Riemann1}
R_{\beta\gamma\delta}^{\alpha}V^{\beta}\eta^{\gamma}V^{\delta} = \frac{1}{2f'(R)}\Biggr[\kappa(\rho+p)E^2   + \epsilon\biggl(\kappa p + \frac{\kappa\rho}{3} + \frac{f(R)}{3}+ \square f'(R)\biggr)\Biggr]\eta^{\alpha}\\
+ \frac{1}{2f'(R)}\biggl[\bigl[(\delta_{\gamma}^{\alpha}\mathcal{D}_{\delta\beta} - \delta_{\delta}^{\alpha}\mathcal{D}_{\gamma\beta} +
g_{\beta\delta}\mathcal{D}_{\gamma}^{\, \, \alpha}- g_{\beta\gamma}\mathcal{D}_{\delta}^{\, \,  \alpha})f'(R)\bigr]V^{\beta}V^{\delta}\biggr]\eta^{\gamma},
\end{multline}
Usando de nuevo la expresión (\ref{RWscalar})
\begin{equation}\label{RWscalar}
R = 6\biggl[\frac{\ddot{a}}{a}+ \biggl(\frac{\dot{a}}{a}\biggr)^2 + \frac{k}{a^2}\biggr] = 6\biggl[\dot{H} + 2H^2 + \frac{k}{a^2}\biggr],
\end{equation}
Usando ahora los resultados del Apéndice \ref{AppendixC} tenemos que los únicos operadores $\mathcal{D}_{\alpha\beta}$ que son diferentes de cero son
\begin{align}\label{square}
\square f'(R) &=  - \partial_{0}^2 f'(R) - 3H \partial_{0} f'(R),\notag \\
 &=  - f''(R) \ddot{R} - f'''(R) \dot{R}^2 - 3H f''(R) \dot{R},
\end{align}
\begin{align}
\mathcal{D}_{00} &= -3 H \partial_{0} f'(R),\notag \\
& = -3 H f''(R) \dot{R},
\end{align}
\begin{align}
\mathcal{D}_{ij} &= 2 H g_{ij}\partial_{0} f'(R) + g_{ij}\partial_{0}^2 f'(R), \notag \\
& = 2 H g_{ij}f''(R) \dot{R} + g_{ij}(f''(R) \ddot{R} + f'''(R) \dot{R}^2),
\end{align}
con $g_{ij}$ las componentes espaciales de la métrica de Robertson-Walker y $\dot{R} = \partial_{0} R$. Con estos resultados podemos obtener la contribución total de los operadores en (\ref{Riemann1}), ver Apéndice \ref{AppendixE}
\begin{equation}
\bigl(\delta_{\gamma}^{\alpha}\mathcal{D}_{\delta\beta} - \delta_{\delta}^{\alpha}\mathcal{D}_{\gamma\beta} + g_{\beta\delta}\mathcal{D}_{\gamma}^{\, \, \alpha}
- g_{\beta\gamma}\mathcal{D}_{\delta}^{\, \,  \alpha}\bigr)f'(R)V^{\beta}V^{\delta}\eta^{\gamma} = \epsilon \bigl(5H f''(R) \dot{R}  + f''(R) \ddot{R} + f'''(R) \dot{R}^2\bigr) \eta^{\alpha},
\end{equation}
entonces la expresión para $R_{\beta\gamma\delta}^{\alpha}V^{\beta}\eta^{\gamma}V^{\delta}$ se reduce a
\begin{equation}\label{Pirani2}
\boxed{R_{\beta\gamma\delta}^{\alpha}V^{\beta}\eta^{\gamma}V^{\delta} = \frac{1}{2f'(R)}\Biggl[\kappa(\rho+p)E^2   + \epsilon\biggl(\kappa p + \frac{\kappa\rho}{3} + \frac{f(R)}{3} + 2 H  f''(R) \dot{R}\biggr)\Biggr]\eta^{\alpha},}
\end{equation}
la cual es la generalización de la ecuación de Pirani. En el caso particular $f(R) = R -2\Lambda$ la anterior expresión se reduce a (\ref{Pirani1}). \\\\
Entonces podemos escribir la EDG en gravedad $f(R)$ a partir de la expresión (\ref{GDE})
\begin{equation}\label{GDeFR}
\boxed{\frac{D^2 \eta^{\alpha}}{D \nu^2} = - \frac{1}{2f'(R)}\Biggl[\kappa(\rho+p)E^2   + \epsilon\biggl(\kappa p + \frac{\kappa\rho}{3} + \frac{f(R)}{3} + 2 H  f''(R) \dot{R}\biggr)\Biggr]\eta^{\alpha},}
\end{equation}
como esperábamos la EDG induce solo un cambio en la magnitud del vector desviación $\eta^{\alpha}$, lo cual ocurre también en RG. Este resultado era de esperarse debido a la forma de la métrica, que describe un universo homogéneo e isotrópico. Para universos anisotrópicos, como por ejemplo los de Bianchi I la EDG induce también un
\textit{cambio en la dirección} del vector desviación, como se muestra en \cite{Caceres1},\cite{Caceres}.
\subsection{EDG para Observadores Fundamentales}
\noindent En este caso tenemos $V^{\alpha}$ como la cuadri-velocidad del fluido $u^{\alpha}$. El parámetro afín  $\nu$ coincide con el tiempo propio de los observadores fundamentales $\nu = t$. Puesto que tenemos geodésicas temporales entonces $\epsilon=-1$ y también $E = 1$, entonces a partir de (\ref{Pirani2})
\begin{equation}\label{GDEfR1}
\boxed{R_{\beta\gamma\delta}^{\alpha}u^{\beta}\eta^{\gamma}u^{\delta} = \frac{1}{2f'(R)}\biggl[\frac{2\kappa\rho}{3}  -\frac{f(R)}{3} - 2 H  f''(R) \dot{R}\biggr]\eta^{\alpha},}
\end{equation}
si el vector desviación es $\eta_{\alpha} = \ell e_{\alpha}$, la isotropía implica
\begin{equation}
\frac{D e^{\alpha}}{D t} = 0,
\end{equation}
y
\begin{equation}
\frac{D^2 \eta^{\alpha}}{D t^2} = \frac{d^2\ell}{dt^2} e^{\alpha},
\end{equation}
usando este resultado en la EDG (\ref{GDE}) con (\ref{GDEfR1}) tenemos
\begin{equation}
\frac{d^2\ell}{dt^2} = - \frac{1}{2f'(R)}\biggl[\frac{2\kappa\rho}{3}  -\frac{f(R)}{3} - 2 H  f''(R) \dot{R}\biggr]\, \ell ,
\end{equation}
en el caso $\ell = a(t)$ tendremos
\begin{equation}\label{Raycha}
\boxed{\frac{\ddot{a}}{a} = \frac{1}{f'(R)}\biggl[\frac{f(R)}{6}  + H f''(R) \dot{R} -\frac{\kappa\rho}{3} \biggr].}
\end{equation}
Esta ecuación puede ser obtenida como un caso particular de la ecuación de Raychaudhuri generalizada dada en \cite{Rippl}. Es posible obtener como ejemplo, la forma estándar de las ecuaciones modificadas de campo a partir de la ecuación de Raychaudhuri. Primero restando la expresión
$R/6$ en  (\ref{Raycha}) con $R$ dado en (\ref{RWscalar}) tendremos
\begin{align}
\frac{\ddot{a}}{a} - \frac{\ddot{a}}{a} - \biggl(\frac{\dot{a}}{a}\biggr)^2 - \frac{k}{a^2} & = \frac{1}{f'(R)}\biggl[\frac{f(R)}{6}  + H f''(R) \dot{R} -\frac{\kappa\rho}{3} \biggr] - \frac{R}{6}, \notag \\
- \biggl(\frac{\dot{a}}{a}\biggr)^2 - \frac{k}{a^2} & = -\frac{1}{f'(R)}\biggl[\frac{\kappa\rho}{3} - \frac{f(R)}{6}  - H f''(R) \dot{R} + \frac{R f'(R)}{6}\biggr], \notag \\
- H^2 - \frac{k}{a^2} & = -\frac{1}{3f'(R)}\biggl[\kappa\rho + \frac{(R f'(R) - f(R))}{2} - 3H f''(R) \dot{R} \biggr], \notag \\
 H^2 + \frac{k}{a^2} & = \frac{1}{3f'(R)}\biggl[\kappa\rho + \frac{(R f'(R) - f(R))}{2} - 3H f''(R) \dot{R} \biggr],
\end{align}
que corresponde a la ecuación (\ref{Friedmod1}). Para obtener la segunda ecuación sumamos a ambos lados de la ecuación  (\ref{Raycha}) la expresión $R/6$, y usamos también la forma para la traza de las ecuaciones de campo
(\ref{scalar}) con el operador $\square f'(R)$ dado en (\ref{square})
\begin{align}\label{ModFried2}
\frac{\ddot{a}}{a} + \frac{\ddot{a}}{a} + \biggl(\frac{\dot{a}}{a}\biggr)^2 + \frac{k}{a^2} &= \frac{1}{f'(R)}\biggl[\frac{f(R)}{6}  + H f''(R) \dot{R} -\frac{\kappa\rho}{3} \biggr] + \frac{R}{6}, \notag \\
2\frac{\ddot{a}}{a} + \biggl(\frac{\dot{a}}{a}\biggr)^2 + \frac{k}{a^2} & = \frac{1}{f'(R)}\biggl[\frac{f(R)}{6} -\frac{\kappa\rho}{3} + \frac{R f'(R)}{3}-\kappa p + \frac{\kappa \rho}{3}- \frac{2 f(R)}{3} - f''(R) \ddot{R}, \notag \\
& \hspace{5.3cm} - f'''(R)\dot{R}^2 - 2H f''(R) \dot{R}  + \frac{R F(R)}{6}\biggr], \notag \\
2\dot{H} + 3H^2 + \frac{k}{a^2} & = \frac{1}{f'(R)}\biggl[-\kappa p - 2H f''(R) \dot{R} - \frac{(f(R)- R f'(R))}{2}  -f''(R) \ddot{R} - f'''(R) \dot{R}^2 \biggr], \notag \\
2\dot{H} + 3H^2 + \frac{k}{a^2} & = -\frac{1}{f'(R)}\biggl[\kappa p + 2H f''(R) \dot{R} + \frac{(f(R)- R f'(R))}{2}  + f''(R) \ddot{R} + f'''(R) \dot{R}^2\biggr],
\end{align}
que corresponde a la segunda ecuación de Friedmann modificada (\ref{Friedmod2}).
\subsection{EDG para Campos Vectoriales Nulos}
\noindent Consideramos ahora la EDG para campos vectoriales nulos dirigidos al pasado. En este caso tenemos $V^{\alpha}=k^{\alpha}$, $k_{\alpha}k^{\alpha}=0$, entonces la ecuación (\ref{Pirani2}) se reduce a
\begin{equation}\label{RicciFoc}
\boxed{R_{\beta\gamma\delta}^{\alpha}k^{\beta}\eta^{\gamma}k^{\delta} = \frac{1}{2f'(R)}\kappa(\rho+p)E^2\,\eta^{\alpha},}
\end{equation}
expresión que puede ser interpretada como el \textit{Enfocamiento de Ricci} en teorías $f(R)$. Escribiendo $\eta^{\alpha}= \eta e^{\alpha}$,  $e_{\alpha}e^{\alpha}=1$, $e_{\alpha}u^{\alpha}=e_{\alpha}k^{\alpha}=0$ y escogiendo una base paralelamente propagada
$\frac{D e^{\alpha}}{D \nu}=k^{\beta}\nabla_{\beta}e^{\alpha}=0$, la EDG (\ref{GDeFR}) nos queda
\begin{equation}\label{GDE3}
\frac{d^2\eta}{d\nu^2} = - \frac{1}{2f'(R)}\kappa(\rho+p)E^2\, \eta.
\end{equation}
En el caso de RG discutido en  \cite{Ellis1}, todas las familias de geodésicas nulas dirigidas al pasado experimentan enfocamiento, siempre y cuando se cumpla $\kappa(\rho+p) > 0$,
y para un fluido con ecuación de estado $p = - \rho$ (constante cosmológica) no hay influencia en el enfocamiento. A partir de (\ref{GDE3}) la condición para enfocamiento en teorías $f(R)$ es
\begin{equation}
\frac{\kappa(\rho + p)}{f'(R)} > 0.
\end{equation}
Una condición similar sobre la función $f(R)$ fue establecida para evitar entre otras la aparición de fantasmas \footnote{En donde fantasmas se refieren a estados con norma negativa o campos con signo negativo en el termino cinético.} \cite{Felice},\cite{Starobinsky}, en la cual $f'(R) > 0$, ver también Sección \ref{Seccion5.1}. Queremos ahora escribir la ecuación (\ref{GDE3}) en función del parámetro de redshift $z$. Siguiendo la Sección \ref{Seccion3.4} podemos escribir entonces
\begin{equation}
\frac{d\nu}{dz} = \frac{1}{E_0 H (1+z)^2},
\end{equation}
y
\begin{equation}
\frac{d^2\nu}{dz^2} =- \frac{1}{E_0 H (1+z)^3}\biggl[\frac{1}{H}(1+z)\frac{dH}{dz}+2\biggr],
\end{equation}
expresamos nuevamente $\frac{dH}{dz}$ como
\begin{equation}
\frac{dH}{dz} = \frac{d\nu}{dz}\frac{dt}{d\nu}\frac{dH}{dt} = - \frac{1}{H(1+z)} \frac{dH}{dt}.
\end{equation}
De la definición para $H$ tendremos
\begin{equation}
\dot{H} \equiv \frac{dH}{dt} = \frac{d}{dt}\frac{\dot{a}}{a} = \frac{\ddot{a}}{a} - H^2,
\end{equation}
and usando la ecuación de Raychaudhuri (\ref{Raycha})
\begin{equation}
\dot{H} = \frac{1}{f'(R)}\biggl[\frac{f(R)}{6}  + H f''(R) \dot{R} -\frac{\kappa\rho}{3} \biggr]- H^2,
\end{equation}
entonces
\begin{equation}
\frac{d^2\nu}{dz^2} = -\frac{3}{E_0 H (1+z)^3}\biggl[1+ \frac{1}{3H^2 f'(R)}\biggl(\frac{\kappa \rho}{3}- \frac{f(R)}{6} - H f''(R) \dot{R}\biggr)\biggr],
\end{equation}
el operador $\frac{d^2\eta}{d\nu^2}$ toma la forma
\begin{equation}
\frac{d^2\eta}{d\nu^2} = \bigl(EH(1+z)\bigr)^2\Biggl[\frac{d^2\eta}{dz^2} + \frac{3}{(1+z)}\biggl[1+ \frac{1}{3H^2 f'(R)}\biggl
(\frac{\kappa \rho}{3}- \frac{f(R)}{6} - H f''(R) \dot{R}\biggr)\biggr]\frac{d\eta}{dz}\Biggr],
\end{equation}
y la EDG (\ref{GDE3}) se reduce a
\begin{equation}
\boxed{\frac{d^2\eta}{dz^2} + \frac{3}{(1+z)}\Biggl[1+ \frac{1}{3H^2 f'(R)}\biggl
(\frac{\kappa \rho}{3}- \frac{f(R)}{6} - H f''(R) \dot{R}\biggr)\Biggr]\, \frac{d\eta}{dz} + \frac{\kappa(\rho+p)}{2H^2(1+z)^2f'(R)}\, \eta = 0,}
\end{equation}
Ahora, la densidad de energía $\rho$ y la presión $p$ considerando las contribuciones de materia y radiación, se pueden escribir como
\begin{equation}
\kappa\rho = 3H_0^2 \Omega_{m0}(1 + z)^3 + 3H_0^2\Omega_{r0}(1 + z)^4, \qquad \kappa p = H_0^2\Omega_{r0}(1 + z)^4,
\end{equation}
donde hemos usado $p_m=0$ y $p_r = \frac{1}{3}\rho_r$. Podemos ahora escribir la EDG en una forma más compacta
\begin{equation}\label{MattigGen}
\frac{d^2\eta}{dz^2} + \mathcal{P}(H,R,z)\frac{d\eta}{dz} + \mathcal{Q}(H,R,z)\eta = 0,
\end{equation}
con
\begin{equation}
\boxed{\mathcal{P}(H,R,z) = \frac{4\Omega_{m0}(1 + z)^3 + 4\Omega_{r0}(1 + z)^4 +  3f'(R)\Omega_{k0}(1 + z)^2  + 4\Omega_{DE} -\frac{Rf'(R)}{6 H_0^2}}{(1+z)\bigl(\Omega_{m0}(1 + z)^3 + \Omega_{r0}(1 + z)^4 + f'(R)\Omega_{k0}(1 + z)^2 + \Omega_{DE}\bigr)},}
\end{equation}
\begin{equation}
\boxed{\mathcal{Q}(H,R,z) = \frac{3\Omega_{m0}(1 + z) + 4\Omega_{r0}(1 + z)^2}{2\bigl(\Omega_{m0}(1 + z)^3+\Omega_{r0}(1 + z)^4 +  f'(R)\Omega_{k0}(1 + z)^2 + \Omega_{DE}\bigr)}.}
\end{equation}
y $H$ dado por las ecuaciones de campo modificadas (\ref{Friedmod1})
\begin{align}\label{Friedmod}
H^2 &= \frac{1}{f'(R)}\biggl[H_0^2 \Omega_{m0}(1 + z)^3 + H_0^2\Omega_{r0}(1 + z)^4 + \frac{(R f'(R)-f(R))}{6} - H f''(R) \dot{R}\biggr] - \frac{k}{a^2}, \notag \\
H^2 &= H_0^2\biggl[\frac{1}{f'(R)}\bigl(\Omega_{m0}(1 + z)^3 + \Omega_{r0}(1 + z)^4 + \Omega_{DE} \bigr) + \Omega_{k}(1+z)^2\biggr],
\end{align}
donde
\begin{equation}\label{OmegaDE1}
\boxed{\Omega_{DE} \equiv  \frac{1}{H_0^2 }\biggl[\frac{(R f'(R)-f(R))}{6} - H f''(R) \dot{R}\biggr],}
\end{equation}
que es equivalente a la expresión (\ref{OmegaDE}) y
\begin{equation}
\Omega_{k0}=-\frac{k}{H_0^2 a_0^2}.
\end{equation}
Para poder resolver la ecuación (\ref{MattigGen}) es necesario escribir $R$ y $H$ en función del redshift.
definimos entonces
\begin{equation}
\frac{d}{dt} = \frac{dz}{da}\frac{da}{dt}\frac{d}{dz} = -(1+z)H \frac{d}{dz},
\end{equation}
y entonces para el escalar de Ricci tendremos \cite{Capozziello}
\begin{align*}
R &= 6\biggl[\frac{\ddot{a}}{a}+ \biggl(\frac{\dot{a}}{a}\biggr)^2 + \frac{k}{a^2}\biggr], \notag \\
&=6\biggl[2H^2 + \dot{H} + \frac{k}{a^2}\biggr], \notag \\
&=6\biggl[2H^2  -(1+z)H \frac{dH}{dz} + k(1+z)^2 \biggr],
\end{align*}
Ahora, para tener $H = H(z)$ o bien sea fijar de entrada la forma $H(z)$ o fijar una forma específica de la función $f(R)$. Este punto ha sido discutido en detalle en  \cite{Capozziello2} y el método de fijar $H(z)$ para hallar la forma de la función $f(R)$ por observaciones en  \cite{Capoziello3}.\\\\
En el caso particular $f(R) = R -2 \Lambda$ tenemos
$f'(R) = 1$, $f''(R) =0$. la expresión para $\Omega_{DE}$ se reduce a
\begin{equation}
\Omega_{DE} = \frac{1}{H_0^2 }\biggl[\frac{(R-R + 2\Lambda)}{6}\biggr] = \frac{\Lambda}{3H_0^2} \equiv \Omega_{\Lambda},
\end{equation}
y por lo tanto, como se habia anticipado en el Capitulo \ref{Capitulo5} el parámetro $\Omega_{DE}$ generaliza el parámetro de energía oscura \footnote{Cabe aclarar que la generalización es para contribuciones netamente geométricas.}. La ecuación de Friedmann modificada (\ref{Friedmod1}) se convierte en la expresión conocida en RG (\ref{fried7})
\begin{equation}
H^2 = H_0^2\bigl[\Omega_{m0}(1 + z)^3 + \Omega_{r0}(1 + z)^4 + \Omega_{\Lambda}  + \Omega_{k}(1+z)^2\bigr],
\end{equation}
las expresiones $\mathcal{P}$, y $\mathcal{Q}$ se reducen a
\begin{equation}
\mathcal{P}(z) = \frac{4\Omega_{r0}(1 + z)^4 + (7/2)
\Omega_{m0}(1 + z)^3 + 3\Omega_{k0}(1 + z)^2 + 2\Omega_{\Lambda}
}{(1+z)\bigl(\Omega_{r0}(1 + z)^4 + \Omega_{m0}(1 + z)^3 + \Omega_{k0}(1 + z)^2 + \Omega_{\Lambda}\bigr)},
\end{equation}
\begin{equation}
\mathcal{Q}(z) = \frac{2\Omega_{r0}(1 + z)^2 + (3/2)
\Omega_{m0}(1 + z)}{\Omega_{r0}(1 + z)^4 + \Omega_{m0}(1 + z)^3 + \Omega_{k0}(1 + z)^2 + \Omega_{\Lambda}}.
\end{equation}
y la EDG para campos vectoriales nulos es
\begin{multline}
\frac{d^2\eta}{dz^2} + \frac{4\Omega_{r0}(1 + z)^4 + (7/2)
\Omega_{m0}(1 + z)^3 + 3\Omega_{k0}(1 + z)^2 + 2\Omega_{\Lambda}
}{(1+z)\bigl(\Omega_{r0}(1 + z)^4 + \Omega_{m0}(1 + z)^3 + \Omega_{k0}(1 + z)^2 + \Omega_{\Lambda}\bigr)}\, \frac{d\eta}{dz}\\ + \frac{2\Omega_{r0}(1 + z)^2 + (3/2)
\Omega_{m0}(1 + z)}{\Omega_{r0}(1 + z)^4 + \Omega_{m0}(1 + z)^3 + \Omega_{k0}(1 + z)^2 + \Omega_{\Lambda}}\, \eta = 0.
\end{multline}
que coincide con la expresión (\ref{GDEnull1}). Nuevamente la relación de Mattig es obtenida en el caso $\Omega_{\Lambda} = 0$ y escribiendo $\Omega_{k0}=1-\Omega_{m0} - \Omega_{r0}$
\begin{equation}
\frac{d^2\eta}{dz^2} + \frac{6 +
\Omega_{m0}(1 + 7z) + \Omega_{r0}(1 + 8z + 4z^2)}{2(1 + z)(1 + \Omega_{m0}z +
\Omega_{r0}z(2 + z))}\, \frac{d\eta}{dz} + \frac{3\Omega_{m0} + 4\Omega_{r0}(1 + z)}{2(1 + z)(1 + \Omega_{m0}z +
\Omega_{r0}z(2 + z))}\, \eta = 0,
\end{equation}
de este modo la ecuación (\ref{MattigGen}) nos da una generalización a la relación de Mattig en teorías $f(R)$.\\\\
Para relacionar ahora la magnitud del vector desviación $\eta$ con la distancia diametral angular, consideramos nuevamente que  la magnitud del vector desviación $\eta$ se relaciona con el área propia $dA$ de una fuente a redshift $z$ por $d\eta \propto \sqrt{dA}$, y de esta expresión, la definición de la distancia diametral angular $d_A$ puede ser escrita como
\begin{equation}
d_A = \sqrt{\frac{dA}{d\Omega}},
\end{equation}
con $d\Omega$ el ángulo sólido. De modo que la forma de la relación de Mattig generalizada es la misma cambiando $\eta$ por $d_A$ (el factor de proporcionalidad se puede descartar de la ecuación diferencial), de modo que
\begin{multline}\label{diametrald}
\frac{d^2\, d_{A}^{f(R)}}{dz^2} + \frac{4\Omega_{m0}(1 + z)^3 + 4\Omega_{r0}(1 + z)^4 +  3f'(R)\Omega_{k0}(1 + z)^2
+ 4\Omega_{DE} -\frac{Rf'(R)}{6 H_0^2}}{(1+z)\bigl(\Omega_{m0}(1 + z)^3 + \Omega_{r0}(1 + z)^4 + f'(R)\Omega_{k0}(1 + z)^2 + \Omega_{DE}\bigr)}\, \frac{d \, d_{A}^{f(R)}}{dz}\\ +
\frac{3\Omega_{m0}(1 + z) + 4\Omega_{r0}(1 + z)^2}{2\bigl(\Omega_{m0}(1 + z)^3+\Omega_{r0}(1 + z)^4 +  f'(R)\Omega_{k0}(1 + z)^2 + \Omega_{DE}\bigr)}\, d_{A}^{f(R)} = 0, \end{multline}
donde denotamos la distancia diametral angular por $d_{A}^{f(R)}$, para enfatizar que alguna solución de la ecuación anterior necesita una forma específica de la función $f(R)$, o bien la función $H(z)$.  Esta ecuación satisface las condiciones iniciales (para $z \geq z_0$)
\begin{equation}\label{condition1}
d_A^{f(R)}(z,z_0)\biggl|_{z=z_0} = 0,
\end{equation}
\begin{equation}\label{condition2}
\frac{d\, d_A^{f(R)}}{dz}(z,z_0)\biggl|_{z=z_0} = \frac{H_0}{H(z_0) (1+z_0)},
\end{equation}
con $H(z_0)$ siendo la ecuación de Friedmann modificada (\ref{Friedmod1}) evaluada en $z=z_0$.
\section{Relaciones entre Distancias Cosmológicas en Teorías $f(R)$}
\noindent En el Capitulo \ref{Capitulo3} mostramos las diferentes relaciones entre las distancias cosmológicas en RG. Nos preguntamos entonces si tales relaciones se mantienen validas en las teorías $f(R)$. Si bien tales relaciones entre las distancias las mostramos definiendo en sí cada una de ellas, existe un teorema denominado \textit{teorema de Reciprocidad} mostrado por Etherington en (1933) \cite{Etherington} y que nos dice
\begin{thm}[Teorema de Reciprocidad (Etherington)]
La distancia área observador $r_0$ y la distancia área galaxia $r_G$ son iguales salvo
un factor de redshift cosmológico
\begin{equation}
r_G^2 = r_0^2 (1+z)^2.
\end{equation}
\end{thm}
\noindent El punto fundamental de este teorema es el hecho de que varias propiedades geométricas son invariantes cuando los roles de fuente y observador se intercambian.  La distancia $r_G$ se relaciona con la distancia de Luminosidad por la relación
\begin{equation}
d_L = r_G (1+z),
\end{equation}
y por lo tanto
\begin{equation}
d_L = r_0(1+z)^2,
\end{equation}
y entonces usando que $dA = r_0^2 d\Omega$ podemos relacionar la distancia área observador con la distancia diametral angular, para obtener
\begin{equation}
d_L = d_A(1+z)^2.
\end{equation}
Una prueba de este teorema de reciprocidad esta dado en \cite{Ellis1},\cite{Caceres} a partir de la primera integral en la EDG. Mostraremos a continuación muy brevemente estos resultados.
\subsection{Primera Integral en la EDG}
\noindent La primera integral nos relaciona básicamente los
vectores desviación $\eta_{1}^{\alpha}$, $\eta_{2}^{\alpha}$ de
dos geodésicas que comparten el mismo campo vectorial geodésico
$V^{\alpha}$. Es decir, ambas cumplen a partir de (\ref{GDE})
\begin{align}
&\frac{D^{2}\eta_{1}^{\alpha}}{D\nu^{2}}=-R^{\alpha}_{\ \beta\gamma\delta}V^{\beta}\eta_{1}^{\gamma}V^{\delta}\label{eta1},\\
&\frac{D^{2}\eta_{2}^{\alpha}}{D\nu^{2}}=-R^{\alpha}_{\
\beta\gamma\delta}V^{\beta}\eta_{2}^{\gamma}V^{\delta}\label{eta2}.
\end{align}
Multiplicando \eqref{eta1} por $\eta_{2\alpha}$ y
\eqref{eta2} por $\eta_{1\alpha}$ y teniendo en cuenta que podemos
subir y bajar los índices que se suman
\begin{align}
&\eta_{2\alpha}\frac{D^{2}\eta_{1}^{\alpha}}{D\nu^{2}}=-\eta_{2\alpha}R^{\alpha}_{\ \beta\gamma\delta}V^{\beta}\eta_{1}^{\gamma}V^{\delta}=-\eta_{2}^{\alpha}R_{\alpha\beta\gamma\delta}V^{\beta}\eta_{1}^{\gamma}V^{\delta}\label{eta1riemann},\\
&\eta_{1\alpha}\frac{D^{2}\eta_{2}^{\alpha}}{D\nu^{2}}=-\eta_{1\alpha}R^{\alpha}_{\
\beta\gamma\delta}V^{\beta}\eta_{2}^{\gamma}V^{\delta}=-\eta_{1}^{\alpha}R_{\alpha\beta\gamma\delta}V^{\beta}\eta_{2}^{\gamma}V^{\delta}\label{eta2riemann},
\end{align}
teniendo en cuenta que las derivadas covariantes cumplen la regla de Leibnitz podemos escribir
\begin{align}
&\eta_{2\alpha}\frac{D^{2}\eta_{1\alpha}^{\alpha}}{D\nu^{2}}=\frac{D}{D\nu}\left(\eta_{2\alpha}\frac{D\eta_{1}^{\alpha}}{D\nu}\right)-\frac{D\eta_{2\alpha}}{D\nu}\frac{D\eta_{1}^{\alpha}}{D\nu},\\
&\eta_{1\alpha}\frac{D^{2}\eta_{2\alpha}^{\alpha}}{D\nu^{2}}=\frac{d}{D\nu}\left(\eta_{1\alpha}\frac{D\eta_{2}^{\alpha}}{D\nu}\right)-\frac{D\eta_{1\alpha}}{D\nu}\frac{D\eta_{2}^{\alpha}}{D\nu}.
\end{align}
restando estas expresiones
\begin{align}
\eta_{1\alpha}\frac{D^{2}\eta_{2}^{\alpha}}{D\nu^{2}}-\eta_{2\alpha}\frac{D^{2}\eta_{1}^{\alpha}}{D\nu^{2}}&=
\frac{D}{D\nu}\left(\eta_{1\alpha}\frac{D\eta_{2}^{\alpha}}{D\nu}-\eta_{2\alpha}\frac{D\eta_{1}^{\alpha}}{D\nu}\right)
-\left(\frac{D\eta_{2\alpha}}{D\nu}\frac{D\eta_{1}^{\alpha}}{D\nu}
-\frac{D\eta_{1\alpha}}{D\nu}\frac{D\eta_{2}^{\alpha}}{D\nu}\right)\notag\\
&=\frac{D}{D\nu}\left(\eta_{1\alpha}\frac{D\eta_{2}^{\alpha}}{D\nu}-\eta_{2\alpha}\frac{D\eta_{1}^{\alpha}}{D\nu}\right).
\end{align}
Por otro lado, restando \eqref{eta1riemann} de
\eqref{eta2riemann} y teniendo en cuenta las propiedades de
simetría del tensor de Riemann tendremos
\begin{align}
\eta_{1\alpha}\frac{D^{2}\eta_{2}^{\alpha}}{D\nu^{2}}-\eta_{2\alpha}\frac{D^{2}\eta_{1}^{\alpha}}{D\nu^{2}}&=R_{\alpha\beta\gamma\delta}V^{\beta}V^{\delta}(\eta_{2}^{\alpha}\eta_{1}^{\gamma})-R_{\alpha\beta\gamma\delta}V^{\beta}V^{\delta}(\eta_{1}^{\alpha}\eta_{2}^{\gamma})\notag\\
&=R_{\alpha\beta\gamma\delta}V^{\beta}V^{\delta}(\eta_{2}^{\alpha}\eta_{1}^{\gamma})-R_{\gamma\delta\alpha\beta}V^{\beta}V^{\delta}(\eta_{1}^{\alpha}\eta_{2}^{\gamma})\notag\\
&=R_{\alpha\beta\gamma\delta}V^{\beta}V^{\delta}(\eta_{2}^{\alpha}\eta_{1}^{\gamma})-R_{\alpha\beta\gamma\delta}V^{\delta}V^{\beta}(\eta_{1}^{\gamma}\eta_{2}^{\alpha})\notag\\
&=0,
\end{align}
de modo que
\begin{equation}
\frac{D}{D\nu}\left(\eta_{1\alpha}\frac{D\eta_{2}^{\alpha}}{D\nu}-\eta_{2\alpha}\frac{D\eta_{1}^{\alpha}}{D\nu}\right)=0,
\end{equation}
con lo cual finalmente obtenemos la primera integral de la EDG
\begin{equation}\label{primeraintegral}
\boxed{\eta_{1a}\frac{D\eta_{2}^{a}}{D\nu}-\eta_{2a}\frac{D\eta_{1}^{a}}{D\nu}=\text{const.}}
\end{equation}
Esta ecuación es válida para cualquier modelo cosmológico que supone válida la Relatividad General. Esta primera integral se puede investigar
para el caso de rayos nulos. Consideremos geodésicas nulas que
divergen de una fuente $S$ y llegan a un observador $O$, con vector
desviación $\eta_{1}$ y rayos nulos que divergen de la fuente $S$ y
llegan al observador $O$, con vector desviación $\eta_{2}$. La
primera integral está expresada en términos del parámetro afín
$\nu$, pero expresémosla ahora en términos de la magnitud $\ell$
definida por $d\ell:=S dr$. Para encontrar la relación entre $dr$ y
$d\nu$, consideremos una geodésica radial de la siguiente manera
\begin{equation}
V^{\alpha}:=Eu^{\alpha}+Pe^{\alpha},
\end{equation}
donde
$e^{\alpha}=S^{-1}(\partial_{r})^{\alpha}=S^{-1}\delta^{\alpha}_{\
1}$, $e_{\alpha}e^{\alpha}=1$, $e_{\alpha}u^{\alpha}=0$. Para una
geodésica radial nula tenemos además
\begin{equation}
\frac{dt}{d\nu}=V^{0}=E=-(V_{\alpha}u^{\alpha})=E_{0}\frac{S_{0}}{S},
\end{equation}
\begin{equation}
\frac{dr}{d\nu}=V^{1}=\frac{P(S)}{S}=\frac{E(S)}{S}.
\end{equation}
De esta forma
\begin{equation}
d\ell=Sdr=E_{0}\left(\frac{S_{0}}{S}\right)d\nu,
\end{equation}
podemos escribir entonces
\begin{align}
&\frac{d\eta_{1}}{d\nu}=\frac{d\ell}{d\nu}\frac{d\eta_{1}}{d\ell}=E_{0}\frac{S_{0}}{S}\frac{d\eta_{1}}{d\ell}=E_{0}(1+z)\frac{d\eta_{1}}{d\ell},\\
&\frac{d\eta_{2}}{d\nu}=E_{0}(1+z)\frac{d\eta_{2}}{d\ell}.
\end{align}
Teniendo en cuenta que para el observador $z=0$ tenemos
\begin{equation}
\eta_{2}|_{0}\frac{d\eta_{1}}{d\ell}|_{0}=\eta_{1}|_{S}\frac{d\eta_{2}}{d\ell}|_{S}(1+z),
\end{equation}
donde los términos $\frac{d\eta}{d\ell}$ son los ángulos
subtendidos por los pares de rayos nulos y corresponden a los
vectores desviación. Expresado en términos de las distancias área
y luminosidad, definidas por
\begin{align}
&\eta_{1}|_{S}:=r_{0}\frac{d\eta_{1}}{d\ell}|_{0},\\
&\eta_{2}|_{0}:=r_{G}\frac{d\eta_{2}}{d\ell}|_{S}.
\end{align}
encontramos el famoso teorema de reciprocidad para
los modelos de FLRW
\begin{equation}
r_{G}=r_{0}(1+z).
\end{equation}
Ahora, puesto que usamos como métrica la de Robertson-Walker y las expresión para el redshift siguen siendo las mismas que en RG, \textbf{las relaciones entre las distancias cosmológicas se preservan también en las teorías $\boldsymbol{f(R)}$}. Así, tendremos entonces para la distancia de Luminosidad $d_L$ en teorías $f(R)$ que
\begin{equation}
\boxed{d_{L}^{f(R)} = d_{A}^{f(R)}(1+z)^2.}
\end{equation}
El proceso entonces para determinar las distancias cosmológicas en teorías $f(R)$ se puede describir como
\begin{enumerate}
\item Fijar una forma para $H(z)$ o bien una forma específica de la función $f(R)$.
\item Encontrar a partir de las ecuaciones de Friedmann modificadas las expresiones para $H(z)$ y $R(z)$, en general a partir de un cálculo numérico
\item Determinar $\Omega_{DE}$ (\ref{OmegaDE1}) en función de $z$.
\item Resolver la ecuación (\ref{diametrald}) para $d_A^{f(R)}$ con las condiciones iniciales (\ref{condition1}) y (\ref{condition2}).
\item Obtener a partir de esta expresión las demás distancias cosmológicas.
\end{enumerate}
Si bien, nuestro método para la determinación de distancias a partir de la EDG para campos vectoriales nulos en teorías $f(R)$ necesita una forma específica para la función, es una herramienta elegante y que nos brinda la posibilidad de comparar directamente con resultados observacionales, siendo así una buena herramienta para restringir modelos $f(R)$.
\section{Ecuación de tipo Dyer-Roeder en Teorías $f(R)$}
\noindent Finalmente daremos una importante relación que nos permitirá estudiar las distancias cosmológicas también en universos con inhomogeneidades. La \textit{ecuación de Dyer-Roeder} nos da una ecuación diferencial para la distancia diametral angular $d_A$ en función del redshift $z$ introduciendo un parámetro que nos permite introducir inhomogeneidades en la densidad de materia \cite{Dyer1},\cite{Dyer2}. La forma estándar de la ecuación de Dyer-Roeder en RG esta dada por \cite{Castaneda},\cite{Okamura}
\begin{equation}
(1+z)^2 \mathcal{F}(z) \frac{d^2\, d_A}{dz^2} + (1+z) \mathcal{G}(z) \frac{d\, d_A}{dz^2} + \mathcal{H}(z) d_A = 0,
\end{equation}
con
\begin{equation}
\mathcal{F}(z) = H^2(z),
\end{equation}
\begin{equation}
\mathcal{G}(z) = (1+z)H(z) \frac{dH}{dz} + 2 H^2(z),
\end{equation}
\begin{equation}
\mathcal{H}(z) = \frac{3 \tilde{\alpha}(z)}{2}\Omega_{m0}(1+z)^3,
\end{equation}
siendo $\tilde{\alpha}(z)$ el \textit{parámetro de suavidad}, que nos da el carácter de las inhomogeneidades en la densidad de energía.\\\\
Con el fin de obtener una ecuación de tipo Dyer-Roeder en teorías $f(R)$ seguimos \cite{Schneider}. Primero debemos notar que los términos que contienen las derivadas de $d_{A}^{f(R)}$ en la ecuación (\ref{diametrald}) vienen de la transformación $\frac{d}{d\nu} \longrightarrow \frac{d}{dz}$ y el término que acompaña a
$d_{A}^{f(R)}$ viene del enfocamiento de Ricci (\ref{RicciFoc}). Entonces siguiendo \cite{Castaneda},\cite{Dyer2}, introducimos la fracción de masa $\tilde{\alpha}$ (parámetro de suavidad), y entonces reemplazamos
\textit{solamente} en el enfocamiento de Ricci $\rho \longrightarrow \tilde{\alpha}\rho$. Siguiendo estos argumentos en (\ref{MattigGen}), y considerando el caso $\Omega_{r0}=0$ tendremos
\begin{multline}\label{DyerRoederMod}
(1+z)^2\frac{d^2\, d_{A}^{f(R),\text{ DR}}}{dz^2} + (1+z)\frac{4\Omega_{m0}(1 + z)^3 + 3f'(R)\Omega_{k0}(1 + z)^2
+ 4\Omega_{DE} -\frac{Rf'(R)}{6 H_0^2}}{\bigl(\Omega_{m0}(1 + z)^3 + f'(R)\Omega_{k0}(1 + z)^2 + \Omega_{DE}\bigr)}\, \frac{d \, d_{A}^{f(R), \text{ DR}}}{dz}\\ +
\frac{3\tilde{\alpha}(z)\Omega_{m0}(1 + z)^3 }{2\bigl(\Omega_{m0}(1 + z)^3 +  f'(R)\Omega_{k0}(1 + z)^2 + \Omega_{DE}\bigr)}\, d_{A}^{f(R), \text{ DR}} = 0.
\end{multline}
donde denotamos la distancia de Dyer-Roeder en gravedad $f(R)$ por $d_{A}^{f(R),\text{ DR}}$, y además a diferencia de la expresión (\ref{diametrald}), cada uno de los términos esta dividido sobre $(1+z)^2$. Esta ecuación generalizada de Dyer-Roeder también satisface las condiciones (\ref{condition1}) y (\ref{condition2}), y como es de esperar, se reduce a la forma estándar en RG para el caso particular $f(R) = R - 2\Lambda$.

\newpage

{\color{white} .} 
\chapter{Conclusiones y Perspectivas}\label{Capitulo9}
\drop{C}omo vimos a lo largo de este trabajo, las teorías de gravedad modificada $f(R)$ proveen una interesante alternativa para atacar el problema de la actual expansión acelerada del universo.  Su interés creciente en los últimos 15 años radica en parte por ser una de las generalizaciones más simples de la Relatividad General, consistiendo de una función arbitraria del escalar de Ricci, en el lagrangiano que describe la gravitación.\\\\
En el Capítulo \ref{Capitulo4} obtuvimos las ecuaciones de campo en el formalismo métrico de las teorías $f(R)$ destacando la importancia de los términos de frontera en la acción modificada. Las ecuaciones de campo obtenidas nos muestran claramente que son de cuarto orden la métrica, y también como pueden interpretarse los términos adicionales, como la contribución a un tensor de energía-momentum efectivo compuesto de componentes netamente geométricas. Discutimos también las ecuaciones de campo en los formalismos de Palatini y métrico-afín, y la equivalencia de las teorías $f(R)$ con las de Brans-Dicke.\\\\
Los modelos cosmológicos fueron estudiados en el Capitulo \ref{Capitulo5}, partiendo para esto de las ecuaciones de campo modificadas y de la validez del principio cosmológico, lo que nos permitió usar como espacio-tiempo el de Robertson-Walker. El estudio de las ecuaciones de Friedmann modificadas modeladas como un sistema dinámico autónomo, permite encontrar criterios para obtener modelos $f(R)$ cosmológicamente viables, y obteniendo además criterios para las funciones que conecten una época de dominio de materia, con una de expansión acelerada. \\\\
El estudio de la Ecuación de Desvío Geodésico EDG en las teorías $f(R)$ nos permitió tratar el problema de la medición de distancias, realizado en el Capítulo \ref{Capitulo6}. Obtuvimos a partir de la EDG la generalización para la ecuación de Pirani, y en el caso de observadores fundamentales las ecuaciones de Friedmann modificadas escritas en la forma estándar. El caso de la EDG para campos vectoriales nulos permitió encontrar expresiones generales para el enfocamiento de Ricci, la relación de Mattig, y la ecuación general que determina la distancia diametral angular $d_A$ en teorías $f(R)$, lo que permite determinar las distancias cosmológicas (en particular la distancia de luminosidad $d_L$ y la distancia comóvil $\chi(z)$). Finalmente obtuvimos una expresión para la ecuación de tipo Dyer-Roeder en teorías $f(R)$, abriendo la posibilidad del estudio de distancias cosmológicas en universos con inhomogeneidades.\\\\
Si bien las teorías $f(R)$ sufren algunos problemas en cuanto a su viabilidad en pruebas del sistema solar \cite{Berry}, tiene pruebas a grandes escalas que las hacen ver como una posible explicación a la actual expansión acelerada. Últimamente se han hecho extensiones de trabajos en RG a teorías $f(R)$ (desde lentes gravitacionales \cite{Ruggeiro}, agujeros negros \cite{Cruz}, estabilidad para formación estelar \cite{Upadye}, límites Newtonianos \cite{Stabile}, etc.), lo que permite encontrar varias formas para la prueba observacional de las teorías.

\newpage

{\color{white} . } 
\appendix
\chapter{Evaluación del término $g^{\alpha\beta}(\delta\Gamma_{\beta\alpha}^{\sigma}) - g^{\alpha\sigma}
(\delta\Gamma_{\alpha\gamma}^{\gamma})$}\label{Appendix A}
\noindent Hemos calculado ya la variación $\delta \Gamma_{\beta\alpha}^{\sigma}$
\begin{equation}
\delta \Gamma_{\beta\alpha}^{\sigma} = \frac{1}{2}\delta g^{\sigma\gamma}\bigl[\partial_{\beta}g_{\gamma\alpha} + \partial_{\alpha}g_{\gamma\beta} - \partial_{\gamma}g_{\beta\alpha}\bigr] + \frac{1}{2}g^{\sigma\gamma}\bigl[\partial_{\beta}(\delta g_{\gamma\alpha}) + \partial_{\alpha}(\delta g_{\gamma\beta}) -  \partial_{\gamma}(\delta g_{\beta\alpha})\bigr],
\end{equation}
escribiendo las derivadas parciales de las variaciones en la métrica con la expresión para la derivada covariante
\begin{equation}
\nabla_{\gamma}\delta g_{\alpha\beta} = \partial_{\gamma}\delta g_{\alpha\beta} - \Gamma_{\gamma\alpha}^{\sigma}\delta g_{\sigma\beta} - \Gamma_{\gamma\beta}^{\sigma}\delta g_{\alpha\sigma},
\end{equation}
y usando también que estamos en una variedad libre de torsión, i.e., que el símbolo de Christoffel es simétrico $\Gamma_{\beta\gamma}^{\alpha}=\Gamma_{\gamma\beta}^{\alpha}$, podemos escribir
\begin{align}
\delta \Gamma_{\beta\alpha}^{\sigma} &=  \frac{1}{2}\delta g^{\sigma\gamma}\bigl[\partial_{\beta}g_{\gamma\alpha} + \partial_{\alpha}g_{\gamma\beta} - \partial_{\gamma}g_{\beta\alpha}\bigr] \nonumber\\
& \hspace{4cm} + \frac{1}{2}g^{\sigma\gamma}\bigl[\nabla_{\beta}(\delta g_{\gamma\alpha})+ \nabla_{\alpha}(\delta g_{\gamma\beta})-  \nabla_{\gamma}(\delta g_{\beta\alpha})  + \Gamma_{\beta\alpha}^{\lambda} \delta g_{\gamma\lambda} + \Gamma_{\alpha\beta}^{\lambda} \delta g_{\lambda\gamma} \bigr],\nonumber \\
&=  \frac{1}{2}\delta g^{\sigma\gamma}\bigl[\partial_{\beta}g_{\gamma\alpha} + \partial_{\alpha}g_{\gamma\beta} - \partial_{\gamma}g_{\beta\alpha}\bigr] + g^{\sigma\gamma}\Gamma_{\beta\alpha}^{\lambda}\delta g_{\gamma\lambda}+ \frac{1}{2}g^{\sigma\gamma}\bigl[\nabla_{\beta}(\delta g_{\gamma\alpha})+ \nabla_{\alpha}(\delta g_{\gamma\beta})-  \nabla_{\gamma}(\delta g_{\beta\alpha}) \bigr],
\end{align}
usando la ecuación (\ref{varmet1}) en el segundo término
\begin{align}
\delta \Gamma_{\beta\alpha}^{\sigma} &=  \frac{1}{2}\delta g^{\sigma\gamma}\bigl[\partial_{\beta}g_{\gamma\alpha} + \partial_{\alpha}g_{\gamma\beta} - \partial_{\gamma}g_{\beta\alpha}\bigr] - \delta g^{\mu\nu}g^{\sigma\gamma}g_{\gamma\mu}g_{\lambda\nu}\Gamma_{\beta\alpha}^{\lambda} \nonumber \\
& \hspace{6cm}+  \frac{1}{2}g^{\sigma\gamma}\bigl[\nabla_{\beta}(\delta g_{\gamma\alpha})+ \nabla_{\alpha}(\delta g_{\gamma\beta}) -  \nabla_{\gamma}(\delta g_{\beta\alpha}) \bigr],\nonumber \\
&=  \delta g^{\sigma\nu}g_{\lambda\nu}\Gamma_{\beta\alpha}^{\lambda} - \delta g^{\mu\nu}\delta_{\mu}^{\sigma}g_{\lambda\nu}\Gamma_{\beta\alpha}^{\lambda} + \frac{1}{2}g^{\sigma\gamma}\bigl[\nabla_{\beta}(\delta g_{\gamma\alpha})+ \nabla_{\alpha}(\delta g_{\gamma\beta})-  \nabla_{\gamma}(\delta g_{\beta\alpha}) \bigr],\nonumber \\
&=  \delta g^{\sigma\nu}g_{\lambda\nu}\Gamma_{\beta\alpha}^{\lambda} - \delta g^{\sigma\nu}g_{\lambda\nu}\Gamma_{\beta\alpha}^{\lambda} + \frac{1}{2}g^{\sigma\gamma}\bigl[\nabla_{\beta}(\delta g_{\gamma\alpha})+ \nabla_{\alpha}(\delta g_{\gamma\beta})-  \nabla_{\gamma}(\delta g_{\beta\alpha}) \bigr].
\end{align}
Entonces tenemos
\begin{equation}
\delta \Gamma_{\beta\alpha}^{\sigma} = \frac{1}{2}g^{\sigma\gamma}\bigl[\nabla_{\beta}(\delta g_{\alpha\gamma}) + \nabla_{\alpha}(\delta g_{\beta\gamma}) - \nabla_{\gamma}(\delta g_{\beta\alpha})\bigr],
\end{equation}
y similarmente
\begin{equation}
\delta \Gamma_{\alpha\gamma}^{\gamma} = \frac{1}{2}g^{\sigma\gamma}\bigl[\nabla_{\alpha}(\delta g_{\sigma\gamma})\bigr].
\end{equation}
Sin embargo, es conveniente expresar el resultado previo en función de las variaciones  $\delta g^{\alpha\beta}$, de nuevo usando (\ref{varmet1})
\begin{align}
\delta \Gamma_{\beta\alpha}^{\sigma} &= \frac{1}{2}g^{\sigma\gamma}\bigl[\nabla_{\beta}(- g_{\alpha\mu}g_{\gamma\nu}\delta g^{\mu\nu}) + \nabla_{\alpha}(- g_{\beta\mu}g_{\gamma\nu}\delta g^{\mu\nu}) - \nabla_{\gamma}(- g_{\beta\mu}g_{\alpha\nu}\delta g^{\mu\nu})\bigr],\nonumber \\
&= -\frac{1}{2}g^{\sigma\gamma}\bigl[g_{\alpha\mu}g_{\gamma\nu}\nabla_{\beta}(\delta g^{\mu\nu}) + g_{\beta\mu}g_{\gamma\nu}\nabla_{\alpha}(\delta g^{\mu\nu}) - g_{\beta\mu}g_{\alpha\nu}\nabla_{\gamma}(\delta g^{\mu\nu})\bigr],\nonumber \\
&= -\frac{1}{2}\bigl[\delta_{\nu}^{\sigma}g_{\alpha\mu}\nabla_{\beta}(\delta g^{\mu\nu}) + \delta_{\nu}^{\sigma}g_{\beta\mu}\nabla_{\alpha}(\delta g^{\mu\nu}) - g_{\beta\mu}g_{\alpha\nu}g^{\gamma\sigma}\nabla_{\gamma}(\delta g^{\mu\nu})\bigr],\nonumber \\
&= -\frac{1}{2}\bigl[g_{\alpha\gamma}\nabla_{\beta}(\delta g^{\sigma\gamma}) + g_{\beta\gamma}\nabla_{\alpha}(\delta g^{\sigma\gamma}) - g_{\beta\mu}g_{\alpha\nu}\nabla^{\sigma}(\delta g^{\mu\nu})\bigr],
\end{align}
donde usamos $\nabla^{\sigma} = g^{\sigma\gamma}\nabla_{\gamma}$. De forma análoga
\begin{equation}
\delta \Gamma_{\alpha\gamma}^{\gamma} = -\frac{1}{2}g_{\mu\nu}\nabla_{\alpha}(\delta g^{\mu\nu}).
\end{equation}
Calculamos ahora el término $g^{\alpha\beta}(\delta\Gamma_{\beta\alpha}^{\sigma}) - g^{\alpha\sigma}
(\delta\Gamma_{\alpha\gamma}^{\gamma})$
\begin{align}
g^{\alpha\beta}(\delta\Gamma_{\beta\alpha}^{\sigma}) - g^{\alpha\sigma}
(\delta\Gamma_{\alpha\gamma}^{\gamma}) =& -\frac{1}{2}\Bigl(\bigl[g^{\alpha\beta}g_{\alpha\gamma}\nabla_{\beta}(\delta g^{\sigma\gamma}) + g^{\alpha\beta}g_{\beta\gamma}\nabla_{\alpha}(\delta g^{\sigma\gamma}) - g^{\alpha\beta}g_{\beta\mu}g_{\alpha\nu}\nabla^{\sigma}(\delta g^{\mu\nu})\bigr] \nonumber \\
& \hspace{7cm}-  \bigl[g^{\alpha\sigma}g_{\mu\nu}\nabla_{\alpha}(\delta g^{\mu\nu})\bigr]\Bigr),\nonumber \\
=& -\frac{1}{2}\Bigl(\bigl[\delta_{\gamma}^{\beta}\nabla_{\beta}(\delta g^{\sigma\gamma}) + \delta_{\gamma}^{\alpha}\nabla_{\alpha}(\delta g^{\sigma\gamma}) - \delta_{\mu}^{\alpha}g_{\alpha\nu}\nabla^{\sigma}(\delta g^{\mu\nu})\bigr] \nonumber \\
& \hspace{7cm} - \bigl[g_{\mu\nu}g^{\alpha\sigma}\nabla_{\alpha}(\delta g^{\mu\nu})\bigr]\Bigr),\nonumber \\
=& -\frac{1}{2}\Bigl(\bigl[\nabla_{\gamma}(\delta g^{\sigma\gamma}) + \nabla_{\gamma}(\delta g^{\sigma\gamma}) - g_{\mu\nu}\nabla^{\sigma}(\delta g^{\mu\nu})\bigr]
- \bigl[g_{\mu\nu}\nabla^{\sigma}(\delta g^{\mu\nu})\bigr]\Bigr),\nonumber \\
=& -\frac{1}{2}\Bigl(2\nabla_{\gamma}(\delta g^{\sigma\gamma}) - 2g_{\mu\nu}\nabla^{\sigma}(\delta g^{\mu\nu})\Bigr),
\end{align}
tenemos entonces,
\begin{equation}\label{vargamma}
g^{\alpha\beta}(\delta\Gamma_{\beta\alpha}^{\sigma}) - g^{\alpha\sigma}
(\delta\Gamma_{\alpha\gamma}^{\gamma}) = g_{\mu\nu}\nabla^{\sigma}(\delta g^{\mu\nu})-\nabla_{\gamma}(\delta g^{\sigma\gamma}).
\end{equation}

\newpage

{\color{white} . }

\chapter{Integrales con $M_{\tau}$ y $N^{\sigma}$ }\label{Appendix B}
\noindent Tomando la derivada covariante en $M_{\tau}$
\begin{align}
\nabla^{\tau}M_{\tau} =& \nabla^{\tau}\bigl(f'(R)g_{\alpha\beta}\nabla_{\tau}(\delta g^{\alpha\beta})\bigr) - \nabla^{\tau}\bigl(\delta g^{\alpha\beta} g_{\alpha\beta}\nabla_{\tau}(f'(R))\bigr),\nonumber \\
=& \nabla^{\tau}(f'(R))g_{\alpha\beta}\nabla_{\tau}(\delta g^{\alpha\beta}) + f'(R)g_{\alpha\beta}\square(\delta g^{\alpha\beta})  - \nabla^{\tau}(\delta g^{\alpha\beta}) g_{\alpha\beta}\nabla_{\tau}(f'(R))\\
&-\delta g^{\alpha\beta} g_{\alpha\beta}\square(f'(R)),\nonumber \\
=& f'(R)g_{\alpha\beta}\square(\delta g^{\alpha\beta})-\delta g^{\alpha\beta} g_{\alpha\beta}\square(f'(R)).
\end{align}
en donde hemos usado la compatibilidad métrica $\nabla^{\tau}g_{\alpha\beta}=0$, integrando esta expresión
\begin{equation}
\int_{\mathcal{V}} d^4x \, \sqrt{-g}\nabla^{\tau}M_{\tau} = \int_{\mathcal{V}} d^4x \, \sqrt{-g}f'(R)g_{\alpha\beta}\square(\delta g^{\alpha\beta})-\int_{\mathcal{V}} d^4x \, \sqrt{-g}\delta g^{\alpha\beta} g_{\alpha\beta}\square(f'(R)),
\end{equation}
usando el teorema de Gauss-Stokes (\ref{gauss}), la primera integral puede escribirse como un término de frontera
\begin{equation}
\oint_{\partial \mathcal{V}} d^{3}y\, \varepsilon\sqrt{|h|}n^{\tau}M_{\tau} = \int_{\mathcal{V}} d^4x \, \sqrt{-g}f'(R)g_{\alpha\beta}\square(\delta g^{\alpha\beta})\bigr)-\int_{\mathcal{V}} d^4x \, \sqrt{-g}\delta g^{\alpha\beta} g_{\alpha\beta}\square(f'(R)),
\end{equation}
podemos escribir entonces
\begin{equation}
\int_{\mathcal{V}} d^4x \, \sqrt{-g}f'(R)g_{\alpha\beta}\square(\delta g^{\alpha\beta})=\int_{\mathcal{V}} d^4x \, \sqrt{-g}\delta g^{\alpha\beta} g_{\alpha\beta}\square(f'(R)) + \oint_{\partial \mathcal{V}} d^{3}y\, \varepsilon\sqrt{|h|}n^{\tau}M_{\tau}.
\end{equation}
De forma similar, tomando la derivada covariante en  $N^{\sigma}$
\begin{align}
\nabla_{\sigma}N^{\sigma} &= \nabla_{\sigma}\bigl(f'(R)\nabla_{\gamma}(\delta g^{\sigma\gamma})\bigr) - \nabla_{\sigma}\bigl(\delta g^{\sigma\gamma}\nabla_{\gamma}(f'(R))\bigr),\nonumber \\
&= \nabla_{\sigma}(f'(R))\nabla_{\gamma}(\delta g^{\sigma\gamma})+ f'(R)\nabla_{\sigma}\nabla_{\gamma}(\delta g^{\sigma\gamma}) - \nabla_{\sigma}(\delta g^{\sigma\gamma})\nabla_{\gamma}(f'(R))-\delta g^{\sigma\gamma}\nabla_{\sigma}\nabla_{\gamma}(f'(R)),\nonumber \\
&= f'(R)\nabla_{\sigma}\nabla_{\beta}(\delta g^{\sigma\beta}) -\delta g^{\sigma\beta}\nabla_{\sigma}\nabla_{\beta}(f'(R)),
\end{align}
integrando
\begin{equation}
\int_{\mathcal{V}} d^4x \, \sqrt{-g}\nabla_{\sigma}N^{\sigma} = \int_{\mathcal{V}} d^4x \, \sqrt{-g}f'(R)\nabla_{\sigma}\nabla_{\beta}(\delta g^{\sigma\beta}) -\int_{\mathcal{V}} d^4x \, \sqrt{-g}\delta g^{\sigma\beta}\nabla_{\sigma}\nabla_{\beta}(f'(R)),
\end{equation}
usando de nuevo el teorema de Gauss-Stokes podemos escribir
\begin{equation}
\int_{\mathcal{V}} d^4x \, \sqrt{-g}f'(R)\nabla_{\sigma}\nabla_{\beta}(\delta g^{\sigma\beta}) =\int_{\mathcal{V}} d^4x \, \sqrt{-g}\delta g^{\sigma\beta}\nabla_{\sigma}\nabla_{\beta}(f'(R))+\oint_{\partial \mathcal{V}} d^{3}y\, \varepsilon\sqrt{|h|}n_{\sigma}N^{\sigma}.
\end{equation} 

\newpage

{\color{white} . } 
\chapter{Ecuaciones de Friedmann Modificadas}\label{AppendixC}
\noindent Primero escribimos los símbolos de Christoffel no nulos para la métrica (\ref{robw})
\begin{equation}\label{gamma1}
\Gamma_{ij}^0 = a\dot{a}\; \tilde{g}_{ij},
\end{equation}
donde definimos la parte espacial de la métrica por
\begin{equation}\label{gamma2}
\tilde{g}_{ij} = \diag\biggl(\dfrac{1}{1-kr^2}, r^2, r^2\sin^2\theta\biggr),
\end{equation}
\begin{equation}\label{gamma3}
\Gamma_{0j}^{i} = \frac{\dot{a}}{a}\delta_{ij},
\end{equation}
Aunque hay más símbolos de Christoffel no nulos, una propiedad de los espacios maximalmente simétricos nos evitará ese cálculo. Usemos ahora
la definición para las componentes del tensor de Ricci
\begin{equation}
R_{\alpha\beta} = \partial_{\sigma}\Gamma_{\beta\alpha}^{\sigma}-\partial_{\beta} \Gamma_{\sigma\alpha}^{\sigma} +\Gamma_{\sigma\mu}^{\sigma}\Gamma_{\beta\alpha}^{\mu}- \Gamma_{\mu\beta}^{\sigma}\Gamma_{\sigma\alpha}^{\mu},
\end{equation}
con lo cual
\begin{equation}
R_{00} =-3\frac{\ddot{a}}{a}, \qquad R_{0j} =0, \qquad R_{ij}=\bigl(a\ddot{a} + 2\dot{a}^2 + 2k\bigr)\tilde{g}_{ij},
\end{equation}
La curvatura total está dada por
\begin{equation}
R = g^{00}R_{00} + \frac{1}{a^2}\, ^{(3)}R = -3\frac{\ddot{a}}{a}(-1)  +\frac{1}{a^2}\bigl(3a\ddot{a} + 6\dot{a}^2 + 6k\bigr) = 6\biggl[\frac{\ddot{a}}{a}+ \biggl(\frac{\dot{a}}{a}\biggr)^2 + \frac{k}{a^2}\biggr].
\end{equation}
Primero vamos a evaluar los términos $\nabla_{\alpha}\nabla_{\beta}f'(R)$ y $\square f'(R)$
\begin{align}\label{covar1}
(\nabla_{\alpha}\nabla_{\beta})f'(R) &= \nabla_{\alpha}\bigl(\partial_{\beta} f'(R)\bigr),\notag \\
&= \partial_{\alpha}\partial_{\beta} f'(R) - \Gamma_{\alpha\beta}^{\gamma}\partial_{\gamma} f'(R),
\end{align}
y
\begin{align}
\square f'(R) &= g^{\rho\sigma} (\nabla_{\rho}\nabla_{\sigma})f'(R), \notag \\
&= g^{\rho\sigma}\bigl(\partial_{\rho}\partial_{\sigma}f'(R)- \Gamma_{\rho\sigma}^{\gamma}\partial_{\gamma} f'(R)\bigr), \notag \\
&= g^{\rho\sigma}\partial_{\rho}\partial_{\sigma} f'(R)- g^{\rho\sigma}\Gamma_{\rho\sigma}^{\gamma}\partial_{\gamma} f'(R).
\end{align}
Entonces la componente temporal-temporal de las ecuaciones de campo (\ref{ecuacionescampo}) es
\begin{equation}
f'(R)R_{00} -\frac{1}{2}f(R) g_{00} -[\nabla_0\nabla_0-g_{00}\square]f'(R) = \kappa T_{00},
\end{equation}
\begin{equation}
-3\frac{\ddot{a}}{a}f'(R) +\frac{f(R)}{2} -[\nabla_0\nabla_0+\square]f'(R) = \kappa \rho,
\end{equation}
el término $\nabla_0\nabla_0$ es simplemente
\begin{equation}
(\nabla_0\nabla_0)f'(R)=\partial^2_{0} f'(R),
\end{equation}
y para el término $\square f'(R)$ tendremos
\begin{align}
\square f'(R)  &= g^{00}\partial_0^2 f'(R) + g^{ij}\partial_{i}\partial_{j}f'(R) - g^{11}\Gamma_{11}^0 \partial_0 f'(R) -  g^{22}\Gamma_{22}^0\partial_0 f'(R) -  g^{33}\Gamma_{33}^0 \partial_0 f'(R),\notag\\
&= -\partial_0^2 f'(R) + \cancelto{0}{\tilde{g}^{ij}\partial_{i}\partial_{j}f'(R)} - \frac{1-kr^2}{a^2}(a\dot{a}) \frac{1}{1-kr^2}\partial_0 f'(R)\\
& \hspace{4cm} -  \frac{1}{a^2 r^2}(a\dot{a})r^2\partial_0 f'(R) -  \frac{1}{r^2\sin^2\theta}(a\dot{a})r^2\sin^2\theta \partial_0 f'(R),\notag \\
&= -\partial_0^2 f'(R) - 3\frac{\dot{a}}{a}\partial_0 f'(R),
\end{align}
en donde usamos que el escalar $R$, NO depende de las coordenadas espaciales, de modo que cualquier derivada con respecto a las coordenadas es nula. Así tendremos que
\begin{align}
[\nabla_0\nabla_0+\square]f'(R) &= -3\frac{\dot{a}}{a}\partial_0 f'(R), \notag\\
[\nabla_0\nabla_0+\square]f'(R) &= -3\frac{\dot{a}}{a}\partial_R f'(R)\partial_0 R, \notag\\
[\nabla_0\nabla_0+\square]f'(R) &= -3\frac{\dot{a}}{a}f''(R) \dot{R},
\end{align}
con $\dot{R} =\partial_0 R$ y $f''(R) = \partial_R f'(R)$, con esto la ecuación de campo nos queda
\begin{equation}
-3\frac{\ddot{a}}{a} +\frac{f(R)}{2f'(R)} +3\frac{\dot{a}}{a}\frac{f''(R) \dot{R}}{f'(R)} = \frac{\kappa \rho}{f'(R)},
\end{equation}
sumando y restando el término
\begin{equation}
3\biggl[\biggl(\frac{\dot{a}}{a}\biggr)^2 + \frac{k}{a^2}\biggr],
\end{equation}
tendremos
\begin{equation}
-\frac{R}{2}+3\biggl(\frac{\dot{a}}{a}\biggr)^2 + 3 \frac{k}{a^2} +\frac{f(R)}{2f'(R)} +3\frac{\dot{a}}{a}\frac{f''(R) \dot{R}}{f'(R)} = \frac{\kappa \rho}{f'(R)},
\end{equation}
obtenemos finalmente que
\begin{equation}
H^2 + \frac{k}{a^2}= \frac{1}{3 f'(R)}\biggl[\kappa\rho + \frac{(R f'(R)-f(R))}{2} - 3Hf''(R)\dot{R}\biggr],
\end{equation}
en donde usamos que $H = \frac{\dot{a}}{a}$ y la expresión para el escalar de curvatura $R$.
Para la componente espacial $ij$ tendremos la ecuación de campo
\begin{equation}
f'(R)R_{ij} -\frac{1}{2}f(R) g_{ij} -[\nabla_i\nabla_j-g_{jj}\square]f'(R) = \kappa T_{ij},
\end{equation}
\begin{equation}
\bigl(a\ddot{a} + 2\dot{a}^2 + 2k\bigr)\tilde{g}_{ij}f'(R) -\frac{1}{2}f(R) g_{ij}-\bigl[\nabla_i\nabla_j-g_{ij}\square\bigr]f'(R) = \kappa g_{ij}T_i^i,
\end{equation}
de la expresión (\ref{covar1}) el término $\nabla_i\nabla_j$ es
\begin{equation}
(\nabla_i\nabla_j)f'(R)= -a\dot{a}\tilde{g}_{ij}\partial_t f'(R),
\end{equation}
donde de nuevo usamos que las derivadas espaciales de $f'(R)$ son nulas. Para el término $\square f'(R)$ tendremos nuevamente
\begin{equation}
\square f'(R)  = -\partial_0^2 f'(R) - 3\frac{\dot{a}}{a}\partial_0 f'(R).
\end{equation}
Así tendremos que
\begin{align}
[\nabla_i\nabla_j-g_{ij}\square]f'(R) &=  -a\dot{a}\tilde{g}_{ij}\partial_0 f'(R) + g_{ij}\partial_0^2 f'(R) + 3g_{ij}\frac{\dot{a}}{a}\partial_0 f'(R),\notag \\
[\nabla_i\nabla_j-g_{ij}\square]f'(R) &= 2a\dot{a}\tilde{g}_{ij}\partial_0 f'(R) + a^2 \tilde{g}_{ij}\partial_0^2 f'(R), \notag \\
[\nabla_i\nabla_j-g_{ij}\square]f'(R) &= 2a\dot{a}\tilde{g}_{ij}f''(R)\dot{R} + a^2 \tilde{g}_{ij}(f'''(R)(\dot{R})^2+f''(R)\ddot{R}),
\end{align}
de modo que la ecuación de campo nos queda
\begin{equation}
\bigl(a\ddot{a} + 2\dot{a}^2 + 2k\bigr)\tilde{g}_{ij}f'(R) -\frac{1}{2}a^2 \tilde{g}_{ij}f(R) - 2a\dot{a}\tilde{g}_{ij}f'(R)\dot{R} - a^2 \tilde{g}_{ij}\bigl(f'''(R)(\dot{R})^2+f''(R)\ddot{R}\bigr) = \kappa a^2 \tilde{g}_{ij}T_i^i,
\end{equation}
\begin{equation}
\biggl[\frac{\ddot{a}}{a}+ 2\biggl(\frac{\dot{a}}{a}\biggr)^2+\frac{2k}{a^2}\biggr] -\frac{f(R)}{2f'(R)} - 2\frac{\dot{a}}{a}\frac{f''(R)\dot{R}}{f'(R)} - \frac{f'''(R)(\dot{R})^2}{f'(R)}-\frac{f''(R)\ddot{R}}{f'(R)} = \frac{\kappa p}{f'(R)},
\end{equation}
sumando y restando el término
\begin{equation}
2\biggl(\frac{\ddot{a}}{a}\biggr) + \biggl(\frac{\dot{a}}{a}\biggr)^2 + \frac{k}{a^2},
\end{equation}
tendremos finalmente
\begin{equation}
2\dot{H} + 3H^2 + \frac{k}{a^2} = -\frac{1}{f'(R)}\biggl[\kappa p+2H\dot{R}f''(R) +\frac{(f(R) - Rf'(R))}{2}  + \ddot{R}f''(R)  + (\dot{R})^2f'''(R)\biggl],
\end{equation}
donde usamos de nuevo la expresión para $H$.

\newpage

{\color{white} . }

\chapter{Contribución de los operadores $\mathcal{D}_{\alpha\beta}$ en la EDG}\label{AppendixE}
\noindent Vamos a considerar la contribución
\begin{equation}\label{contribucion}
\bigl(\delta_{\gamma}^{\alpha}\mathcal{D}_{\delta\beta} - \delta_{\delta}^{\alpha}\mathcal{D}_{\gamma\beta} + g_{\beta\delta}\mathcal{D}_{\gamma}^{\, \, \alpha}
- g_{\beta\gamma}\mathcal{D}_{\delta}^{\, \,  \alpha}\bigr)f'(R)V^{\beta}V^{\delta}\eta^{\gamma},
\end{equation}
usando para esto lo resultados para los operadores $\mathcal{D}_{\alpha\beta}$ usando como espacio-tiempo el de Robertson-Walker
\begin{align}\label{square}
\square f'(R) &=  - \partial_{0}^2 f'(R) - 3H \partial_{0} f'(R),\notag \\
&=  - f''(R) \ddot{R} - f'''(R) \dot{R}^2 - 3H f''(R) \dot{R},
\end{align}
\begin{align}\label{00}
\mathcal{D}_{00} &= -3 H \partial_{0} f'(R),\notag \\
& = -3 H f''(R) \dot{R},
\end{align}
\begin{align}\label{ij}
\mathcal{D}_{ij} &= 2 H g_{ij}\partial_{0} f'(R) + g_{ij}\partial_{0}^2 f'(R), \notag \\
& = 2 H g_{ij}f''(R) \dot{R} + g_{ij}(f''(R) \ddot{R} + f'''(R) \dot{R}^2).
\end{align}
De estas expresiones es claro que los operadores mixtos $\mathcal{D}_{0i}=\mathcal{D}_{i0}$ son nulos, y además lo son también contribuciones de la forma $\mathcal{D}_{0}^{\, i}$ y $\mathcal{D}_{i}^{\, 0}$. Reescribamos la expresión (\ref{contribucion}) en la siguiente forma
\begin{equation}\label{contribucion1}
\bigl(\delta_{\gamma}^{\alpha}\mathcal{D}_{\delta\beta}+ g_{\beta\delta}\mathcal{D}_{\gamma}^{\, \, \alpha}\bigr)f'(R)V^{\beta}V^{\delta}\eta^{\gamma} - \bigl(\delta_{\delta}^{\alpha}\mathcal{D}_{\gamma\beta}
+ g_{\beta\gamma}\mathcal{D}_{\delta}^{\, \,  \alpha}\bigr)f'(R)V^{\beta}V^{\delta}\eta^{\gamma}.
\end{equation}
Un método simple es fijar $\alpha$ para las componentes temporales y espaciales, y determinar la contribución de cada operador. Consideremos entonces cada uno de los dos casos.\\\\\\
$\bullet$ \, Contribución a $\alpha=0$\\\\
Con $\alpha=0$ la expresión (\ref{contribucion1}) se reduce a
\begin{equation}
\bigl(\delta_{\gamma}^{0}\mathcal{D}_{\delta\beta}+ g_{\beta\delta}\mathcal{D}_{\gamma}^{\, \, 0}\bigr)f'(R)V^{\beta}V^{\delta}\eta^{\gamma} - \bigl(\delta_{\delta}^{0}\mathcal{D}_{\gamma\beta}
+ g_{\beta\gamma}\mathcal{D}_{\delta}^{\, \,  0}\bigr)f'(R)V^{\beta}V^{\delta}\eta^{\gamma},
\end{equation}
puesto que las deltas de Kronecker $\delta_{\gamma}^{0}$ y los operadores mixtos ${D}_{\gamma}^{\, \, 0}$ se hacen cero en ambos paréntesis para  $\gamma = \delta = i$, con $i$ representando las componentes espaciales, la relación anterior se reduce a
\begin{equation}
\bigl(\mathcal{D}_{\delta\beta}+ g_{\beta\delta}\mathcal{D}_{0}^{\, \, 0}\bigr)f'(R)V^{\beta}V^{\delta}\eta^{0} - \bigl(\mathcal{D}_{\gamma\beta}
+ g_{\beta\gamma}\mathcal{D}_{0}^{\, \,  0}\bigr)f'(R)V^{\beta}V^{0}\eta^{\gamma},
\end{equation}
expandiendo explícitamente las sumas, de nuevo recordando que solo los operadores $\mathcal{D}_{00}$ y $\mathcal{D}_{ij}$ son diferentes de cero tendremos
\begin{multline}
\bigl(\mathcal{D}_{\delta\beta}+ g_{\beta\delta}\mathcal{D}_{0}^{\, \, 0}\bigr)f'(R)V^{\beta}V^{\delta}\eta^{0} - \bigl(\mathcal{D}_{\gamma\beta}
+ g_{\beta\gamma}\mathcal{D}_{0}^{\, \,  0}\bigr)f'(R)V^{\beta}V^{0}\eta^{\gamma} = \\\bigl(\mathcal{D}_{00}+ g_{00}\mathcal{D}_{0}^{\, \, 0}\bigr)f'(R)V^{0}V^{0}\eta^{0} - \bigl(\mathcal{D}_{00}
+ g_{00}\mathcal{D}_{0}^{\, \,  0}\bigr)f'(R)V^{0}V^{0}\eta^{0} + \bigl(\mathcal{D}_{ij}+ g_{ij}\mathcal{D}_{0}^{\, \, 0}\bigr)f'(R)V^{i}V^{j}\eta^{0}\\ - \bigl(\mathcal{D}_{ij} + g_{ij}\mathcal{D}_{0}^{\, \,  0}\bigr)f'(R)V^{i}V^{0}\eta^{j},
\end{multline}
\begin{multline}
\bigl(\mathcal{D}_{\delta\beta}+ g_{\beta\delta}\mathcal{D}_{0}^{\, \, 0}\bigr)f'(R)V^{\beta}V^{\delta}\eta^{0} - \bigl(\mathcal{D}_{\gamma\beta}
+ g_{\beta\gamma}\mathcal{D}_{0}^{\, \,  0}\bigr)f'(R)V^{\beta}V^{0}\eta^{\gamma} = \\ \bigl(\mathcal{D}_{ij}+ g_{ij}\mathcal{D}_{0}^{\, \, 0}\bigr)f'(R)V^{i}V^{j}\eta^{0} - \bigl(\mathcal{D}_{ij} + g_{ij}\mathcal{D}_{0}^{\, \,  0}\bigr)f'(R)V^{i}V^{0}\eta^{j},
\end{multline}
escribiendo la forma explícita de los operadores (\ref{00}) y (\ref{ij}), con $\mathcal{D}_0^{\, \, 0} = g^{00}\mathcal{D}_{00}=-\mathcal{D}_{00}$ tendremos
\begin{multline}
\bigl(\mathcal{D}_{\delta\beta}+ g_{\beta\delta}\mathcal{D}_{0}^{\, \, 0}\bigr)f'(R)V^{\beta}V^{\delta}\eta^{0} - \bigl(\mathcal{D}_{\gamma\beta}
+ g_{\beta\gamma}\mathcal{D}_{0}^{\, \,  0}\bigr)f'(R)V^{\beta}V^{0}\eta^{\gamma} = \\ \bigl(2 H g_{ij}f''(R) \dot{R} + g_{ij}(f''(R) \ddot{R} + f'''(R) \dot{R}^2)+ +3g_{ij}H f''(R) \dot{R}\bigr)f'(R)V^{i}V^{j}\eta^{0}\\ - \bigl(2 H g_{ij}f''(R) \dot{R} + g_{ij}(f''(R) \ddot{R} + f'''(R) \dot{R}^2) + 3g_{ij}H f''(R) \dot{R}\bigr)f'(R)V^{i}V^{0}\eta^{j},
\end{multline}
\begin{multline}
\bigl(\mathcal{D}_{\delta\beta}+ g_{\beta\delta}\mathcal{D}_{0}^{\, \, 0}\bigr)f'(R)V^{\beta}V^{\delta}\eta^{0} - \bigl(\mathcal{D}_{\gamma\beta}
+ g_{\beta\gamma}\mathcal{D}_{0}^{\, \,  0}\bigr)f'(R)V^{\beta}V^{0}\eta^{\gamma} = \\ g_{ij}\bigl(5H f''(R) \dot{R} + (f''(R) \ddot{R} + f'''(R) \dot{R}^2)+ \bigr)f'(R)V^{i}V^{j}\eta^{0}\\ - g_{ij}\bigl(5H f''(R) \dot{R} + (f''(R) \ddot{R} + f'''(R) \dot{R}^2)\bigr)f'(R)V^{i}V^{0}\eta^{j}.
\end{multline}
Ahora, escribimos $\eta_{\alpha}V^{\alpha}=0$
\begin{equation}
\eta_{\alpha}V^{\alpha} = g_{00}\eta^{0}V^0 + g_{ij}\eta^{i}V^{j} = 0, \qquad \Longrightarrow \qquad g_{00}\eta^{0}V^0 = - g_{ij}\eta^{i}V^{j},
\end{equation}
de modo que
\begin{multline}
\bigl(\mathcal{D}_{\delta\beta}+ g_{\beta\delta}\mathcal{D}_{0}^{\, \, 0}\bigr)f'(R)V^{\beta}V^{\delta}\eta^{0} - \bigl(\mathcal{D}_{\gamma\beta}
+ g_{\beta\gamma}\mathcal{D}_{0}^{\, \,  0}\bigr)f'(R)V^{\beta}V^{0}\eta^{\gamma} = \\ g_{ij}\bigl(5H f''(R) \dot{R} + (f''(R) \ddot{R} + f'''(R) \dot{R}^2)+ \bigr)f'(R)V^{i}V^{j}\eta^{0}\\ + g_{00}\bigl(5H f''(R) \dot{R} + (f''(R) \ddot{R} + f'''(R) \dot{R}^2)\bigr)f'(R)V^{0}V^{0}\eta^{0},
\end{multline}
\begin{multline}
\bigl(\mathcal{D}_{\delta\beta}+ g_{\beta\delta}\mathcal{D}_{0}^{\, \, 0}\bigr)f'(R)V^{\beta}V^{\delta}\eta^{0} - \bigl(\mathcal{D}_{\gamma\beta}
+ g_{\beta\gamma}\mathcal{D}_{0}^{\, \,  0}\bigr)f'(R)V^{\beta}V^{0}\eta^{\gamma} = \\ \bigl(5H f''(R) \dot{R} + (f''(R) \ddot{R} + f'''(R) \dot{R}^2)+ \bigr)f'(R)\bigl[g_{ij}V^{i}V^{j} + g_{00}V^{0}V^{0}\eta^{0}\bigl]\eta^{0}.
\end{multline}
Nuevamente la expresión $V_{\alpha}V^{\alpha}=\epsilon$ puede expandirse como
\begin{equation}
V_{\alpha}V^{\alpha} = g_{00}V^{0}V^0 + g_{ij}V^{i}V^{j} = \epsilon,
\end{equation}
de modo que el término entre paréntesis es simplemente $\epsilon$, y así la contribución para $\alpha=0$ es finalmente
\begin{multline}
\bigl(\delta_{\gamma}^{0}\mathcal{D}_{\delta\beta}+ g_{\beta\delta}\mathcal{D}_{\gamma}^{\, \, 0}\bigr)f'(R)V^{\beta}V^{\delta}\eta^{\gamma} - \bigl(\delta_{\delta}^{0}\mathcal{D}_{\gamma\beta}
+ g_{\beta\gamma}\mathcal{D}_{\delta}^{\, \,  0}\bigr)f'(R)V^{\beta}V^{\delta}\eta^{\gamma}\\ = \epsilon\bigl(5H f''(R) \dot{R} + (f''(R) \ddot{R} + f'''(R) \dot{R}^2)+ \bigr)f'(R)\eta^{0}.
\end{multline}\\
$\bullet$ \, Contribución a $\alpha=i$\\\\
Con $\alpha=i$ la expresión (\ref{contribucion1}) se reduce a
\begin{equation}
\bigl(\delta_{\gamma}^{i}\mathcal{D}_{\delta\beta}+ g_{\beta\delta}\mathcal{D}_{\gamma}^{\, \, i}\bigr)f'(R)V^{\beta}V^{\delta}\eta^{\gamma} - \bigl(\delta_{\delta}^{i}\mathcal{D}_{\gamma\beta}
+ g_{\beta\gamma}\mathcal{D}_{\delta}^{\, \,  i}\bigr)f'(R)V^{\beta}V^{\delta}\eta^{\gamma},
\end{equation}
de nuevo,  las deltas de Kronecker $\delta_{\gamma}^{i}$ y los operadores mixtos ${D}_{\gamma}^{\, \, i}$ se hacen cero en ambos paréntesis para  $\gamma = \delta = 0$, de modo que
\begin{equation}
\bigl(\mathcal{D}_{\delta\beta}+ g_{\beta\delta}\mathcal{D}_{i}^{\, \, i}\bigr)f'(R)V^{\beta}V^{\delta}\eta^{i} - \bigl(\mathcal{D}_{\gamma\beta}
+ g_{\beta\gamma}\mathcal{D}_{i}^{\, \,  i}\bigr)f'(R)V^{\beta}V^{i}\eta^{\gamma},
\end{equation}
aquí $\mathcal{D}_{i}^{\, \, i}$ es el operador para \textit{una sola componente}, no debe confundirse con una suma. Expandiendo explícitamente las sumas tendremos
\begin{multline}
\bigl(\mathcal{D}_{\delta\beta}+ g_{\beta\delta}\mathcal{D}_{i}^{\, \, i}\bigr)f'(R)V^{\beta}V^{\delta}\eta^{i} - \bigl(\mathcal{D}_{\gamma\beta}
+ g_{\beta\gamma}\mathcal{D}_{i}^{\, \,  i}\bigr)f'(R)V^{\beta}V^{i}\eta^{\gamma} = \\ \bigl(\mathcal{D}_{00}+ g_{00}\mathcal{D}_{i}^{\, \, i}\bigr)f'(R)V^{0}V^{0}\eta^{i} - \bigl(\mathcal{D}_{00}
+ g_{00}\mathcal{D}_{i}^{\, \,  i}\bigr)f'(R)V^{0}V^{i}\eta^{0} + \bigl(\mathcal{D}_{jk}+ g_{jk}\mathcal{D}_{i}^{\, \, i}\bigr)f'(R)V^{j}V^{k}\eta^{i}\\ - \bigl(\mathcal{D}_{jk} + g_{jk}\mathcal{D}_{i}^{\, \,  i}\bigr)f'(R)V^{j}V^{i}\eta^{k},
\end{multline}
\begin{multline}
\bigl(\mathcal{D}_{\delta\beta}+ g_{\beta\delta}\mathcal{D}_{i}^{\, \, i}\bigr)f'(R)V^{\beta}V^{\delta}\eta^{i} - \bigl(\mathcal{D}_{\gamma\beta}
+ g_{\beta\gamma}\mathcal{D}_{i}^{\, \,  i}\bigr)f'(R)V^{\beta}V^{i}\eta^{\gamma} = \\ \bigl(\mathcal{D}_{00}+ g_{00}\mathcal{D}_{i}^{\, \, i}\bigr)f'(R)V^{0}V^{0}\eta^{i} - \bigl(\mathcal{D}_{00}
+ g_{00}\mathcal{D}_{i}^{\, \,  i}\bigr)f'(R)V^{0}V^{i}\eta^{0},
\end{multline}
escribiendo la forma de los operadores (\ref{00}) y (\ref{ij}), con $\mathcal{D}_i^{\, \, i} = g^{ij}\mathcal{D}_{ij}$ tendremos
\begin{multline}
\bigl(\mathcal{D}_{\delta\beta}+ g_{\beta\delta}\mathcal{D}_{i}^{\, \, i}\bigr)f'(R)V^{\beta}V^{\delta}\eta^{i} - \bigl(\mathcal{D}_{\gamma\beta}
+ g_{\beta\gamma}\mathcal{D}_{i}^{\, \,  i}\bigr)f'(R)V^{\beta}V^{i}\eta^{\gamma} = \\ \bigl(-3 H f''(R) \dot{R} - g^{ij}\bigl[2 H g_{ij}f''(R) \dot{R} + g_{ij}(f''(R) \ddot{R} + f'''(R) \dot{R}^2)\bigr]\bigr)f'(R)V^{0}V^{0}\eta^{i}\\ - \bigl(3 H f''(R) \dot{R}
- g^{ij}\bigl[2 H g_{ij}f''(R) \dot{R} + g_{ij}(f''(R) \ddot{R} + f'''(R) \dot{R}^2)\bigr]\bigr)f'(R)V^{0}V^{i}\eta^{0},
\end{multline}
\begin{multline}
\bigl(\mathcal{D}_{\delta\beta}+ g_{\beta\delta}\mathcal{D}_{i}^{\, \, i}\bigr)f'(R)V^{\beta}V^{\delta}\eta^{i} - \bigl(\mathcal{D}_{\gamma\beta}
+ g_{\beta\gamma}\mathcal{D}_{i}^{\, \,  i}\bigr)f'(R)V^{\beta}V^{i}\eta^{\gamma} = \\ -\bigl(5 H f''(R) \dot{R} + (f''(R) \ddot{R} + f'''(R) \dot{R}^2)\bigr]\bigr)f'(R)V^{0}V^{0}\eta^{i}\\ + \bigl(5 H f''(R) \dot{R}
 + (f''(R) \ddot{R} + f'''(R) \dot{R}^2)\bigr]\bigr)f'(R)V^{0}V^{i}\eta^{0},
\end{multline}
\begin{multline}
\bigl(\mathcal{D}_{\delta\beta}+ g_{\beta\delta}\mathcal{D}_{i}^{\, \, i}\bigr)f'(R)V^{\beta}V^{\delta}\eta^{i} - \bigl(\mathcal{D}_{\gamma\beta}
+ g_{\beta\gamma}\mathcal{D}_{i}^{\, \,  i}\bigr)f'(R)V^{\beta}V^{i}\eta^{\gamma} = \\ g_{00}\bigl(5 H f''(R) \dot{R} + (f''(R) \ddot{R} + f'''(R) \dot{R}^2)\bigr]\bigr)f'(R)V^{0}V^{0}\eta^{i}\\ -g_{00}\bigl(5 H f''(R) \dot{R}
 + (f''(R) \ddot{R} + f'''(R) \dot{R}^2)\bigr]\bigr)f'(R)V^{0}V^{i}\eta^{0},
\end{multline}
en donde consideramos que el producto $g^{ij}g_{ij}=1$, pues es para una sola componente y usamos que $g_{00}=-1$. Usando de nuevo que $g_{00}\eta^{0}V^0 = - g_{ij}\eta^{i}V^{j}$ tendremos
\begin{multline}
\bigl(\mathcal{D}_{\delta\beta}+ g_{\beta\delta}\mathcal{D}_{i}^{\, \, i}\bigr)f'(R)V^{\beta}V^{\delta}\eta^{i} - \bigl(\mathcal{D}_{\gamma\beta}
+ g_{\beta\gamma}\mathcal{D}_{i}^{\, \,  i}\bigr)f'(R)V^{\beta}V^{i}\eta^{\gamma} = \\ g_{00}\bigl(5 H f''(R) \dot{R} + (f''(R) \ddot{R} + f'''(R) \dot{R}^2)\bigr]\bigr)f'(R)V^{0}V^{0}\eta^{i}\\ +g_{ij}\bigl(5 H f''(R) \dot{R}
 + (f''(R) \ddot{R} + f'''(R) \dot{R}^2)\bigr]\bigr)f'(R)V^{j}V^{i}\eta^{i},
\end{multline}
\begin{multline}
\bigl(\mathcal{D}_{\delta\beta}+ g_{\beta\delta}\mathcal{D}_{i}^{\, \, i}\bigr)f'(R)V^{\beta}V^{\delta}\eta^{i} - \bigl(\mathcal{D}_{\gamma\beta}
+ g_{\beta\gamma}\mathcal{D}_{i}^{\, \,  i}\bigr)f'(R)V^{\beta}V^{i}\eta^{\gamma} = \\ \bigl(5 H f''(R) \dot{R} + (f''(R) \ddot{R} + f'''(R) \dot{R}^2)\bigr]\bigr)f'(R)\bigl[g_{00}V^{0}V^{0} +g_{ij}V^{j}V^{i}\bigl]\eta^{i}.
\end{multline}
Con la expansión para  $V_{\alpha}V^{\alpha}=\epsilon$ obtenemos finalmente que la contribución para una componente $\eta^{i}$ es
\begin{multline}
\bigl(\mathcal{D}_{\delta\beta}+ g_{\beta\delta}\mathcal{D}_{i}^{\, \, i}\bigr)f'(R)V^{\beta}V^{\delta}\eta^{i} - \bigl(\mathcal{D}_{\gamma\beta}
+ g_{\beta\gamma}\mathcal{D}_{i}^{\, \,  i}\bigr)f'(R)V^{\beta}V^{i}\eta^{\gamma} = \\ \epsilon\bigl(5 H f''(R) \dot{R} + (f''(R) \ddot{R} + f'''(R) \dot{R}^2)\bigr]\bigr)f'(R)\eta^{i},
\end{multline}
Así tenemos entonces que la contribución es la misma considerando los casos por separado para $\alpha=0$ y $\alpha=i$, y que además, como era de esperarse, no existen términos cruzados (con cambio en la dirección del vector $\eta^{\alpha}$). Por lo tanto la contribución total de los operadores es
\begin{multline}\label{contribucionF}
\bigl(\delta_{\gamma}^{\alpha}\mathcal{D}_{\delta\beta} - \delta_{\delta}^{\alpha}\mathcal{D}_{\gamma\beta} + g_{\beta\delta}\mathcal{D}_{\gamma}^{\, \, \alpha}
- g_{\beta\gamma}\mathcal{D}_{\delta}^{\, \,  \alpha}\bigr)f'(R)V^{\beta}V^{\delta}\eta^{\gamma}\\ = \epsilon\bigl(5 H f''(R) \dot{R} + (f''(R) \ddot{R} + f'''(R) \dot{R}^2)\bigr]\bigr)f'(R)\eta^{\alpha}.
\end{multline}
\newpage

{\color{white} . }

\newpage

{\color{white} .}

\newpage

\thispagestyle{empty}%
    \noindent
    \begin{minipage}[t][5.5cm]{\textwidth}
    \begin{center}
        \hyphenpenalty=10000\LARGE\textbf{\expandafter{\@titulo}}
    \end{center}
    \ifsubtitu@lo
       \begin{center}
          \hyphenpenalty=10000\large\textbf{\uppercase\expandafter{\@subtitulo}}
       \end{center}
    \fi
    \vfill
    \begin{center}
        \Large\textsc{\expandafter{\@autor}}
    \end{center}
    \end{minipage}
    \vfill
    \iflogo@grav
        \logograv@include
        \vfill
    \fi
    \begin{center}
        \textsc{\expandafter{\@universidad}} \\
        \textsc{\expandafter{\@facultad}} \\
        \textsc{\expandafter{\@departamento}} \\
        \textsc{\expandafter{\@direccion}} \\
        \textsc{\expandafter{\@theyear}}\\
        $\copyright$
    \end{center}\newpage

\end{document}